\documentclass[aps,preprintnumbers,reprint,showkeys,eqsecnum,nofootinbib,floatfix,amsmath,asmfonts,latexsym,amssymb]{revtex4-1}
\usepackage{hyperref,graphics,slashed,color,multirow,float,amsfonts,bbold,mathtools}
\usepackage{booktabs,tabularx,comment}
\graphicspath{{sections/figures/}}
\pdfoutput=1
\allowdisplaybreaks

\begin{document}
\preprint{IP/BBSR/2021-XX}
\title{Revisiting Type-II see-saw: Present Limits and Future Prospects at LHC}
\author{Saiyad Ashanujjaman \href{https://orcid.org/0000-0001-5643-2652}{\includegraphics[scale=0.4]{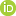}}}
\email[]{saiyad.a@iopb.res.in} 
\author{Kirtiman Ghosh}
\email[]{kirti.gh@gmail.com}
\affiliation{Institute of Physics, Bhubaneswar, Sachivalaya Marg, Sainik School, Bhubaneswar 751005, India}%                                                           
\affiliation{Homi Bhabha National Institute, Training School Complex, Anushakti Nagar, Mumbai 400094, India}%                                                                                           
\date{\today}

                                                                                                                 %=============================================================================
\begin{abstract}
%=============================================================================

\noindent The type-II see-saw mechanism based on the annexation of the Standard Model by weak gauge triplet scalar field proffers a natural explanation for the very minuteness of neutrino masses. Noting that the phenomenology for the non-degenerate triplet Higgs spectrum is substantially contrasting than that for the degenerate one, we perform a comprehensive study for an extensive model parameter space parametrised by the triplet scalar vacuum expectation value (VEV), the mass-splitting between the triplet-like doubly and singly charged scalars and the mass of the doubly charged scalar. Considering all Drell-Yan production mechanisms for the triplet-like scalars and taking into account the all-encompassing complexity of their decays, we derive the most stringent $95\%$ CL lower limits on the mass of the doubly charged scalar for a vast model parameter space by implementing already existing direct collider searches by CMS and ATLAS. These estimated limits are beyond those from the existing LHC searches by approximately 50--230 GeV. However, we also find that a specific region of the parameter space is not constrained by the LHC searches. Then, we forecast future limits by extending an ATLAS search at high-luminosity, and we propose a search strategy that yields improved limits for a part of the parameter space.
\end{abstract}
\maketitle

%=============================================================================
\section{\label{sec:intro}Introduction}
%=============================================================================
The Standard Model (SM) falls short at offering explanations of tiny neutrino masses and mixings. Although plausible, it seems philosophically displeasing that the tiny neutrino masses are effectuated via the usual Brout-Englert-Higgs mechanism as it entails extremely small Yukawa couplings causing hierarchy among them. Conversely, the widely-studied see-saw mechanisms seem to proffer a natural explanation for the very minuteness of neutrino masses. The type-II see-saw model based on the annexation of the SM by weak gauge triplet of scalar field \cite{Konetschny:1977bn,Cheng:1980qt,Lazarides:1980nt,Schechter:1980gr,Mohapatra:1980yp,Magg:1980ut} is one such variant. Yukawa interaction of the scalar triplet with the SM lepton doublet leads to neutrino masses after its neutral component procure a non-zero vacuum expectation value (VEV). The Yukawa coupling driving the leptonic decays of the non-standard scalars in the model pans out to be determined by the neutrino oscillation parameters up to the triplet VEV ($v_t$). Though {\it ad hoc}, this prognostic characteristic of the present scenario makes the same a tempting one beyond the SM (BSM). Not only this model holds out a riveting rationale for the minuscule neutrino masses, but it also put forward an elaborated electroweak symmetry breaking (EWSB) mechanism and well-to-do phenomenology at the Large Hadron Collider (LHC).

Copious production of the triplet-like scalars owing to their gauge interactions at the LHC caters an up-and-coming way to probe this model. Phenomenological outcome of this model at the LHC has been studied all-encompassingly in the literature, in particular, emphasising the doubly-charged scalars \cite{Huitu:1996su,Chakrabarti:1998qy,Chun:2003ej,Akeroyd:2005gt,Garayoa:2007fw,Kadastik:2007yd,Akeroyd:2007zv,Perez:2008ha,delAguila:2008cj,Akeroyd:2009hb,Melfo:2011nx,Aoki:2011pz,Akeroyd:2011zza,Chiang:2012dk,Chun:2012jw,Akeroyd:2012nd,Chun:2012zu,Dev:2013ff,Banerjee:2013hxa,delAguila:2013mia,Chun:2013vma,Kanemura:2013vxa,Kanemura:2014goa,Kanemura:2014ipa,kang:2014jia,Han:2015hba,Han:2015sca,Das:2016bir,Babu:2016rcr,Mitra:2016wpr,Cai:2017mow,Ghosh:2017pxl,Crivellin:2018ahj,Du:2018eaw,Dev:2018kpa,Antusch:2018svb,Aboubrahim:2018tpf,deMelo:2019asm,Primulando:2019evb,Padhan:2019jlc,Chun:2019hce}. A wealth of BSM models such as the present model \cite{Konetschny:1977bn,Cheng:1980qt,Lazarides:1980nt,Schechter:1980gr,Mohapatra:1980yp,Magg:1980ut}, left-right symmetric models \cite{Pati:1974yy,Mohapatra:1974hk,Senjanovic:1975rk} Higgs triplet models \cite{Gunion:1989ci,CiezaMontalvo:2006zt}, little Higgs model \cite{ArkaniHamed:2002qx,ArkaniHamed:2002qy,Hektor:2007uu}, Georgi-Machacek model \cite{Chanowitz:1985ug,Georgi:1985nv}, Zee-Babu model \cite{Zee:1985id,Babu:1988ki} and other extensions of SM \cite{Babu:2009aq,Picek:2009is,Kumericki:2012bh,Cepedello:2017lyo,Anamiati:2018cuq,Avnish:2020rhx} envisage presence of doubly charged scalar bosons and their illustrious signatures. This is why, a number of collider searches have been carried out at the LHC by CMS and ATLAS \cite{ATLAS:2012hi,Chatrchyan:2012ya,ATLAS:2014kca,Khachatryan:2014sta,CMS:2016cpz,CMS:2017pet,Aaboud:2017qph,CMS:2017fhs,Aaboud:2018qcu,Aad:2021lzu} to look for the same. In view of the observations being consistent with the SM  background expectations, these analyses derived stringent limits with 95\% confidence level (CL) on the mass of the doubly charged scalar in the context of a simplified model.

In addition to the doubly charged scalars ($H^{\pm \pm}$), the present model contains several other triplet-like physical scalars, namely the singly charged scalars ($H^\pm$) and CP-even and CP-odd neutral scalars ($H^0$ and $A^0$). Phenomenology of this model, by and large, is governed by three parameters only --- $m_{H^{\pm \pm}}$, $\Delta m = m_{H^{\pm \pm}} - m_{H^\pm}$ and $v_t$ (see section \ref{sec:model}). For degenerate scenario ($\Delta m=0$), $H^{\pm \pm}$ decays to same-sign dilepton for $v_t < 10^{-4}$ GeV and to same-sign $W$-boson for $v_t > 10^{-4}$ GeV. 
%For $H^{\pm \pm}$ decaying 100\% into same-sign lepton ($e,\mu, \tau$) pair, the CMS collaboration \cite{CMS:2017pet} has excluded them with mass below 535--820 GeV. Similarly, the ATLAS collaboration \cite{Aaboud:2017qph} has derived a limit of 770--870 ($\gtrsim$450) GeV for 100\% ($\gtrsim$10\%) decay into light lepton ($e,\mu$) pair. For $H^{\pm \pm}$ decaying 100\% into same-sign $W$-boson, the ATLAS collaboration \cite{Aad:2021lzu} has excluded $H^{\pm \pm}$ with masses up to 350 GeV and 230 GeV, respectively, for the pair and associated production modes. 
For $H^{\pm \pm}$ decaying 100\% into same-sign dilepton, a search in three and four lepton final states with an integrated luminosity of 12.9 fb$^{-1}$ of $pp$ collisions at $\sqrt{s}=13$ TeV LHC by the CMS collaboration \cite{CMS:2017pet} has excluded them with mass below 716--761 GeV considering four benchmark points targeting four possible neutrino mass hypotheses. In addition, considering 100\% decay of $H^{\pm \pm}$ into lepton ($e,\mu,\tau$) pair, the same search has set a limit of 535--820 GeV. Another search in multilepton final states with an integrated luminosity of 36.1 fb$^{-1}$ of $pp$ collisions at $\sqrt{s}=13$ TeV LHC by the ATLAS collaboration \cite{Aaboud:2017qph} has set a limit of 770--870 GeV and 450 GeV for $H^{\pm \pm}$ decaying, respectively, 100\% and 10\% into same-sign light lepton ($e,\mu$) pair. A recent search in multilepton final states, optimised for $H^{\pm \pm}$ decaying exclusively into same-sign $W$-boson pair, with an integrated luminosity of 139 fb$^{-1}$ of $pp$ collisions at $\sqrt{s}=13$ TeV LHC by the ATLAS collaboration \cite{Aad:2021lzu} has excluded them with masses up to 350 GeV and 230 GeV, respectively, for the pair and associated production modes assuming $v_t=0.1$ GeV and the mixing between the CP-even scalars to be $10^{-4}$.

Patently, the above-cited limits are not befitting to the entire parameter space, rather valid only for a constrained parameter space of the model. For instance, the CMS search in \cite{CMS:2017pet} is only valid for $\Delta m=0$ and $v_t < 10^{-4}$ GeV, whereas the ATLAS search in ref \cite{Aad:2021lzu} is only valid for $\Delta m=0$ and $v_t > 10^{-4}$ GeV. Though in a realistic type-II see-saw scenario, the branching fractions of the triplet-like scalars into different lepton flavours are dictated by the neutrino oscillation parameters, most of the aforecited limits are derived in the context of simplified scenarios without reckoning the footprints of the low-energy neutrino parameters. Furthermore, these limits are often conservative as these searches do not incorporate all the production channels for the triplet-like scalars. For instance, the ATLAS search in \cite{Aad:2021lzu} considered either pair or associated production modes for the doubly charged scalars, but not both at once. As argued in Section~\ref{sec:prod}, all the channels entail to be incorporated in the analyses. Moreover, the triplet components in this model are conceivably non-degenerate in mass. For moderate $v_t$ and passably large $\Delta m$, cascade decays quickly dominate over the leptonic and diboson decay modes, see Section~\ref{sec:dec}. Thus, in the non-degenerate scenario ($\Delta m \neq 0$), the cascade decays are entitled to play a notable role in the phenomenology, thereby making the phenomenology for the non-degenerate scenario substantially contrasting than that for the degenerate one.

Bearing the aforesaid discussion in mind, we perform a systematic and comprehensive collider study of this model. Incorporating all the Drell-Yan production modes for the triplet-like scalars and taking into account the all-encompassing complexity of their decays, we derive the most stringent $95\%$ CL lower limit on $m_{H^{\pm \pm}}$ for a wide range of $v_t$ and $\Delta m$ by implementing already existing direct collider searches by CMS and ATLAS. Then, we forecast future limits on $m_{H^{\pm \pm}}$ by extending the ATLAS search at high-luminsity, and we propose a search strategy that yields improved limits on $m_{H^{\pm \pm}}$ for a part of the parameter space of $v_t$ and $\Delta m$.

The rest of this work is structured as follows. In Section~\ref{sec:model}, we briefly describe the theoretical structure of the type-II see-saw model. Production of the triplet-like scalars and their decays are discussed in Section~\ref{sec:prod} and \ref{sec:dec}, respectively. In Section~\ref{sec:collider}, we discuss the LHC phenomenology of this model and obtain stringent limits on $m_{H^{\pm \pm}}$ for a wide region of parameter space.

%=============================================================================
\section{\label{sec:model} The Type-II see-saw Model}
%============================================================================= 
The scalar sector of the minimal type-II see-saw model employs a $SU(2)_L$ triplet scalar field with hypercharge 2, $\Delta$ in addition to the SM Higgs doublet, $\Phi$:
\begin{eqnarray*}
\Delta &=\left( \begin{array}{cc}
\Delta^+/\sqrt{2} & \Delta^{++} \\
\Delta^0 & -\Delta^+/\sqrt{2} \\
\end{array} \right) ~~~~{\rm and}~~~~ 
\Phi=\left(\begin{array}{c} \Phi^+ \\ \Phi^0 \\ \end{array} \right)~.
\end{eqnarray*}
The most general renormalizable and $SU(3)_C \times SU(2)_L \times U(1)_Y$ invariant 
%Lagrangian of this sector is given by
%\begin{eqnarray*}
%\mathcal{L} &\supset &
%(D_\mu{\Phi})^\dagger(D^\mu{\Phi})+{\rm Tr}(D_\mu{\Delta})^\dagger(D^\mu{\Delta}) -V(\Phi, \Delta)~,
%\end{eqnarray*}
%where $D_\mu{\Phi} = \partial_\mu{\Phi}+ig\tau^a{W}^a_\mu{\Phi}+i\frac{g^\prime}{2}B_\mu{\Phi}$ and $D_\mu{\Delta} =\partial_\mu{\Delta}+ig[\tau^a{W}^a_\mu,\Delta]+ig^\prime B_\mu{\Delta}$ are the gauge covariant derivatives with $W^a_\mu$ and $B_\mu$ being the gauge bosons of $SU(2)_L$ and $U(1)_Y$, respectively. $g$ and $g^\prime$, respectively, are the $SU(2)_L$ and $U(1)_Y$ gauge couplings. $\tau^a=\sigma^a/2$, where $\sigma^a$ are the Pauli matrices, are the generators of fundamental representation of $SU(2)_L$. The 
scalar potential invloving $\Phi$ and $\Delta$ is given by
\begin{eqnarray*}
V(\Phi, \Delta) &=& -m_\Phi^2{\Phi^\dagger \Phi}+\frac{\lambda}{4}(\Phi^\dagger \Phi)^2+m_\Delta^2{\rm Tr}(\Delta^{\dagger}{\Delta}) +
\\
&& [\mu(\Phi^T{i}\sigma^2\Delta^\dagger \Phi)+{\rm h.c.}] 
+ \lambda_1(\Phi^\dagger \Phi){\rm Tr}(\Delta^{\dagger}{\Delta}) +
\\
&& \lambda_2[{\rm Tr}(\Delta^{\dagger}{\Delta})]^2
+ \lambda_3[{\rm Tr}(\Delta^{\dagger}{\Delta})^2] 
+ \lambda_4{\Phi^\dagger \Delta \Delta^\dagger \Phi} ~,
\end{eqnarray*}
where $m_\Phi^2, m_\Delta^2$ and $\mu$ are the mass parameters, $\lambda$ and $\lambda_i$ ($i=1,...,4$) are the independent dimensionless couplings. %All these parameters can be assumed to be real without loss of generality. 
After the EWSB, $\Phi$ and $\Delta$ acquire the following VEVs
\begin{eqnarray*}
\langle \Delta \rangle &=\left( \begin{array}{cc}
0 & 0 \\ v_t/\sqrt{2} & 0\\
\end{array} \right) ~~~~{\rm and}~~~~
\langle \Phi \rangle =\left(\begin{array}{c} 0 \\ v_d/\sqrt{2}\\ \end{array} \right)~.
\end{eqnarray*}

\noindent Minimization of the scalar potential $V(\Phi, \Delta)$ allows to enunciate $m_\Delta^2$ and $m_\Phi^2$ in terms of the VEVs and the quartic and trilinear couplings as
\begin{eqnarray*}
m_\Delta^2 &=& \frac{\mu v_d^2}{\sqrt{2}v_t} -\frac{(\lambda_1+ \lambda_4)}{2}v_d^2 -(\lambda_2+\lambda_3)v_t^2~, \\
m_\Phi^2 &=& \frac{\lambda{v_d^2}}{4}-\sqrt{2}\mu{v_t}+\frac{(\lambda_1+\lambda_4)}{2}v_t^2~.
\end{eqnarray*}

\noindent For $v_d^2 \gg v_t^2 $, the first tadpole equation shortens to
\begin{equation}
v_t \approx \frac{\mu v_d^2}{\sqrt{2}m_\Delta^2}~.
\end{equation}
This relation withstands the so-called see-saw spirit. For small $\mu$, which is justified by the 't Hooft's naturalness criterion, the triplet mass scale is naturally connected to the physics at the TeV scale (unlike in the original see-saw scenario where the new physics is naturally motivated to be at very large scale), thereby making the present model potentially testable, and hence falsifiable at the LHC.

The neutral components of $\Phi$ and $\Delta$ can be parametrised as $\Phi^0=\frac{1}{\sqrt{2}} (v_d+h+iZ_1)$ and $\Delta^0=\frac{1}{\sqrt{2}} (v_t+\xi+iZ_2)$. %,where $h$ and $\xi$ are the real parts and $Z_1$ and $Z_2$ are the imaginary parts of the fields $\Phi^0$ and $\Delta^0$ shifted by their respective VEVs. 
After the EWSB, the mixing among the scalar fields lead to several Higgs bosons. The gauge eingenstates can be rotated to obtain the mass eigenstates as in the following:
\begin{eqnarray*}
&&\left( \begin{array}{c} \Phi^\pm \\ \Delta^\pm \end{array} \right) = R(\beta_\pm) \left( \begin{array}{c} H^\pm \\ G^\pm \end{array} \right) ~,~ \left( \begin{array}{c} h \\ \xi \end{array} \right) = R(\alpha) \left( \begin{array}{c} h^0 \\ H^0 \end{array} \right)~,
\\
&&\left( \begin{array}{c} Z_1 \\ Z_2 \end{array} \right) = R(\beta_0) \left( \begin{array}{c} A^0 \\ G^0 \end{array} \right) ~,~ R(\theta) =\left( \begin{array}{cc} \cos\theta & -\sin\theta \\ \sin\theta & \cos\theta \\ \end{array} \right)~,
\end{eqnarray*}

\noindent where $\beta_\pm$, $\alpha$ and $\beta_0$ are the rotation angles in the singly-charged, CP-even and CP-odd Higgs sectors with $\sqrt{2} \tan\beta_\pm = \tan\beta_0 = \frac{2v_t}{v_d}$ and $\tan2\alpha = \frac{2 B}{ A - C}$, where $A=\frac{\lambda}{2}{v_d^2}, B=-\sqrt{2}\mu v_d+(\lambda_1+\lambda_4)v_dv_t$ and $C=\frac{\mu v_d^2}{\sqrt{2} v_t} + 2(\lambda_2+\lambda_3)v_t^2$~. $G^\pm$ and $G^0$ are the Nambu-Goldstone bosons, respectively, eaten by the longitudinal modes of $W^\pm$ and $Z$. $H^\pm$ is the singly-charged Higgs boson with mass $m_{H^\pm}$:
\begin{eqnarray*}
m_{H^\pm}^2=\frac{(v_d^2+2 v_t^2)(2\sqrt{2}\mu- \lambda_4 v_t)}{4v_t}~.
\end{eqnarray*}

\noindent $h^0$ and $H^0$ are the CP-even neutral Higgs with masses $m_{h^0}$ and $m_{H^0}$, respectively \cite{Arhrib:2011uy}:
\begin{eqnarray*}
m_{h^0}^2&=&\frac{1}{2}[A+C-\sqrt{(A-C)^2+4B^2}]~, \\
m_{H^0}^2&=&\frac{1}{2}[A+C+\sqrt{(A-C)^2+4B^2}]~.
\end{eqnarray*}
We discern $h^0$ as the 125-GeV resonance observed at the LHC. Though $h^0$ is not necessarily the lighter of these two eigenstates, we presume it to be the lighter one. $A^0$ is the CP-odd neutral Higgs with mass $m_{A^0}$:
\begin{eqnarray*}
 m_{A^0}^2 &=& \frac{\mu(v_d^2+4v_t^2)}{\sqrt{2}v_t}~.
\end{eqnarray*}

\noindent Finally, the doubly-charged Higgs, $\Delta^{\pm \pm}(\equiv H^{\pm \pm})$, has mass $m_{H^{\pm \pm}}$:
\begin{eqnarray*}
m_{H^{\pm\pm}}^2=\frac{\sqrt{2}\mu{v_d^2}- \lambda_4v_d^2v_t-2\lambda_3v_t^3}{2v_t}~. 
\end{eqnarray*}

\noindent As mentoined above, $m_\Delta^2$ and $m_\Phi^2$ can be enunciated in terms of the other scalar potential parameters, namely $\lambda, \lambda_i$'s, $\mu, v_t$ and $v_d$. These parameters, in turn, can be framed in terms of the physical masses ($m_{h^0}, m_{H^0}, m_{A^0}, m_{H^\pm}, m_{H^{\pm \pm}}$), the VEVs ($v_t,v_d$) and the rotation angle of the CP-even Higgs sector ($\alpha$) as \cite{Arhrib:2011uy}
\begin{eqnarray*}
\lambda_1 &=& \frac{-2}{v_d^2+4v_t^2}m_{A^0}^2+\frac{4}{v_d^2+2v_t^2}m_{H^\pm}^2+\frac{\sin2\alpha}{2v_d{v_t}}(m_{h^0}^2-m_{H^0}^2)~,
\\
\lambda_2&=&\frac{1}{v_t^2}\Big\{\frac{\sin^2 \alpha ~m_{h^0}^2+ \cos^2 \alpha~ m_{H^0}^2}{2}+\frac{1}{2}\frac{v_d^2}{v_d^2+4v_t^2}m_{A^0}^2 
\\
&&-\frac{2v_d^2}{v_d^2+2v_t^2}m_{H^\pm}^2+m_{H^{\pm\pm}}^2\Big\}~,
\\
\lambda_3 &=& \frac{1}{v_t^2}\Big\{\frac{-v_d^2}{v_d^2+4v_t^2}m_{A^0}^2+\frac{2v_d^2}{v_d^2+2v_t^2}m_{H^\pm}^2-m_{H^{\pm\pm}}^2\Big\}~,
\\
\lambda_4 &=& \frac{4}{v_d^2+4v_t^2}m_{A^0}^2-\frac{4}{v_d^2+2v_t^2}m_{H^\pm}^2~,
\\
\lambda &=& \frac{2}{v_d^2}\Big\{\cos^2 \alpha~ m_{h^0}^2+\sin^2 \alpha ~m_{H^0}^2\Big\} ~,
\\
\mu &=& \frac{\sqrt{2} v_t}{v_d^2+4v_t^2}m_{A^0}^2~.
\end{eqnarray*}

\noindent For $v_t \ll v_d$, the masses of the physical Higgs states reduce to
\begin{eqnarray*}
&&m_{H^{\pm \pm}}^2 \simeq m_\Delta^2 -\frac{\lambda_4}{2} v_d^2 ~,~
m_{H^\pm}^2 \simeq m_\Delta^2 -\frac{\lambda_4}{4} v_d^2 ~,~ \\
&& m_{h^0}^2 \simeq 2v_d^2\lambda ~,~ \text{and~} m_{H^0}^2 \approx m_{A^0}^2 \simeq m_\Delta^2 ~;
\end{eqnarray*}
and their mass-squared differences are given by 
\begin{equation}
m_{H^{\pm\pm}}^2 - m_{H^\pm}^2 \approx m_{H^\pm}^2 - m_{H^0/A^0}^2 \approx \frac{\lambda_4}{4} v_d^2~.
\end{equation}

\noindent For usefulness, we define the mass-splitting between $H^{\pm \pm}$ and $H^\pm$ as $\Delta m = m_{H^{\pm \pm}}-m_{H^\pm}$~. Thereby, the masses of all the physical Higgs states can be traded in terms of just two parameters --- $m_{H^{\pm \pm}}$ and $\Delta m$. In addition to the tree-level mass-splitting dictated by the dimensionless quartic coupling $\lambda_4$, radiative corrections, dominantly driven by the electroweak gauge bosons at one loop, induce mass-splittings among differently charged triplet scalars: $m_{H^{\pm \pm}}-m_{H^\pm} \sim 885$ MeV and $m_{H^\pm}-m_{H^0/A^0} \sim 540$ MeV \cite{Cirelli:2005uq}. These radiative mass-splittings exclusively are small enough to have considerable effects on the decays of the triplet-like scalars and thereby neglected in the rest of this work \cite{Perez:2008ha}.
Depending on the value (sign) of $\lambda_4$, and hence $\Delta m$, three characteristic mass spectra are predicted --- (i) $\lambda_4 = 0$: $m_{H^{\pm \pm}} \simeq m_{H^\pm} \simeq m_{H^0/A^0}$, (ii) $\lambda_4 > 0$: $m_{H^{\pm \pm}} > m_{H^\pm} > m_{H^0/A^0}$ and (iii) $\lambda_4 < 0$: $m_{H^{\pm \pm}} < m_{H^\pm} < m_{H^0/A^0}$. These characteristic mass spectra will be sometimes referred to as degenerate, positive and negative scenario, respectively. In what follows, we briefly discuss the relevant constraints on $v_t,\Delta m$ and $\alpha$.\\

\noindent {\bf Constraint on $v_t$ from $\rho$ parameter:} Both the doublet and triplet VEVs contribute to the weak gauge boson's masses at tree level:
\begin{equation*}
m_W^2=\frac{1}{4} g^2(v_d^2 + 2 v_t^2) {\rm ~and~} m_Z^2=\frac{1}{4} (g^2+g^{\prime 2})(v_d^2 + 4 v_t^2)~.
\end{equation*}
The VEVs are subject to the constraint $\sqrt{v_d^2 + 2 v_t^2} \equiv v_{\rm SM} = 246$ GeV. The $\rho$ parameter, defined as $\rho = m_W^2/(m_Z^2\cos^2\theta_w)$, takes the form 
\begin{equation*}
\rho = (v_d^2+2v_t^2)/(v_d^2+4v_t^2)~.
\end{equation*} 
The value of the $\rho$ parameter from the electroweak precision data, $\rho=1.00038(20)$ \cite{Zyla:2020zbs}, which is $1.9\sigma$ above the SM expectation at tree level, $\rho(v_t=0)=1$, leads to an upper bound of $\mathcal{O}(1)$ GeV on $v_t$ ($\ll v_d$).\\

\noindent {\bf Constraint on $\Delta m$ from oblique parameters:} As mentioned earlier, the mass-splitting, $\Delta m = m_{H^{\pm \pm}}-m_{H^\pm}$ affects the electroweak precision data observables, also known as the oblique parameters, namely $S,T$ and $U$ parameters. These parameters robustly constrain the mass-splitting $\Delta m$ to be within $|\Delta m| \lesssim 40$ GeV \cite{Aoki:2012jj,Chun:2012jw,Primulando:2019evb,Das:2016bir}.\\

\noindent {\bf Constraint on $\alpha$ from Higgs data:} The rotation angle in the CP-even Higgs sector ($\alpha$) is given by
\begin{equation*}
\tan 2\alpha = \frac{2B}{A-C} = \frac{2\sqrt{2}\mu v_d}{2m_\Delta^2-m_{H^0}^2-m_{h^0}^2+4(\lambda_2+\lambda_3)v_t^2}~.
\end{equation*}
For $v_t \ll v_d$, this reduces to
\begin{equation}
\tan 2\alpha \approx \frac{4v_t}{v_d} \left(1-\frac{m_{h^0}^2}{m_{H^0}^2} \right)^{-1}
\label{eq:tanAlpha}
\end{equation}
which further reduces to $\tan \alpha \sim \frac{2v_t}{v_d}$ for $m_{h^0}^2 \ll m_{H^0}^2$. The 125-GeV Higgs data, in particular, the Higgs-to-diphoton decay rate at the LHC constrains $\alpha$: $|\sin\alpha| \lesssim 0.3$ at 95\% CL \cite{Primulando:2019evb}. This bound is consistent with the above expression of $\tan 2\alpha$ for the allowed values of $v_t$.\\

\noindent {\bf Neutrino masses:} The Yukawa interaction which generates neutrino masses is given by
\begin{equation}
-\mathcal{L}_\nu = Y^{\nu}_{ij} L^T_i C i \sigma^2 \Delta L_j  + {\rm h.c.}~,
\end{equation}
where $Y^\nu$ is a $3\times 3$ symmetric complex matrix, $i$ and $j$ are the generation indices ($i,j=1,2,3$), $L=\left(\nu_L, \ell_L \right)^T$ is the left-handed SM lepton doublet and $C$ is the charge-conjugation matrix. After the scalar triplet acquires a VEV, the following neutrino mass matrix is procured
\begin{equation*}
m_\nu=\sqrt{2}Y^\nu v_t~.
\end{equation*}

\noindent Further, $m_\nu$ can be diagonalised using the Pontecorvo-Maki-Nakagawa-Sakata matrix $U$ as $U^T m_\nu U=D_{m_\nu}={\rm diag}(m_1,m_2,m_3)$. $U$ is usually parametrised by three mixing angles ($\theta_{12},\theta_{23}$ and $\theta_{13}$), one dirac phase ($\delta$) and two Majorana phases ($\phi$ and $\phi^\prime$) with $U\times \mathrm{diag}(e^{i \phi/2},e^{i \phi^\prime/2},1)~=$
\begin{equation*}
\left[ \begin{array}{ccc}
1 & 0 & 0 \\
0 & c^{23} & s^{23} \\
0 & -s^{23} & c^{23} 
\end{array} \right] 
\left[\begin{array}{ccc}
c^{13} & 0 & s^{13}e^{-i\delta}\\
0 & 1 & 0 \\
-s^{13}e^{i\delta} & 0 & c^{13} 
\end{array} \right]
\left[ \begin{array}{ccc}
c^{12} & s^{12} & 0 \\
-s^{12} & c^{12} & 0 \\
0 & 0 & 1
\end{array} \right],
\end{equation*}

\noindent where $c^{ij}\left(s^{ij}\right)=\cos\theta_{ij}\left(\sin\theta_{ij}\right)$. We simply set $\delta, \phi$ and $\phi^\prime$ to be zero throughout this work unless stated otherwise as they are either poorly measured or hitherto not measured at all. For normal hierarchy (NH), $m_2=\sqrt{\Delta m_{21}^2+m_1^2}$, $m_3=\sqrt{\Delta m_{31}^2+m_1^2}$, and for inverted hierarchy (IH), $m_1=\sqrt{|\Delta m_{32}^2+\Delta m_{21}^2-m_3^2|}$, $m_2=\sqrt{|\Delta m_{32}^2-m_3^2|}$, where $\Delta m_{ij}^2$'s are the experimentally measured mass-squared differences, and $m_1(m_3)$ is the lightest neutrino mass for NH(IH). Measurements of large scale structure in the universe by the Planck satellite has put a bound $\sum_i m_i < 0.12$ eV when combined with baryon acoustic oscillation data \cite{Aghanim:2018eyx}. The best fit values for the neutrino oscillation parameters used in this work are taken from Ref.~\cite{Esteban:2020cvm}.\\

\noindent {\bf Constraint from lepton flavour violating decays:} The Yukawa interaction of the scalar triplet with the SM lepton doublet leads to lepton flavour violating decays such as $\ell_\alpha \to \ell_\beta \gamma$ at 1-loop and $\ell_\alpha \to \ell_\beta \ell_\gamma \ell_\delta$ at tree level. The relevant branching fractions are given by \cite{Kakizaki:2003jk,Akeroyd:2009nu,Dinh:2012bp}
\begin{eqnarray*}
&& {\rm Br}(\mu^- \to e^- \gamma) = \frac{\alpha|(Y^{\nu \dagger} Y^\nu)_{e\mu}|^2}{192\pi G_F^2} \left(\frac{1}{m_{H^\pm}^2}+\frac{8}{m_{H^{\pm \pm}}^2} \right)^2~,
\\
&& {\rm Br}(\mu^- \to e^+e^-e^-) = \frac{|(Y^\nu)^\dagger_{ee} Y^\nu_{\mu e}|^2}{4G_F^2 m_{H^{\pm \pm}}^4}~,
\end{eqnarray*}
where $\alpha$ is the electromagnetic fine-structure constant and $G_F$ is the Fermi constant. The upper limits of $4.2\times 10^{-13}$ \cite{TheMEG:2016wtm} and $1.0\times 10^{-12}$ \cite{Bellgardt:1987du}, respectively, on ${\rm Br}(\mu^- \to e^- \gamma)$ and ${\rm Br}(\mu^- \to e^+e^-e^-)$ tightly constrain the $v_t-m_{H^{\pm \pm}}$ parameter space. These limits translates into a lower limit of
\begin{equation*}
v_t \gtrsim 0.78\text{--}1.5 (0.69\text{--}2.3) {\rm ~GeV} \times 10^{-9} \times \frac{1~ \rm TeV}{m_{H^{\pm \pm}}}
\end{equation*}
for NH(IH) with $m_i$'s consistent with the bound from cosmology.\\

\noindent {\bf Phenomenologically relevant parameters:} Before concluding this section, we briefly reckon the phenomenologically relevant parameters. As mentioned earlier, the Yukawa couplings are determined by the neutrino oscillation parameters up to $v_t$ \footnote{Some of the neutrino oscillation parameters, namely the lightest neutrino mass and the CP phases, are either poorly measured or hitherto not measured at all. In this work, we set the phases to be zero for simplicity. However, note that these parameters could substantially change the leptonic decays and thereby the phenomenology of the charged scalars.}, and all the scalar potential parameters can be framed in terms of the masses of the physical Higgs states, the VEVs and the mixing angle between the CP-even scalars. The mixing angle is determined in terms of the others by Eq.~\eqref{eq:tanAlpha}. Moreover, the masses can be traded in terms of just two parameters --- $m_{H^{\pm \pm}}$ and $\Delta m$. Therefore, the phenomenology of this model, by and large, is governed by three parameters only --- $m_{H^{\pm \pm}}$, $\Delta m$ and $v_t$.

%=============================================================================
\section{\label{sec:prod}Production of Triplet Scalars}
%============================================================================= 
The TeV scale triplet-like scalar bosons are pair produced copiously at the LHC by quark-antiquark annihilation via the neutral current and charged current Drell-Yan mechanism, respectively, through $s$-channel $\gamma/Z$ and $W^\pm$ exchanges \footnote{Also, the triplet-like scalars are produced via $t/u$-channel photon-photon fusion process \cite{Babu:2016rcr,Ghosh:2017jbw} and vector boson-fusion process \cite{Huitu:1996su,delAguila:2013mia,Dutta:2014dba} with two associated forward jets at the LHC. However, their production through such processes is sub-dominant for the mass range of our interest, and thereby neglected.}:
\begin{eqnarray*}
&& q\bar{q^\prime} \to W^* \to H^{\pm \pm} H^\mp, H^\pm S^0 {\rm ~with~} S^0 \ni \{H^0, A^0\},
\\
&& q\bar{q} \to \gamma^*/Z^* \to H^{\pm \pm} H^{\mp \mp}, H^\pm H^\mp, H^0 A^0.
\end{eqnarray*}
We implement the model in SARAH \cite{Staub:2013tta,Staub:2015kfa} to generate UFO modules, and use MadGraph5aMC@NLO \cite{Alwall:2011uj,Alwall:2014hca} with the  NNPDF23$\_$lo$\_$as$\_$0130$\_$qed parton distribution function \cite{Ball:2013hta,Ball:2014uwa} for numerical evaluation of the leading order (LO) production cross-sections of the triplet scalars at the 13 TeV LHC.

\begin{widetext}

\begin{figure}[htb!]
\centering
\includegraphics[scale=0.61]{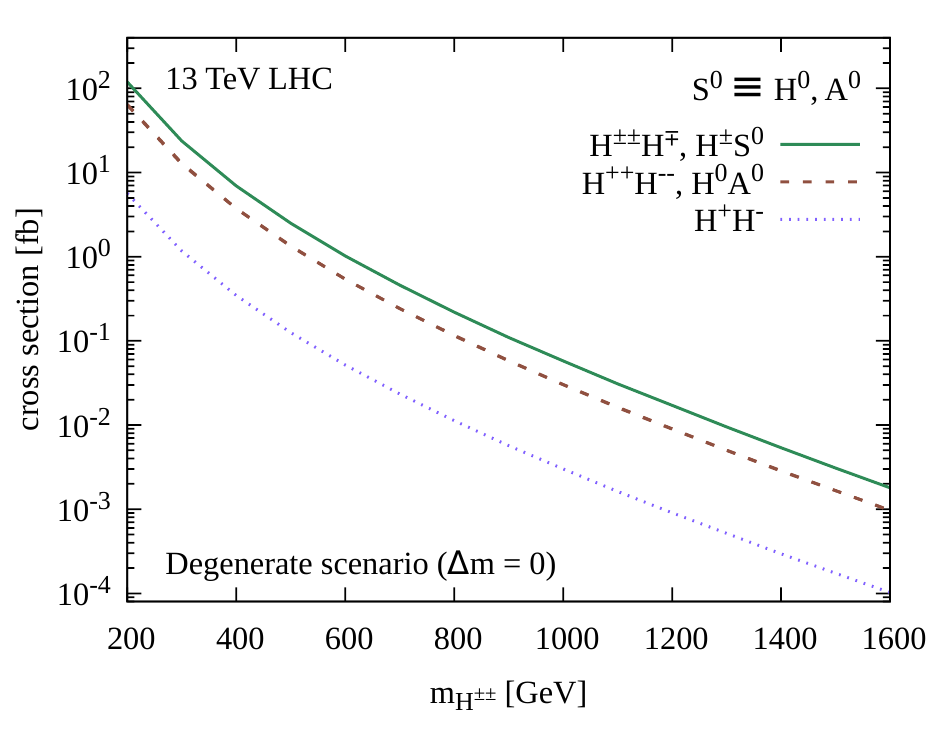}
\includegraphics[scale=0.61]{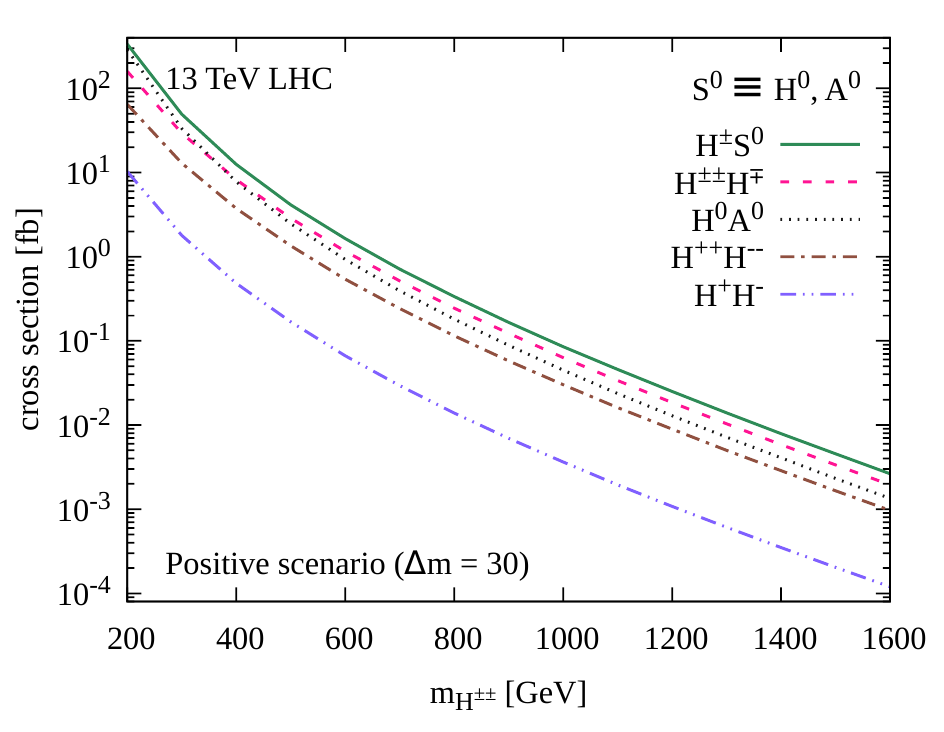}
\includegraphics[scale=0.61]{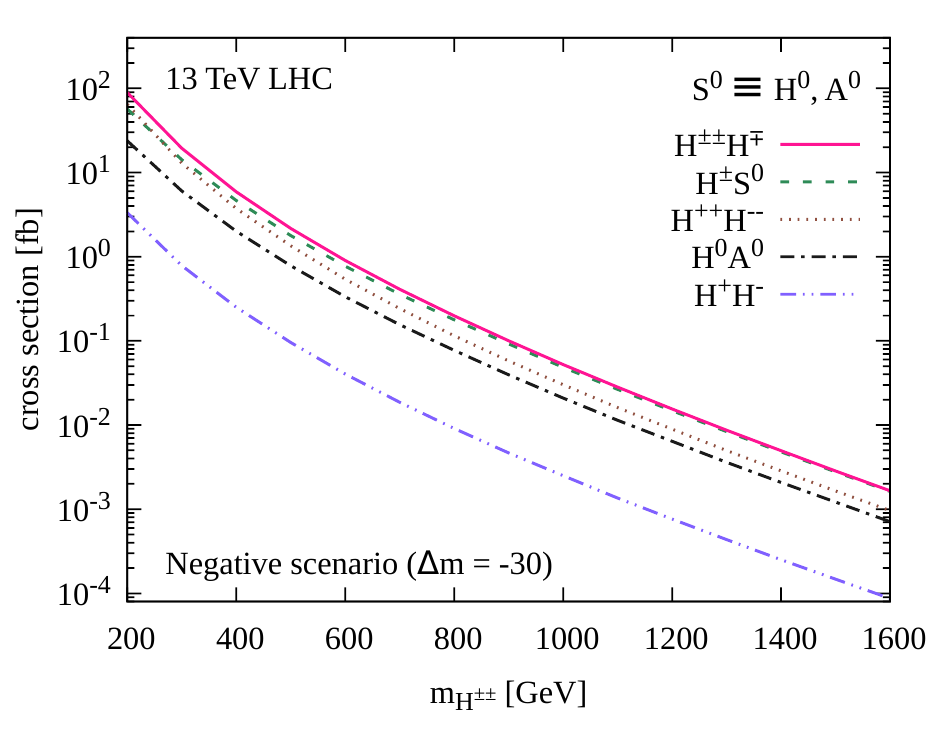}
\caption{Different pair and associated production cross-sections of the triplet-like scalars for degenerate scenario (left), positive scenario with $\Delta m = 30$ GeV (middle) and negative scenario with $\Delta m = -30$ GeV (right).}
\label{cs}
\end{figure}

\end{widetext}

The left, middle and right plots in Figure~\ref{cs} shows different pair and associated production cross-sections at LO as a function of the doubly charged scalar mass $m_{H^{\pm \pm}}$, respectively, for degenerate scenario ({\it i.e.} $\Delta m =0$), positive scenario with $\Delta m = 30$ GeV and negative scenario with $\Delta m = -30$ GeV. All the Drell-Yan production mechanisms are of handsome cross-sections; in particular, production of the doubly charged scalars in association with the singly charged ones, which is sometimes precluded by experimental searches, has the largest cross-section for both degenerate and negative scenarios. Production of the singly charged scalars in association with the neutral ones, which is also forsaken by both CMS and ATLAS, has the largest cross-section for both degenerate and positive scenarios. This substantiates that all the channels entail to be incorporated into the analyses.

Refs.~\cite{Muhlleitner:2003me,Fuks:2019clu} have estimated the QCD corrections to the pair and associated productions of the doubly charged scalars at hadron colliders which result in a next-to-leading (NLO) $K$-factor of 1.2--1.3. Considering that the QCD corrections to the pair and associated productions of the singly charged scalars are, in principle, similar to those of the doubly charged ones, we apply an overall QCD $K$-factor of $1.25$ to the LO cross-section.

\begin{comment}
\begin{figure}[htb!]
\centering
\includegraphics[width=0.9\columnwidth]{cs_DY}
\caption{Total production cross-sections of the triplet-like scalars for different values of $\Delta m$ (in GeV).}
\label{cs_DY}
\end{figure}

Figure~\ref{cs_DY} shows total production cross-sections (LO cross-sections multiplied by a $K$-factor of 1.25) of the triplet-like scalars as a function of $m_{H^{\pm \pm}}$ for different values of the mass-splitting, $\Delta m$: $\Delta m = 0, \pm 10$ and $\pm 30$ GeV. The more negative(positive) the mass-splitting, the more(less) the total production cross-section.
\end{comment}

%=============================================================================
\section{\label{sec:dec} Decays of the Scalar Bosons}
%============================================================================= 
In this section, we study the decays of the triplet-like physical scalar bosons, namely doubly-charged scalars ($H^{\pm \pm}$), singly-charged scalars ($H^\pm$), CP-even and CP-odd neutral scalars ($H^0$ and $A^0$). The partial decay widths of the scalars can be found in the literature \cite{Rizzo:1980gz,Keung:1984hn,Cahn:1990xc,Djouadi:1997rp,Perez:2008ha,Aoki:2011pz}. However, we found few typos in some of those analytical expressions; %\cite{Melfo:2011nx,Aoki:2011pz} 
for the sake of completeness, we provide the same in Appendix~\ref{sec:app}.

%=============================================================================
\subsection{\label{sec:decHpp} Decays of doubly charged scalar bosons}
%============================================================================= 
The doubly charged scalar bosons have three possible decay modes --- (i) leptonic decay, {\it i.e.} $\ell^\pm \ell^\pm$, (ii) gauge boson decay, {\it i.e.} $W^\pm W^\pm$ and (iii) cascade decay, {\it i.e.} $H^\pm W^{\pm *}$. The latter decay mode kinematically opens up only for $\Delta m > 0$. For $m_{H^{\pm \pm}}^2 \gg m_W^2$, the ratio of the braching fractions for these decay modes can be estimated as (see Appendix~\ref{sec:app})
\begin{equation*}
\frac{1}{4} \left(\frac{m_\nu}{m_{H^{\pm \pm}}} \right)^2 \left(\frac{v_d}{v_t}\right)^2 : \left(\frac{v_t}{v_d}\right)^2 : \frac{12}{5\pi^2} \frac{{\rm max}(\Delta m,0)^5}{v_d^2 m_{H^{\pm \pm}}^3}~.
\end{equation*}

\begin{widetext}

\begin{figure}[htb!]
\centering
\includegraphics[scale=0.44]{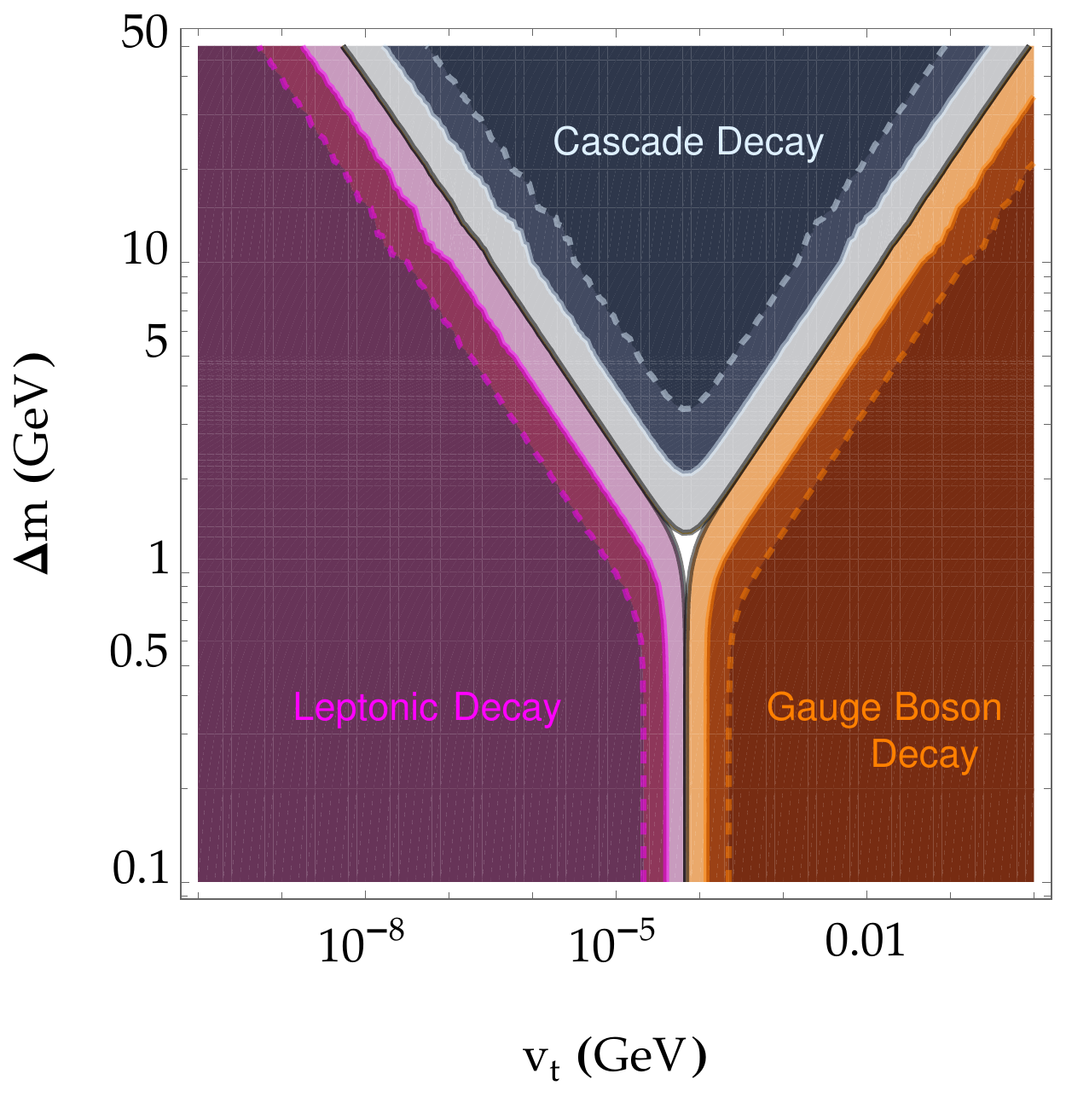} \quad
\includegraphics[scale=0.44]{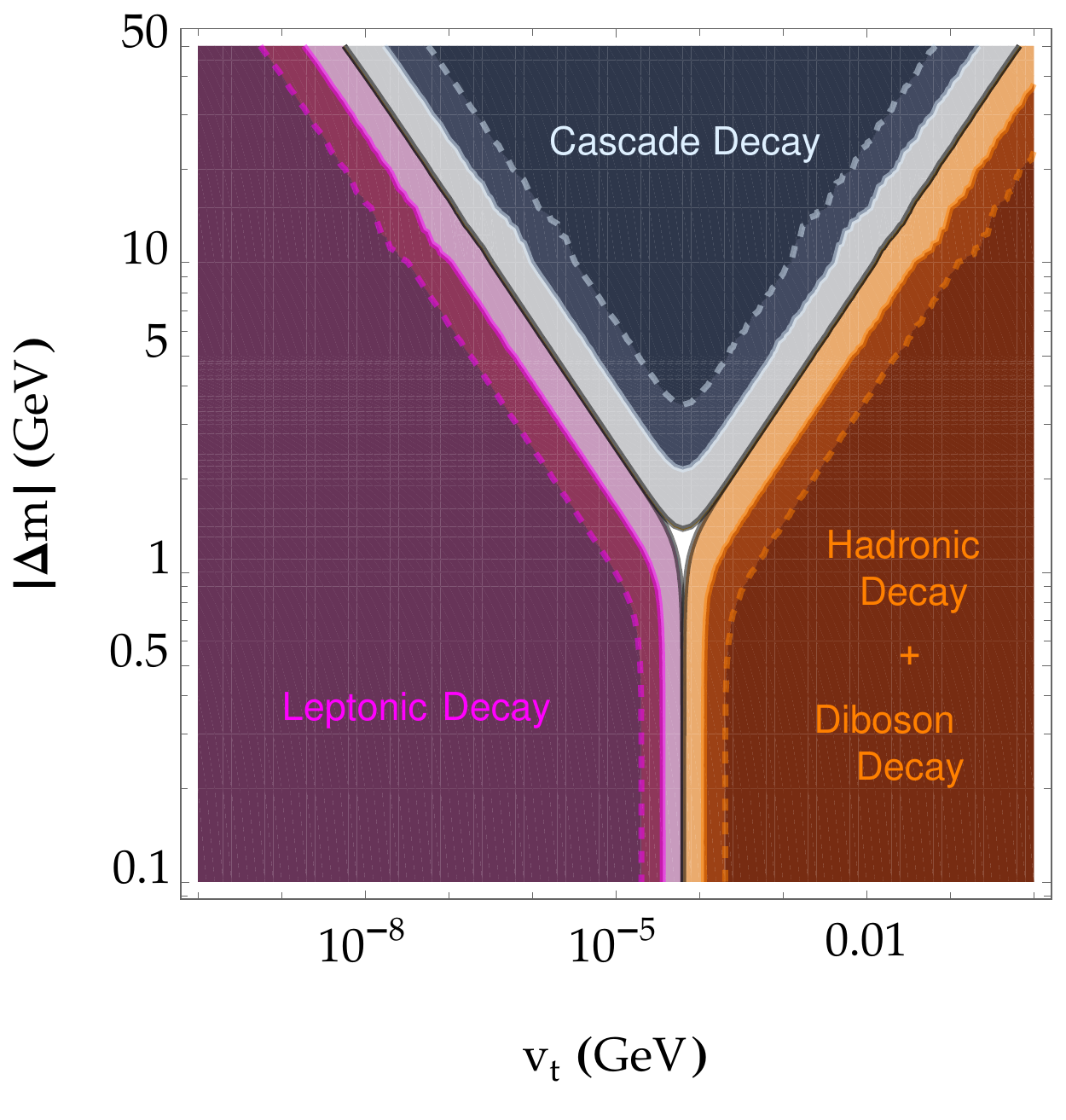}  \quad
\includegraphics[scale=0.44]{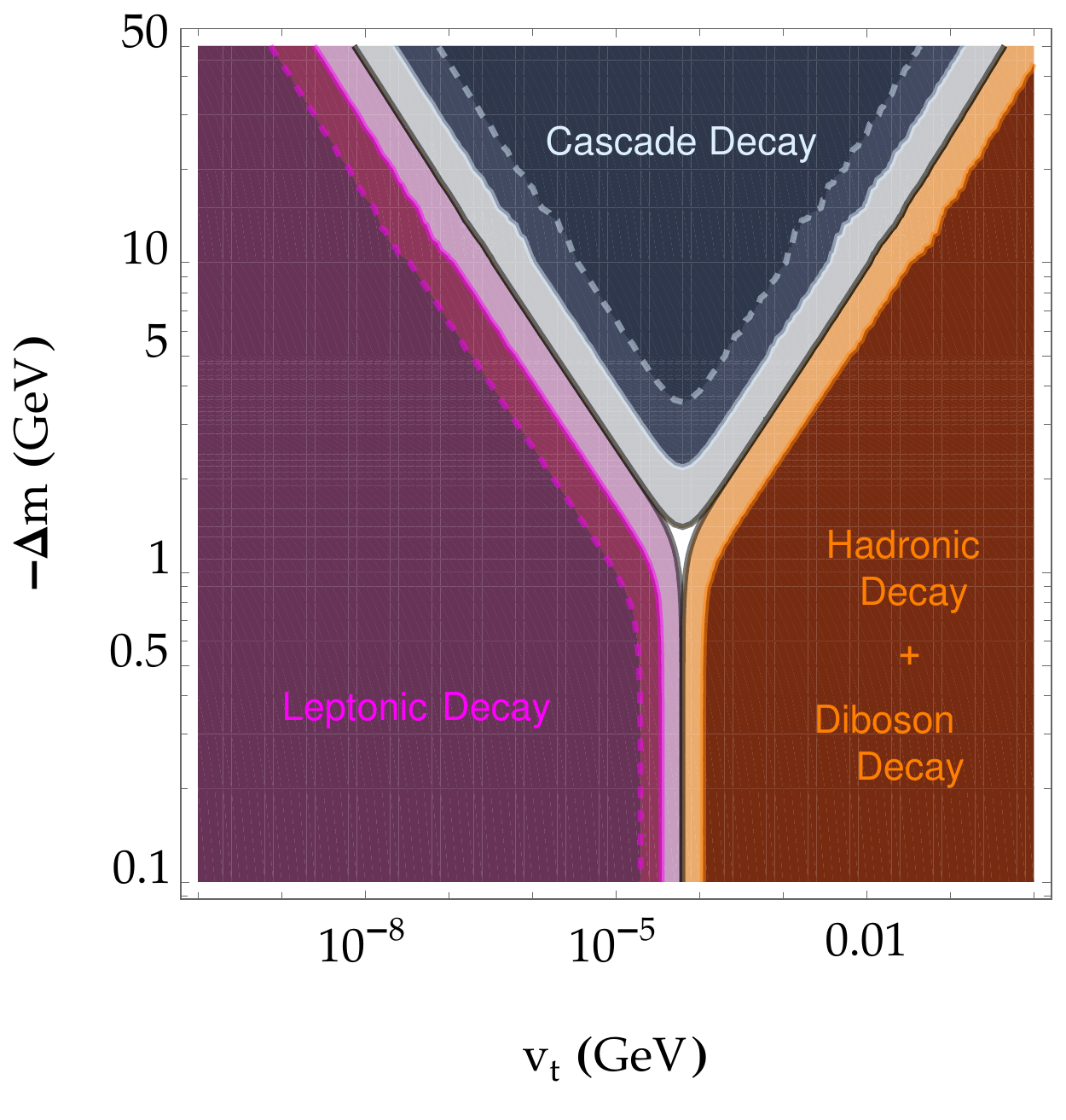}
\caption{Decay phase diagram of $H^{\pm \pm}$ (left), $H^\pm$ (middle) and $H^0/A^0$ (right) with $m_{H^{\pm \pm}}=500$ GeV segregating the $v_t$--$\Delta m$ parameter space with different decay modes' dominance for NH neutrino mass spectrum with $m_1=0.03$ eV. Magenta/Orange/Light blue dashed and solid contours, respectively, correspond to 99 and 90\%, and Black solid contours to 50\% branching ratios in different decay regions.}
\label{Br500}
\end{figure}
\end{widetext}

For $\Delta m < \mathcal{O}(1)$ GeV, $H^{\pm \pm}$ decays into $\ell^\pm \ell^\pm$ and/or $W^\pm W^\pm$. These two decay modes are on a par with each other for $v_t \sim \mathcal{O}(10^{-4})$ GeV, and the former decay mode subjugates the other one for $v_t < 10^{-4}$ GeV and vice versa. One expects the cascade channel to kick off for $\Delta m \gtrsim \mathcal{O}(1)$ GeV, and become prepotent for larger mass-splitting. For $m_{H^{\pm \pm}}=500$ GeV, the cascade decay dominates over the gauge boson one for $\Delta m > 55 \times \left(\frac{v_t}{1 \rm GeV} \right)^{2/5}$, whereas it dominates over the leptonic one for $\Delta m > 40 \times \left(\frac{10^{-8} \rm GeV}{v_t} \right)^{2/5}$ if NH neutrino mass spectrum with $m_1=0.03$ eV is assumed.\footnote{For NH , 0.03 eV is the maximum possible value for the lightest neutrino mass consistent with the bound from cosmology.} This has been reflected in the $H^{\pm \pm}$ decay phase diagram shown in the left most panel of Figure~\ref{Br500}. Figure~\ref{Br500} shows decay phase diagrams of $H^{\pm \pm}$ (left), $H^\pm$ (middle) and $H^0/A^0$ (right) with $m_{H^{\pm \pm}}=500$ GeV segregating the $v_t$--$\Delta m$ parameter space with different decay modes' dominance for NH neutrino mass spectrum with $m_1=0.03$ eV. Magenta/Orange/Light blue dashed and solid contours, respectively, correspond to 99 and 90\%, and Black solid contours to 50\% branching ratios in different decay regions.

\begin{comment}
For $m_{H^{\pm \pm}}^2 \gg m_W^2$, the partial decay widths reduces to
\begin{align}
&\Gamma(H^{\pm\pm} \to \ell^\pm \ell^\pm) \approx \frac{|m_\nu|^2m_{H^{\pm \pm}}}{16\pi v_t^2}~,
\\
&\Gamma(H^{\pm\pm} \to W^\pm W^\pm) \approx \frac{v_t^2m_{H^{\pm \pm}}^3}{4\pi v_d^4}~,
\\
&\Gamma(H^{\pm\pm} \to H^\pm W^{\pm *}) \approx \frac{3(\Delta m)^5}{5\pi^3 v_d^4}~.
\end{align}

\begin{equation*}
\frac{1}{4} \left(\frac{m_\nu}{m_{H^{\pm \pm}}} \right)^2 \left(\frac{v_d}{v_t}\right)^4 : 1 : \frac{12}{5\pi^2} \frac{{\rm max}(\Delta m,0)^5}{v_t^2 m_{H^{\pm \pm}}^3}~.
\end{equation*}
\end{comment}

%=============================================================================
\subsection{\label{sec:decHp} Decays of singly charged scalar bosons}
%============================================================================= 
The singly charged scalar bosons have four types of decay modes --- (i) leptonic decay, {\it i.e.} $\ell^\pm \nu$, (ii) hadronic decay, {\it i.e.} $t\bar{b}$, (iii) diboson decay, {\it i.e.} $W^\pm Z$,$W^\pm h^0$ and (iii) cascade decay, {\it i.e.} $H^0/A^0 W^{\pm *}$ or $H^{\pm \pm} W^{\mp *}$. These two cascade decay modes kinematically open up, respectively, for $\Delta m > 0$ and $\Delta m < 0$. For $m_{H^\pm}^2 \gg m_W^2$, the ratio of the braching fractions for the $\ell^\pm \nu$, $t\bar{b}$, $W^\pm Z$, $W^\pm h^0$, $H^0/A^0 W^{\pm *}$ and $H^{\pm \pm} W^{\mp *}$ decay modes can be evaluated as (see Appendix~\ref{sec:app})
\begin{multline*}
\frac{1}{2} \left(\frac{m_\nu}{m_{H^\pm}} \right)^2 \left(\frac{v_d}{v_t}\right)^2 : 6\left(\frac{m_t}{m_{H^\pm}}\right)^2 \left(\frac{v_t}{v_d}\right)^2 : \left(\frac{v_t}{v_d}\right)^2 : \hfill
\\
(1-2\zeta)^2 \left(\frac{v_t}{v_d}\right)^2 : \frac{12}{5\pi^2} \frac{{\rm max}(\Delta m,0)^5}{v_d^2 m_{H^\pm}^3} : \frac{24}{5\pi^2} \frac{{\rm max}(0,-\Delta m)^5}{v_d^2 m_{H^\pm}^3}~, \hfill
\end{multline*}
where $\zeta = \frac{v_d}{2v_t} \sin\alpha$ and $m_t$ is the top quark mass. For sufficiently small mass-splitting, {\it i.e.} $|\Delta m| < \mathcal{O}(1)$ GeV, $H^\pm$ decays into $\ell^\pm \nu$ and/or $t\bar{b}$, $W^\pm Z$ and $W^\pm h^0$. The leptonic decay mode dominates for $v_t < 10^{-4}$ GeV, whereas the hadronic and diboson decay modes dominate for $v_t > 10^{-4}$ GeV. Furthermore, the diboson decay modes subjugate the hadronic one for $m_{H^\pm}^2 > 3m_t^2$ and vice versa. %For large $v_t$, the $t\bar{b}$ mode has branching fraction of $3m_t^2/(3m_t^2+m_{H^\pm}^2)$, whereas the $W^\pm Z$ and $W^\pm h^0$ modes have branching fraction of $m_{H^\pm}^2/(6m_t^2+2m_{H^\pm}^2)$ each. 
The cascade channels kick off for large enough mass-splitting, {\it i.e.} $|\Delta m| \gtrsim \mathcal{O}(1)$ GeV, and quickly dominate for larger mass-splitting. Dominance of these different decay modes in different regions of $v_t-|\Delta m|$ parameter space has been shown in the $H^\pm$ decay phase diagram in the middle panel of Figure~\ref{Br500}.

\begin{comment}
For $m_{H^\pm}^2 \gg m_W^2$, the partial decay widths reduces to
\begin{align}
&\Gamma(H^\pm \to \ell^\pm \nu)\approx \frac{|m_\nu|^2m_{H^\pm}}{16\pi v_t^2}~,
\\
&\Gamma(H^\pm \to t\bar{b})\approx \frac{3v_t^2 m_{H^\pm} m_t^2}{4\pi v_d^4}~,
\\
&\Gamma(H^\pm \to W^\pm Z) \approx \frac{v_t^2 m_{H^\pm}^3}{8\pi v_d^4}~,
\\
&\Gamma(H^\pm \to W^\pm h^0) \approx \frac{v_t^2 m_{H^\pm}^3}{8\pi v_d^4} (1-2\zeta)^2~,
\\
&\Gamma(H^\pm \to H^0/A^0 W^{\pm *}) \approx \frac{3(\Delta m)^5}{10\pi^3 v_d^4}~,
\\
&\Gamma(H^\pm \to H^{\pm \pm} W^{\mp *}) \approx \frac{3(-\Delta m)^5}{5\pi^3 v_d^4}~,
\end{align}

\begin{multline*}
\frac{1}{2} \left(\frac{m_\nu}{m_{H^\pm}} \right)^2 \left(\frac{v_d}{v_t}\right)^4 : 6\left(\frac{m_t}{m_{H^\pm}}\right)^2 : 1 : (1-2\zeta)^2 : \hfill
\\
\quad \frac{12}{5\pi^2} \frac{{\rm max}(\Delta m,0)^5}{v_t^2 m_{H^\pm}^3} : \frac{24}{5\pi^2} \frac{{\rm max}(0,-\Delta m)^5}{v_t^2 m_{H^\pm}^3}~. \hfill
\end{multline*}
\end{comment}

%=============================================================================
\subsection{\label{sec:decH} Decays of heavy neutral scalar bosons}
%============================================================================= 
The CP-odd (CP-even) heavy neutral scalar boson has four types of decay modes --- (i) leptonic decay, {\it i.e.} $\nu \nu$, (ii) hadronic decay, {\it i.e.} $q\bar{q}$ with $q\ni b,t$, (iii) diboson decay, {\it i.e.} $h^0Z$ ($WW$,$ZZ$,$h^0h^0$) and (iii) cascade decay, {\it i.e.} $H^\pm W^{\mp *}$. The latter decay mode kinematically opens up only for $\Delta m < 0$. For $m_{H^0}^2 \gg m_W^2$, the ratio of the braching fractions for the $\nu \nu$, $q\bar{q}$, $h^0h^0$, $WW$, $ZZ$ and $H^\pm W^{\mp *}$ decay modes of $H^0$ can be estimated as (see Appendix~\ref{sec:app})
\begin{multline*}
\frac{1}{2} \left(\frac{m_\nu}{m_{H^0}} \right)^2 \left(\frac{v_d}{v_t}\right)^2 : 12\zeta^2 \left(\frac{m_q}{m_{H^0}}\right)^2 \left(\frac{v_t}{v_d}\right)^2 : \zeta^2 \left(\frac{v_t}{v_d}\right)^2 : \hfill
\\
2(1-\zeta)^2 \left(\frac{v_t}{v_d}\right)^2 : 4\left(1-\frac{\zeta}{2}\right)^2 \left(\frac{v_t}{v_d}\right)^2 : \frac{12}{5\pi^2} \frac{{\rm max}(0,-\Delta m)^5}{v_d^2 m_{H^0}^3} ~. \hfill
\end{multline*}

\noindent Likewise, the ratio of the braching fractions for the $\nu \nu$, $q\bar{q}$, $h^0Z$ and $H^\pm W^{\mp *}$ decay modes of $A^0$ can be evaluated as
\begin{multline*}
\frac{1}{4} \left(\frac{m_\nu}{m_{A^0}} \right)^2 \left(\frac{v_d}{v_t}\right)^2 : 6\left(\frac{m_q}{m_{A^0}}\right)^2 \left(\frac{v_t}{v_d}\right)^2 : (1-2\zeta)^2 \left(\frac{v_t}{v_d}\right)^2 : 
\\
\quad \frac{6}{5\pi^2} \frac{{\rm max}(0,-\Delta m)^5}{v_d^2 m_{A^0}^3} ~. \hfill
\end{multline*}
For $-\Delta m < \mathcal{O}(1)$ GeV, $A^0(H^0)$ decays into neutrinos and hadrons/dibosons respectively, for $v_t < 10^{-4}$ GeV and $v_t > 10^{-4}$ GeV. Furthermore, the dibosons decay modes of $A^0$($H^0$) dominate over the hadronic one for $m_{A^0(H^0)}^2 > 6m_t^2$ as well as for $m_{A^0(H^0)}^2 < 4m_t^2$.\footnote{For $m_{H^0}^2 < 4m_t^2$, the decay of $A^0/H^0$ into $t\bar{t}$ is kinematically not allowed; thus, the diboson decay modes are dominant.} The cascade channel kicks off for $-\Delta m \gtrsim \mathcal{O}(1)$ GeV, and swiftly dominates for larger mass-splitting as can be seen from the $H^0/A^0$ decay phase diagram in the right panel of Figure~\ref{Br500}.

\section{\label{sec:collider} Collider Searches}
%============================================================================= 
Profuse Drell-Yan production of the triplet-like scalars via $s$-channel $\gamma/Z$ and $W^\pm$ exchanges and their subsequent prompt decays \footnote{For $v_t > 10^{-4}$ GeV and $m_{H^{\pm \pm}} < 2m_W$, $H^{\pm \pm}$ deacys to $W^\pm W^{\pm *}$. For a region of $v_t$--$m_{H^{\pm \pm}}$ parameter space, $H^{\pm \pm}$ could be long-lived and have displaced vertex signatures at collider \cite{Antusch:2018svb}.} to SM particles lead to a variety of final state signatures at the LHC. Possible final states includes smoking gun signatures like two pairs of same-sign lepton or two pairs of same-sign $W$-boson. Phenomenological consequence of the present model at the LHC has been studied extensively in the literature \cite{Huitu:1996su,Chakrabarti:1998qy,Chun:2003ej,Akeroyd:2005gt,Garayoa:2007fw,Kadastik:2007yd,Akeroyd:2007zv,Perez:2008ha,delAguila:2008cj,Akeroyd:2009hb,Melfo:2011nx,Aoki:2011pz,Akeroyd:2011zza,Chiang:2012dk,Chun:2012jw,Akeroyd:2012nd,Chun:2012zu,Dev:2013ff,Banerjee:2013hxa,delAguila:2013mia,Chun:2013vma,Kanemura:2013vxa,Kanemura:2014goa,Kanemura:2014ipa,kang:2014jia,Han:2015hba,Han:2015sca,Das:2016bir,Babu:2016rcr,Mitra:2016wpr,Cai:2017mow,Ghosh:2017pxl,Crivellin:2018ahj,Du:2018eaw,Dev:2018kpa,Antusch:2018svb,Aboubrahim:2018tpf,deMelo:2019asm,Primulando:2019evb,Padhan:2019jlc,Chun:2019hce}. Rightfully, central attention of most of those studies pivots around the doubly charged scalars because of their distinct decay signatures. For the very same reason, both the CMS and ATLAS collaborations have carried out a number of collider searches to look for the same at the LHC \cite{ATLAS:2012hi,Chatrchyan:2012ya,ATLAS:2014kca,Khachatryan:2014sta,CMS:2016cpz,CMS:2017pet,Aaboud:2017qph,CMS:2017fhs,Aaboud:2018qcu,Aad:2021lzu}. Hitherto no significant excess over the SM background expectations has been observed in any of these direct collider seraches. These seraches thereupon have set stringent limits with 95\% CL on the masses of the doubly charged scalars in the context of a simplified model. As argued in Section~\ref{sec:intro}, these limits are not befitting to the entire parameter space, rather valid only for a constrained parameter space of the model. Also, these limits are often conservative as these searches do not incorporate all the production channels for the triplet-like scalars. Furthermore, most of these limits are derived in the context of simplified scenarios without reckoning the footprints of the low-energy neutrino parameters.

The quartic scalar interaction of the triplet with the doublet, $\lambda_4{\Phi^\dagger \Delta \Delta^\dagger \Phi}$, countenances the triplet components to split in mass. For moderate $v_t$ and passably large $\Delta m$, cascade decays quickly dominate over the leptonic and diboson decay modes, see Figure~\ref{Br500}. Conceivably, the cascade decays play notable role in the pheneomenology, thereby making the pheneomenology for the non-degenerate scenario substantially contrasting than that for the degenerate one. %\cite{Chakrabarti:1998qy,Chun:2003ej,Akeroyd:2005gt,Melfo:2011nx,Aoki:2011pz,Akeroyd:2011zza,Chiang:2012dk,Chun:2012zu,Chun:2013vma,Han:2015hba,Han:2015sca,Akeroyd:2012nd,delAguila:2008cj,Mitra:2016wpr,Du:2018eaw,Antusch:2018svb,Primulando:2019evb,Chun:2019hce}. 
Thus, in the non-degenerate scenario ($\Delta m \neq 0$), the cascade decays are entitled to play a pronounced part in constraining the model parameter space. Not only the mass-splitting overwhelms the decays of the triplet-like scalars, but it also affects their production cross-sections at the LHC, see Figure~\ref{cs}. In what follows, we briefly discuss possible final state signatures for different parts of the parameter space and outline already existing direct collider searches by CMS and ATLAS which are potentially sensitive in constraining the model.\\

\noindent \textbf{Degenerate scenario ($\Delta m = 0$)} All the Drell-Yan pair production mechanism of triplet-like scalars except $H^+H^-$ are of sizeable cross-sections, see the left most plot in Figure~\ref{cs}. For small triplet vev, {\it i.e.} $v_t < 10^{-4}$ GeV, $H^{\pm \pm}$, $H^\pm$ and $H^0/A^0$ decay to $\ell^\pm \ell^\pm$, $\ell^\pm \nu$ and $\nu \nu$, respectively. Production of $H^{\pm \pm} H^\mp$ and $H^{++}H^{--}$ lead to, respectively, three and four light leptons ($e,\mu$) in the final state. Though $H^\pm H^0/A^0$ and $H^0A^0$ have large enough cross-sections, they fall through to complement the multilepton final state because of their invisible decays. The already existing multilepton searches by CMS and ATLAS such as \cite{CMS:2017pet,Aaboud:2017qph,Sirunyan:2019bgz,ATLAS:2021eyc} are expected to constrain this part of the parameter space.

For large triplet vev, {\it i.e.} $v_t > 10^{-4}$ GeV, $H^{\pm \pm}$, $H^\pm$, $H^0$ and $A^0$ decay to $W^\pm W^\pm$, $W^\pm Z/h^0$, $ZZ/WW/h^0h^0$ and $h^0Z$, respectively. All the production channels give rise multiboson final state leading to multilepton signatures. Therefore, one anticipates this part of the parameter space to be probed by the existing ATLAS searches \cite{Aaboud:2018qcu,Aad:2021lzu}.\\

\noindent \textbf{Negative scenario ($\Delta m < 0$)} For small mass-splitting, {\it i.e.} $\Delta m \lesssim \mathcal{O}(1)$, this scenario resembles the degenerate one. However, for passably large mass-splitting and moderate $v_t$, the cascade decays $H^0/A^0 \to H^\pm W^{\mp *}$ and $H^\pm \to H^{\pm \pm} W^{\mp *}$ dominate over the other decay modes, thereby enhancing the effective production cross-section for $H^{\pm \pm}$. Then, depending on the value of $v_t$, $H^{\pm \pm}$ decays to either same-sign dilepton pair or same-sign $W$-boson pair, both leading to multilepton signatures in the final state. Thus, this scenario also can be probed either by the multilepton searches like \cite{CMS:2017pet,Aaboud:2017qph,Sirunyan:2019bgz,ATLAS:2021eyc} or multiboson leading to multilepton searches like \cite{Aaboud:2018qcu,Aad:2021lzu}.\\

\noindent \textbf{Positive scenario ($\Delta m > 0$)} Again, this scenario resembles the degenerate one for small $\Delta m$. However, for large enough $\Delta m$ and moderate $v_t$, the cascade decays $H^{\pm \pm} \to H^\pm W^{\pm *}$ and $H^\pm \to H^0/A^0 W^{\pm *}$ dominate over the other decay modes. This enhances the effective production cross-section for $H^0$ and $A^0$. For $v_t > 10^{-4}$ GeV, $H^0$ and $A^0$ decay to $ZZ/WW/h^0h^0$ and $h^0Z$, respectively. This gives rise to multiboson final state signatures. Therefore, one expects that this part of the parameter space can probed by the existing ATLAS searches \cite{Aaboud:2018qcu,Aad:2021lzu}.\footnote{We see that for large enough $\Delta m$ and $v_t \sim 10^{-3}$--$10^{-4}$ GeV, the existing collider searches \cite{Aaboud:2018qcu,Aad:2021lzu} fall short in probing the triplet-like scalars. For this part of the parameter space, $H^0$ dominantly decays to $h^0h^0$ and $ZZ$, and $A^0$ decays to $h^0Z$. The leptonic decays of $Z$ give rise to multilepton final states. However, the usual searches \cite{Aaboud:2018qcu,Aad:2021lzu} use a $Z$-veto to suppress the oversized background from the Drell-Yan processes making these searches insensitive in probing this scenario. On the contrary, for hadronic decays of $h^0$ and $Z$, the signal cross-section is small compared to the overwhelming QCD jets background, and thus probing this scenario is very challenging. }

For $v_t < 10^{-4}$ GeV, both $H^0$ and $A^0$ decay invisibly into neutrinos. The relevant production mechanisms $H^{\pm \pm}H^\mp$ and $H^{++}H^{--}$ yield soft leptons or jets resulting from the off-shell $W$-bosons and neutrinos. Being very soft, these final state leptons/jets are very difficult to reconstruct at the LHC. Therefore, in this scenario, the most optimistic final states would be an energetic jet resulting from initial state radiation plus large missing transverse momentum \cite{Aad:2021egl,ATLAS:2016bek,ATLAS:2017bfj} or two/three soft leptons plus missing transverse momentum \cite{CMS:2021xji,Sirunyan:2018iwl}. For the former final state signature, the recent monojet search by ATLAS \cite{Aad:2021egl} could be potentially sensitive in probing this scenario. Whereas, for the latter, one would expect the soft leptons searches by CMS \cite{CMS:2021xji,Sirunyan:2018iwl} to be sensitive in probing this scenario.\footnote{However, it turns out that both the monojet search by ATLAS and the soft leptons search by CMS cease to constrain this part of the parameter space. Monojet search usually requires a larger signal cross-section to suppress the huge SM background and hence, by and large, applicable to the strongly produced particles decaying into soft leptons/jets plus missing particles. Furthermore, the soft lepton final states are suppressed by $W$-leptonic branching fractions. That said, inconsiderably small signal cross-section in the present model compared to the SM background makes such a nightmare scenario impossible to probe.}

%=============================================================================
\subsection{\label{sec:cms}Multilepton final states search by CMS \cite{Sirunyan:2019bgz}}
%=============================================================================
The CMS collaboration has published a multilepton final states search \cite{Sirunyan:2019bgz} with an integrated luminosity of $137.1$ fb$^{-1}$ of pp collisions at $\sqrt{s}=13$ TeV. This search targeted the triplet fermions in the type-III see-saw model \cite{Foot:1988aq}. However, because of similar multilepton final state signatures, this search is conjectured to be passably sensitive in probing the type-II see-saw model. Hitherto, there is no multilepton search targeting the type-II see-saw model using the full Run-2 dataset by CMS or ATLAS. Thereupon, we set forth to implement this search meticulously.

We simulate the signal events using MadGraph \cite{Alwall:2011uj,Alwall:2014hca} with the  NNPDF23$\_$lo$\_$as$\_$0130$\_$qed parton distribution function \cite{Ball:2013hta,Ball:2014uwa}. The subsequent decays, initial state radiation (ISR), final state radiation (FSR), showering, fragmentation and hadronisation are simulated with PYTHIA \cite{Sjostrand:2014zea}. Hadronized events are passed into Delphes \cite{deFavereau:2013fsa} for object reconstruction and selection, defining signal regions and event selection. In doing so, we rigorously follow the search strategy in Ref.~\cite{Sirunyan:2019bgz}. Lastly, we use a hypothesis tester which uses a library of C++ classes ‘RooFit’ \cite{Verkerke:2003ir} in the ROOT environment to estimate CL.

\begin{widetext}

\begin{figure}[htb!]
\centering
\includegraphics[scale=0.29]{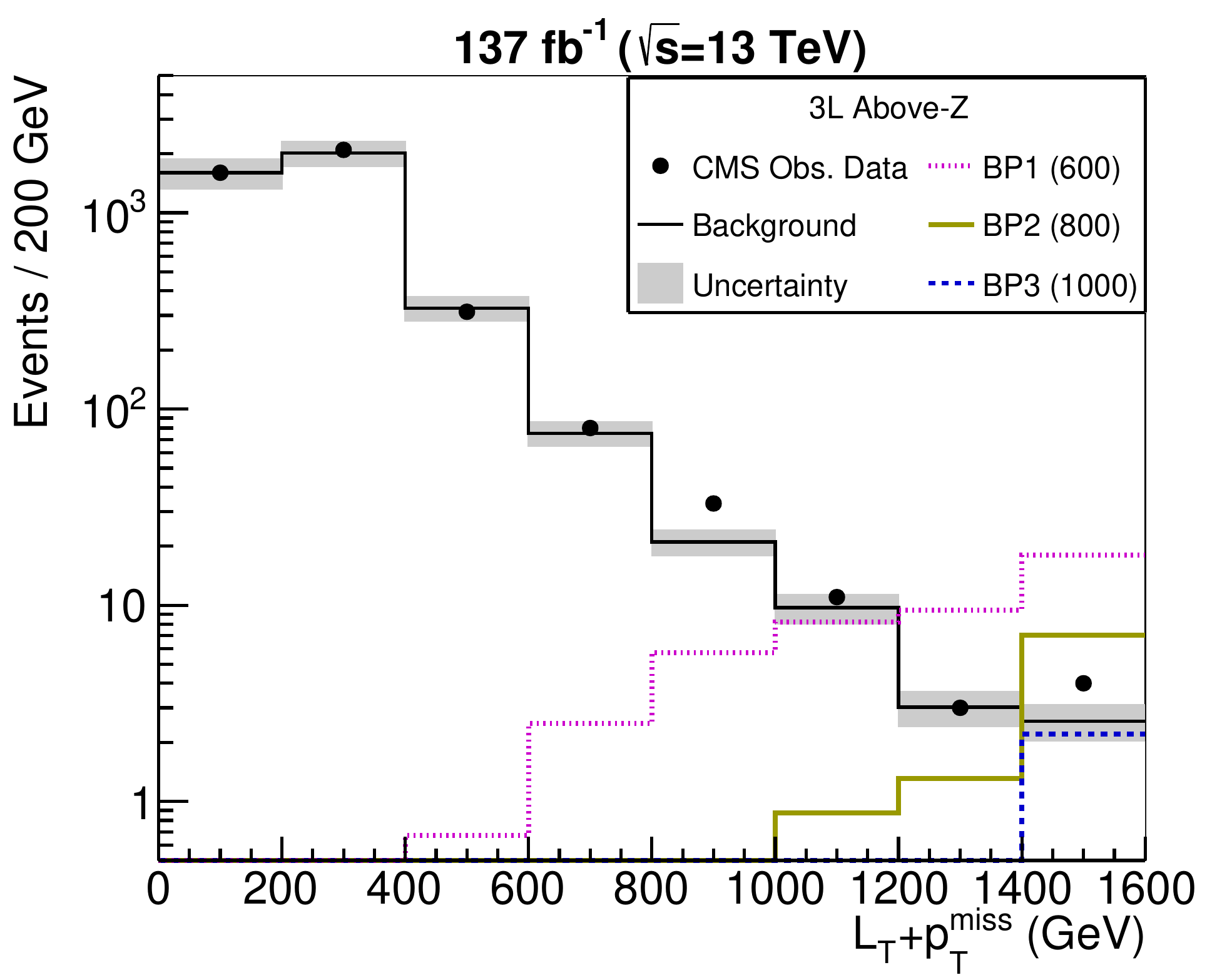}
\includegraphics[scale=0.29]{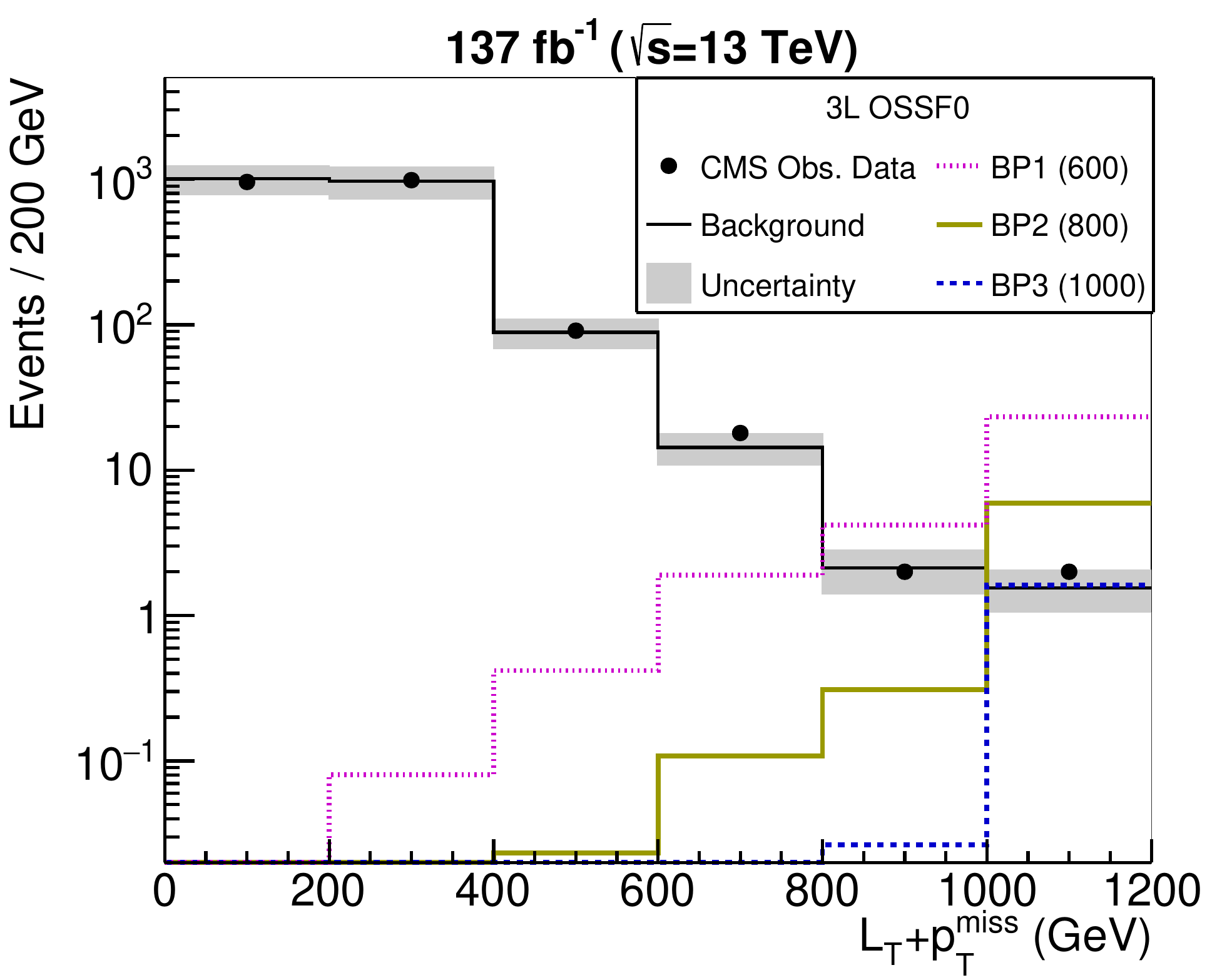}
\includegraphics[scale=0.29]{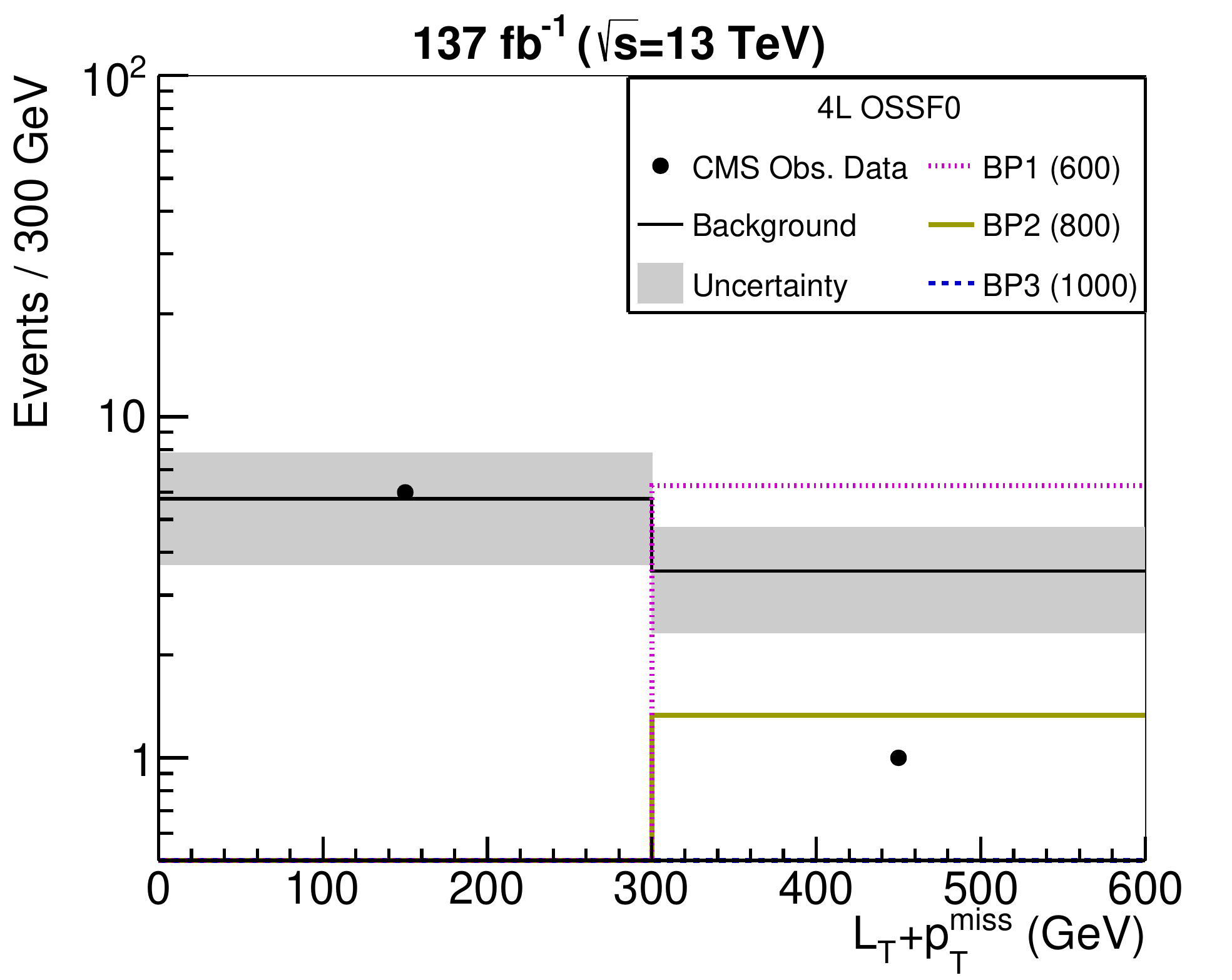}
\caption{$L_T+p_T^{\rm miss}$ distributions of the expected SM background events, the observed events and the expected signal events for {\it 3L above-Z} (left), {\it 3OSSF0} (middle) and {\it 4OSSF0} (right) SRs. The signal predictions are shown for three benchmark points listed in Table~\ref{cms_BP}.}
\label{cms_dist}
\end{figure}
\end{widetext}

The selected events are categorised into several mutually exclusive signal regions (SRs), namely {\it 3LOSSF0, 3LOSSF1, 4LOSSF0, 4LOSSF1} and {\it 4LOSSF2}, based on the multiplicity of light leptons, the multiplicity and mass of opposite-sign same-flavour (OSSF) lepton pairs, $N_{\rm OSSF}$ and $M_{\rm OSSF}$ . The names of the SRs are self-explanatory, see Ref.~\cite{Sirunyan:2019bgz} for details. The events in the {\it 3LOSSF1} SR are further classified as {\it 3L below-Z, 3L on-Z} and {\it 3L above-Z} when $M_{\rm OSSF}$ is below, within and above the Z-boson mass window ($M_Z \pm 15$), respectively. All the SRs are further divided into several independent signal bins using a primary kinematic discriminant, thereby resulting in 40 signal bins in total. For {\it 3L on-Z} SR, this search uses transverse mass ($M_T$) \footnote{The transverse mass is defined as $M_T=\sqrt{2 p_T^{\rm miss} p_T^\ell [1-\cos(\Delta\phi_{\vec p_T^{\rm miss},\vec p_T^\ell})]}~,$ where $\vec p_T^\ell$ is the transverse momentum vector of the lepton which is not a part of the on-Z pair, and $\Delta\phi_{\vec p_T^{\rm miss},\vec p_T^\ell}$ is the azimuthal separation between $\vec p_T^{\rm miss}$ and $\vec p_T^\ell$.} as the primary discriminant, whereas for all other SRs, scalar sum of the transverse momenta of all charged leptons ($L_T$) plus the missing transverse momentum ($p_T^{\rm miss}$) is used. These variables, exploiting the relatively high momenta of the triplet fermions' decay products, are found to be nifty in discriminating the signal from the background. For the detailed description and validation of our implementation of this search \cite{Sirunyan:2019bgz}, see Ref.~\cite{Ashanujjaman:2020tuv,Ashanujjaman:2021jhi}. This well-to-do implementation enables us to use the distributions of expected SM backgrounds and observed events in Ref.~\cite{Sirunyan:2019bgz} to constrain the type-II see-saw scalars in various $v_t$ -- $\Delta m$ regions.
%Before continuing, we validate our implementation of this search by reproducing the CMS 95\% CL limit on the total triplet pair production cross-section in simplified flavour democratic type-III see-saw in Ref.~\cite{Sirunyan:2019bgz}. Our reproduced result is found to be in admissible accordance with the CMS one, thereby validating our implementation of the aforecited search. For the detailed description of our implementation and validation of this search, see \cite{Ashanujjaman:2020tuv,Ashanujjaman:2021jhi}. This well-to-do implementation enables us to use the distributions of expected SM backgrounds and observed events in Ref.~\cite{Sirunyan:2019bgz}.
Figure~\ref{cms_dist} shows the $L_T+p_T^{\rm miss}$ distributions of the expected SM background events (histograms with black line),\footnote{The gray bands represent the total (systematic + statistical) uncertainty on the expected SM background.} the observed events (big black dots) and the expected signal events corresponding to 137.1 fb$^{-1}$ integrated luminosity data at the 13 TeV LHC for three SRs --- {\it 3L above-Z} (left), {\it 3OSSF0} (middle) and {\it 4OSSF0} (right). For brevity, we avert to show similar distributions for the other SRs. The magenta dotted, dark yellow solid and blue dashed histrograms show the expected signal events for three benchmark masses listed in Table~\ref{cms_BP} for $v_t = 10^{-8}$ GeV \footnote{The CMS multilepton search in Ref.~\cite{Sirunyan:2019bgz} is designed to probe final states with hard-$p_T$ leptons, and hence sensitive to small $v_t$ region where the scalars directly decay to leptons, and results into hard signal leptons in the final state.} and $\Delta m=0$ assuming NH neutrino mass spectrum with $m_1 = 0.03$ eV. Table~\ref{cms_BP} shows three benchmark masses along with their exclusion significances from the afore-described CMS multilepton search \cite{Sirunyan:2019bgz}. This shows that {\bf BP1} and {\bf BP2} are excluded with more than 95\% CL significances, whereas {\bf BP3} is allowed.

\begin{table}[htb!]
\centering
\scalebox{1}{
\begin{tabular}{|c|c|c|}
\hline 
Benchmark & $m_{H^{\pm \pm}}$ & Exclusion significance \\
\hline 
\bf BP1 & 600 & 11.9$\sigma$ \\
\hline
\bf BP2 & 800 & 4.32$\sigma$ \\
\hline
\bf BP3 & 1000 & 1.24$\sigma$ \\
\hline 
\end{tabular} 
}
\caption{\label{cms_BP} Benchmark masses and their exclusion significances from the CMS multilepton search\cite{Sirunyan:2019bgz}.}
\end{table}

%===========================================================================================
\subsection{\label{sec:atlas}Multiboson leading to multilepton final states search by ATLAS \cite{Aad:2021lzu}}
%===========================================================================================
The ATLAS collaboration has recently published a search for doubly and singly charged Higgs bosons decaying into vector bosons in multilepton final states with an integrated luminosity of $139$ fb$^{-1}$ of pp collisions at $\sqrt{s}=13$ TeV \cite{Aad:2021lzu}. As mentioned earlier, this ATLAS search considered either pair or associated production modes for the doubly charged scalars, but not both at once. Also, this search does not incorporate the production channels for the singly charged and neutral triplet-like scalars. Thus, the limits set by this ATLAS search are conservative. Furthermore, these limits are not befitting to the entire parameter space, rather valid only for $\Delta m=0$ and $v_t> \mathcal{O}(10^{-4})$ GeV. Therefore, we set forth to implement this search incorporating all the production modes for the triplet-like scalars to constrain the triplet-like scalars in various $v_t$ -- $\Delta m$ regions. For the implementation, we minutely follow the search strategy in Ref.~\cite{Aad:2021lzu}.

After object reconstruction and selection (see Ref.~\cite{Aad:2021lzu}), the events are categorised into three mutually exclusive analysis channels, namely same-sign dilepton ($2\ell^{sc}$), three leptons ($3\ell$) and four leptons ($4\ell$) channels. The event selection proceeds in two steps --- the preselection and the SRs selection. The preselection requirements are based on a number of variables such as the absolute value of the sum of charges of the leptons, their transverse momenta, $p_T^{\rm miss}$, the jet multiplicity, the $b$-jet multiplicity, $M_{\rm OSSF}$, {\it etc}. Four signal regions ({\it SR1, SR2, SR3} and {\it SR4}) are defined for each channel. For defining SRs, several other variables such as the invariant mass of all selected leptons, the invariant mass of all the jets, the distance between two same-sign leptons in the $\eta$-$\phi$ plane, the azimuthal distance between the dilepton system and $p_T^{\rm miss}$, the smallest distance between any lepton and its closest jet in the $\eta$-$\phi$ plane, {\it etc.} are used (see Ref.~\cite{Aad:2021lzu} for details). \footnote{Though different sets of kinematic variables and selection cuts are used to define the SRs ({\it SR1, SR2, SR3} and {\it SR4}), they are not mutually exclusive. They are designed by optimising the sensitivity for the $H^{\pm \pm}$ pair production mode , respectively, for the $m_{H^{\pm \pm}}=$ 200, 300, 400 and 500 GeV mass hypotheses. Ref.~\cite{Aad:2021lzu} considers {\it SR1}({\it SR2})\{{\it SR3}\}[{\it SR4}] for 200--300(300--400)\{400-500\}[$>500$] GeV mass hypothesis. Here, we differ from Ref.~\cite{Aad:2021lzu}; for a given mass hypothesis, we conider all the SRs disjointly, and eventually, choose the most sensitive one. However, for the validation of our implementation, we adhere to the Ref.~\cite{Aad:2021lzu}'s approach.} These variables, making capital out of the boosted decay topology of the triplet-like Higgs bosons as well as the high energy of their decay products, are discerned to discriminate between the signal and the background. Events in the $2\ell^{sc}$ SRs are further divided into $ee,\mu\mu$ and $e\mu$ final states, whereas those in the $3\ell$ SRs are separated into two categories ($3\ell$1 and $3\ell$0) based on whether or not an OSSF lepton pair exists in the event. This ameliorates the sensitivity of this search by contradistinguishing the lepton-flavour composition between signal and background.

We, then, move forward to validate our implementation of this search by reproducing the ATLAS 95\% CL bound on the total pair production cross-section times branching fraction for two scenarios corresponding to the pair production $H^{\pm \pm} H^{\pm \pm}$ and the associated production $H^{\pm \pm} H^\mp$ in Ref.~\cite{Aad:2021lzu}. The left (right) plot in Figure~\ref{ATLAS} shows the ATLAS observed and expected 95\% CL upper limits on the $H^{\pm \pm} H^{\pm \pm}$ ($H^{\pm \pm} H^\mp$) production cross-section times branching fraction. The green and yellow bands represent the expected exclusion curves within one and two standard deviations, respectively. The NLO QCD corrected \cite{Muhlleitner:2003me} theoretical prediction is shown by the solid red curve. The reproduced 95\% CL upper limit is represented by the blue dashed curve. The reproduced result is found to be in passable conformity with the ATLAS one, thereby validating our implementation of this search. Yet again, this entitles us to use the distributions of expected SM backgrounds and observed events in Ref.~\cite{Aad:2021lzu}. Figure~\ref{atlas_dist} shows the expected SM background events(histograms with black line),\footnote{The gray bands represent the total (systematic + statistical) uncertainty on the expected SM background.} the observed events (big black dots) and the expected signal events corresponding to 139 fb$^{-1}$ integrated luminosity data at the 13 TeV LHC for four different SRs --- {\it SR1, SR2, SR3} and {\it SR4}. For each SR, the yields are shown for all the relevant channels, namely $ee$, $e\mu$, $\mu \mu$, $3\ell 0$, $3\ell 1$ and $4\ell$. The magenta dotted, dark yellow solid and blue dashed histograms show the expected signal events for three benchmark masses listed in Table~\ref{atlas_BP} for $v_t = 1$ GeV and $\Delta m=0$ assuming NH neutrino mass spectrum with $m_1 = 0.03$ eV.
\begin{table}[htb!]
\centering
\scalebox{1}{
\begin{tabular}{|c|c|c|}
\hline 
Benchmark & $m_{H^{\pm \pm}}$ & Exclusion significance \\
\hline 
\bf BP1 & 200 & 13.6$\sigma$ \\
\hline
\bf BP2 & 350 & 2.72$\sigma$ \\
\hline
\bf BP3 & 500 & 0.54$\sigma$ \\
\hline 
\end{tabular} 
}
\caption{\label{atlas_BP} Benchmark masses and their exclusion significances from the ATLAS multiboson leading into multilepton search\cite{Aad:2021lzu}.}
\end{table}
Table~\ref{atlas_BP} shows three benchmark masses along with their exclusion significances from the above-described multiboson leading into multilepton search by ATLAS \cite{Aad:2021lzu}. This shows that {\bf BP1} and {\bf BP2} are excluded with more than 95\% CL significances, whereas {\bf BP3} is allowed.

\begin{widetext}

\begin{figure}[htb!]
\centering
\includegraphics[scale=0.8]{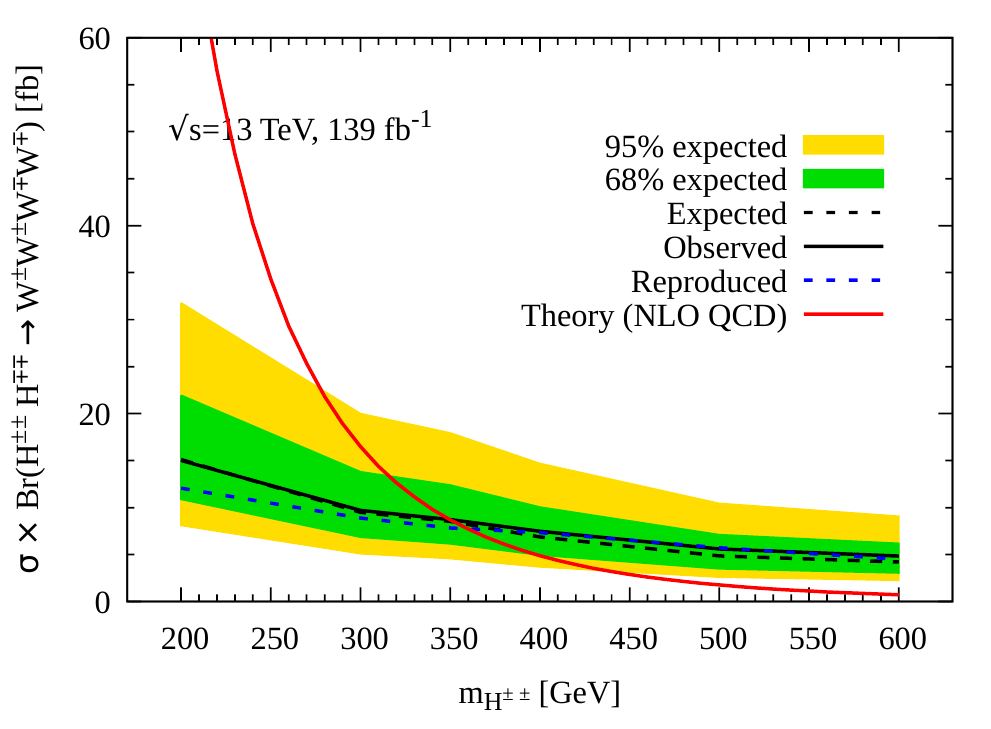} \qquad
\includegraphics[scale=0.8]{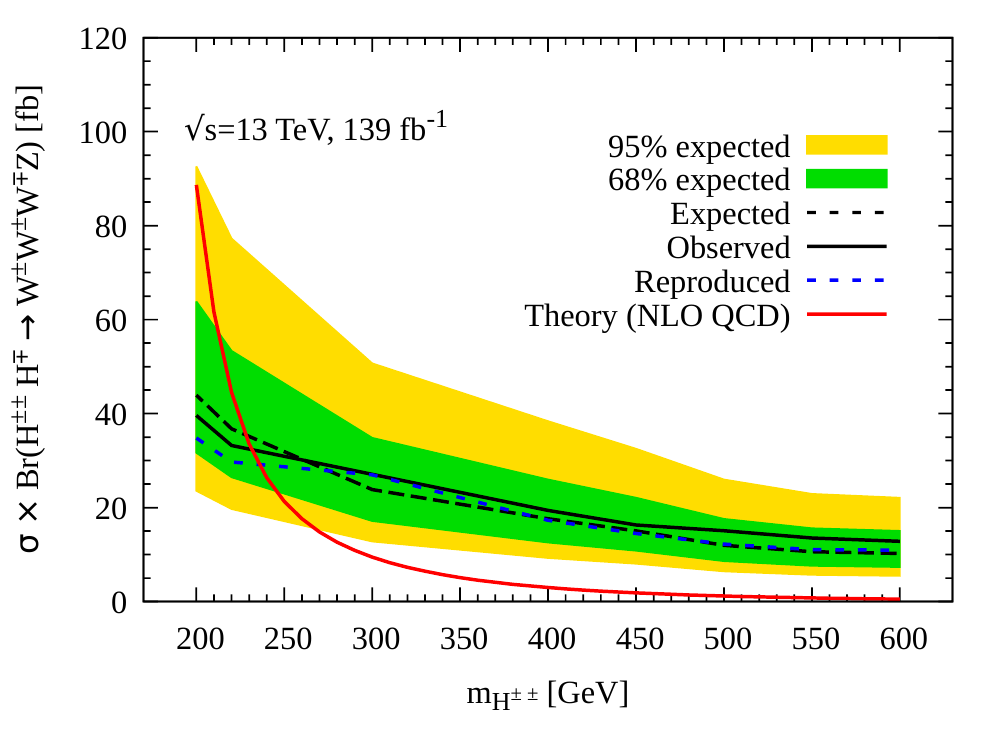}
\caption{Left (right): The ATLAS observed and expected 95\% CL upper limits on the $H^{\pm \pm} H^{\pm \pm}$ ($H^{\pm \pm} H^\mp$) production cross-section times branching fraction. The reproduced 95\% CL upper limit is represented by the blue dashed curve.}
\label{ATLAS}
\end{figure}

\begin{figure}[htb!]
\centering
\includegraphics[scale=0.44]{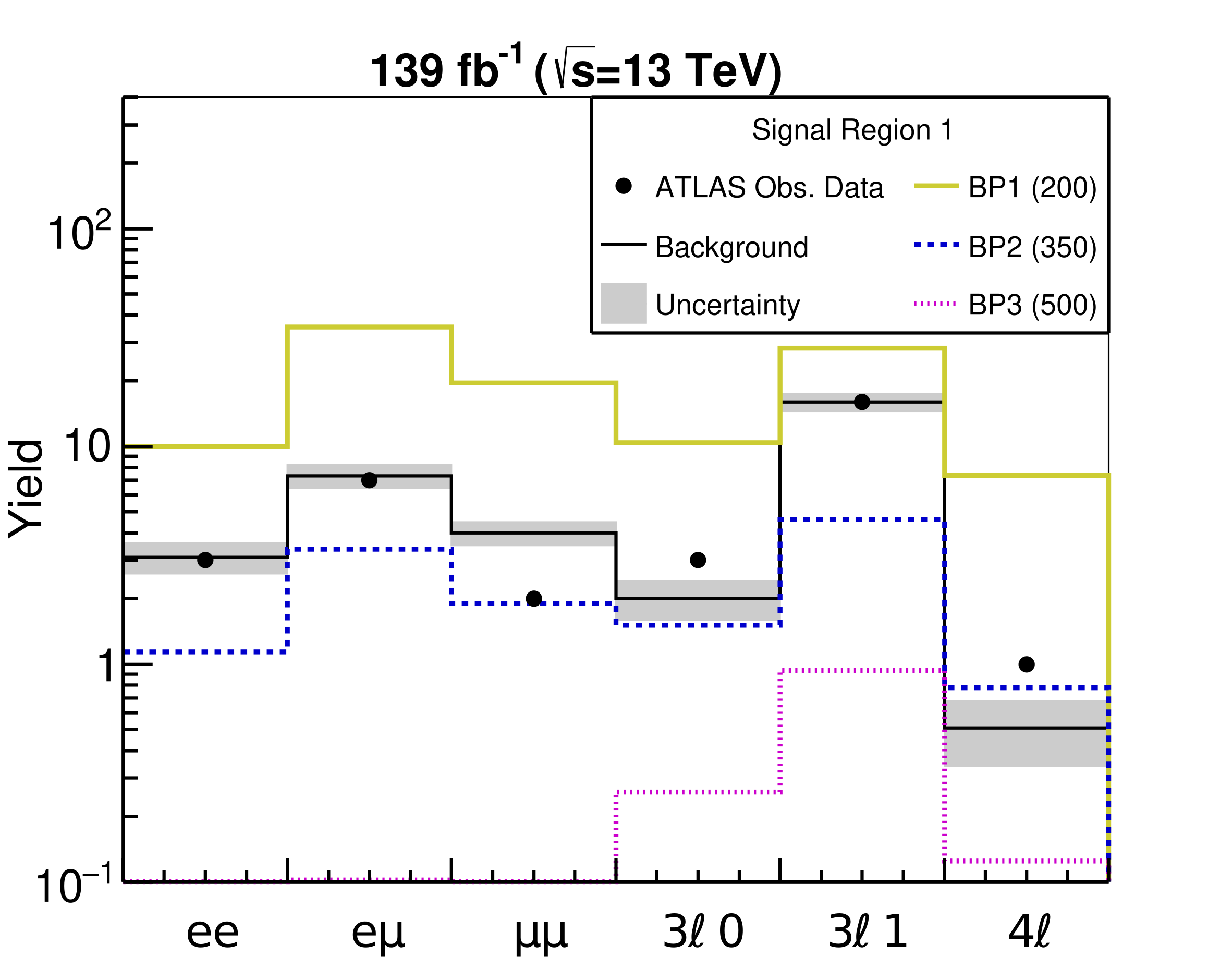} 
\includegraphics[scale=0.44]{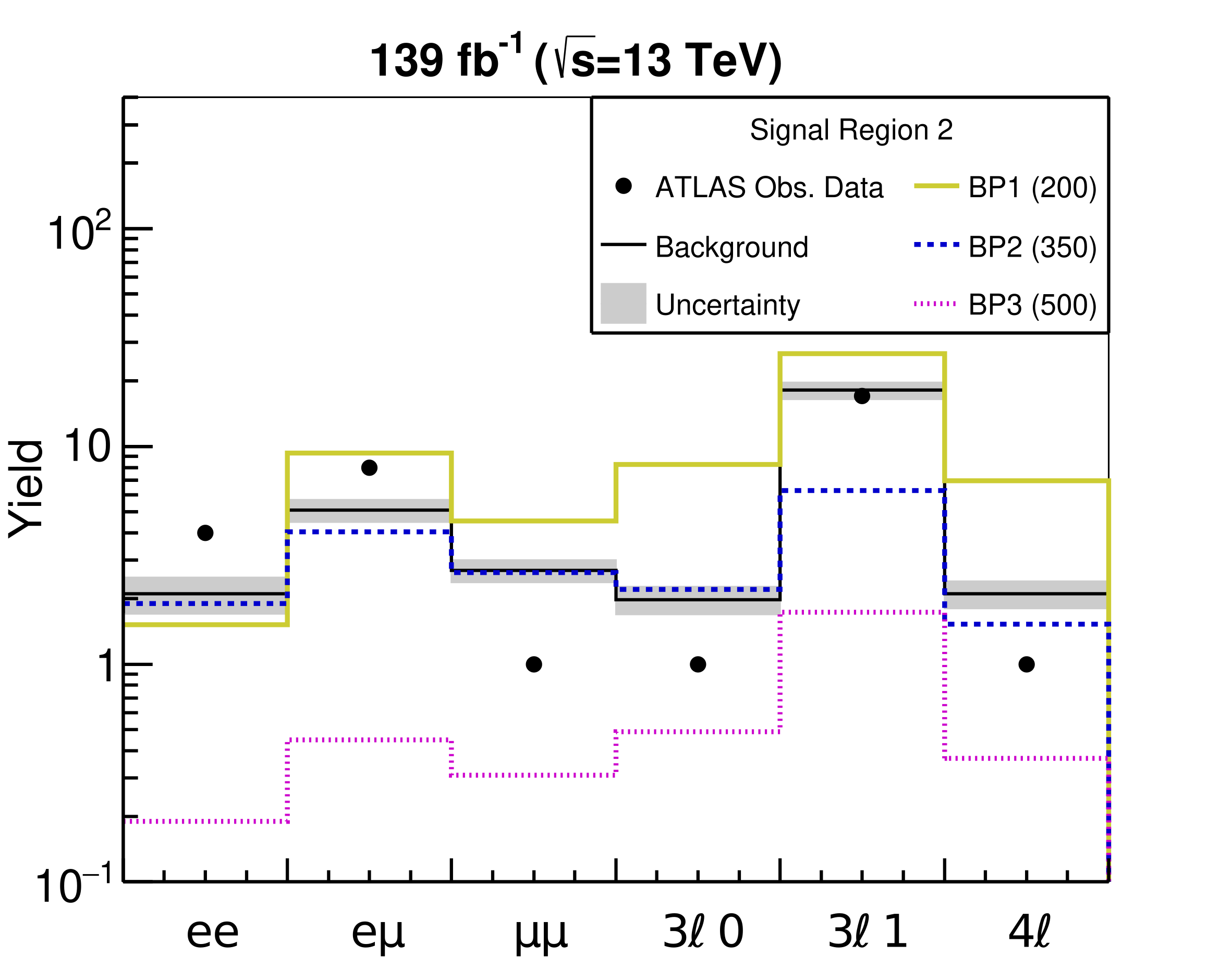}

\includegraphics[scale=0.44]{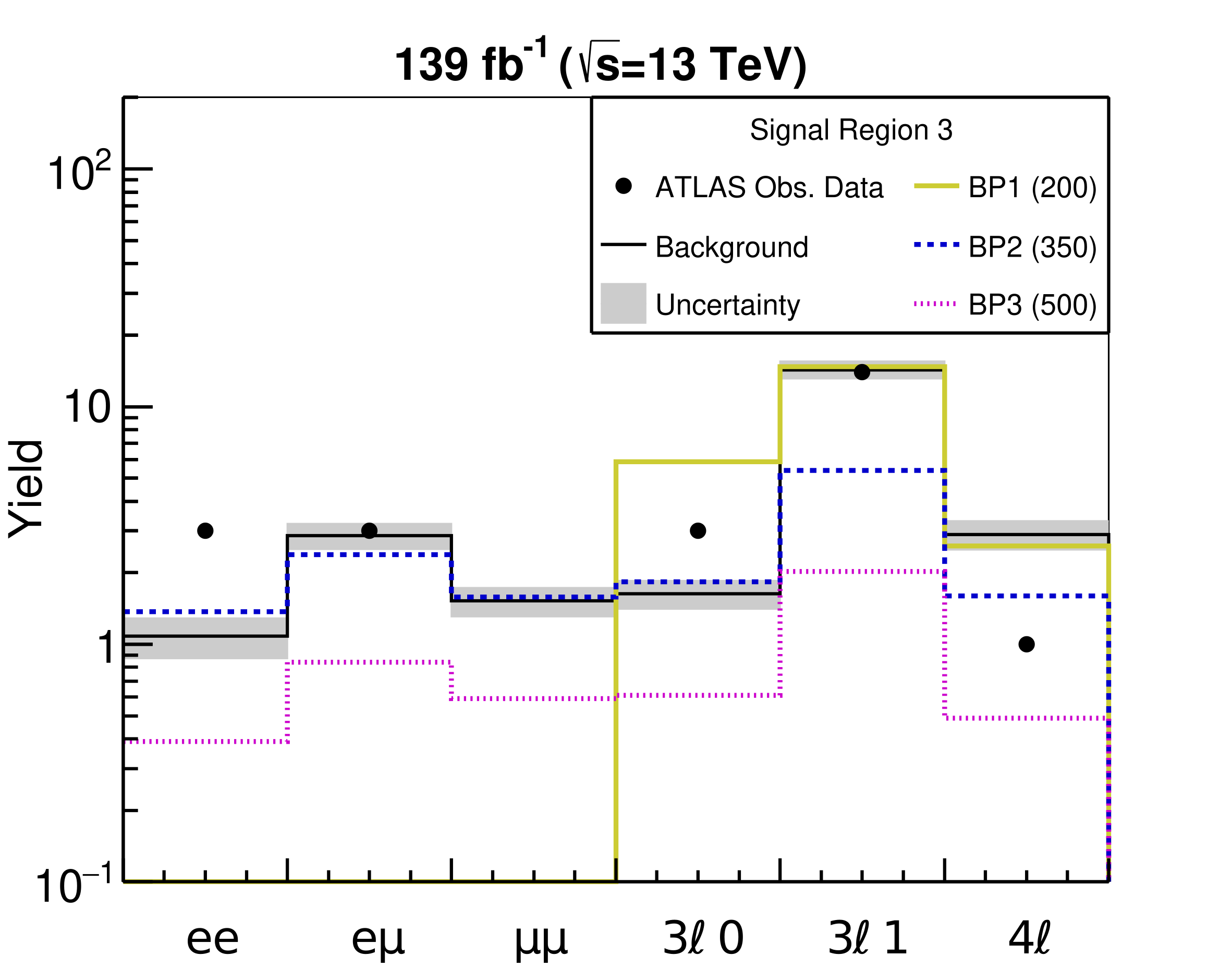}
\includegraphics[scale=0.44]{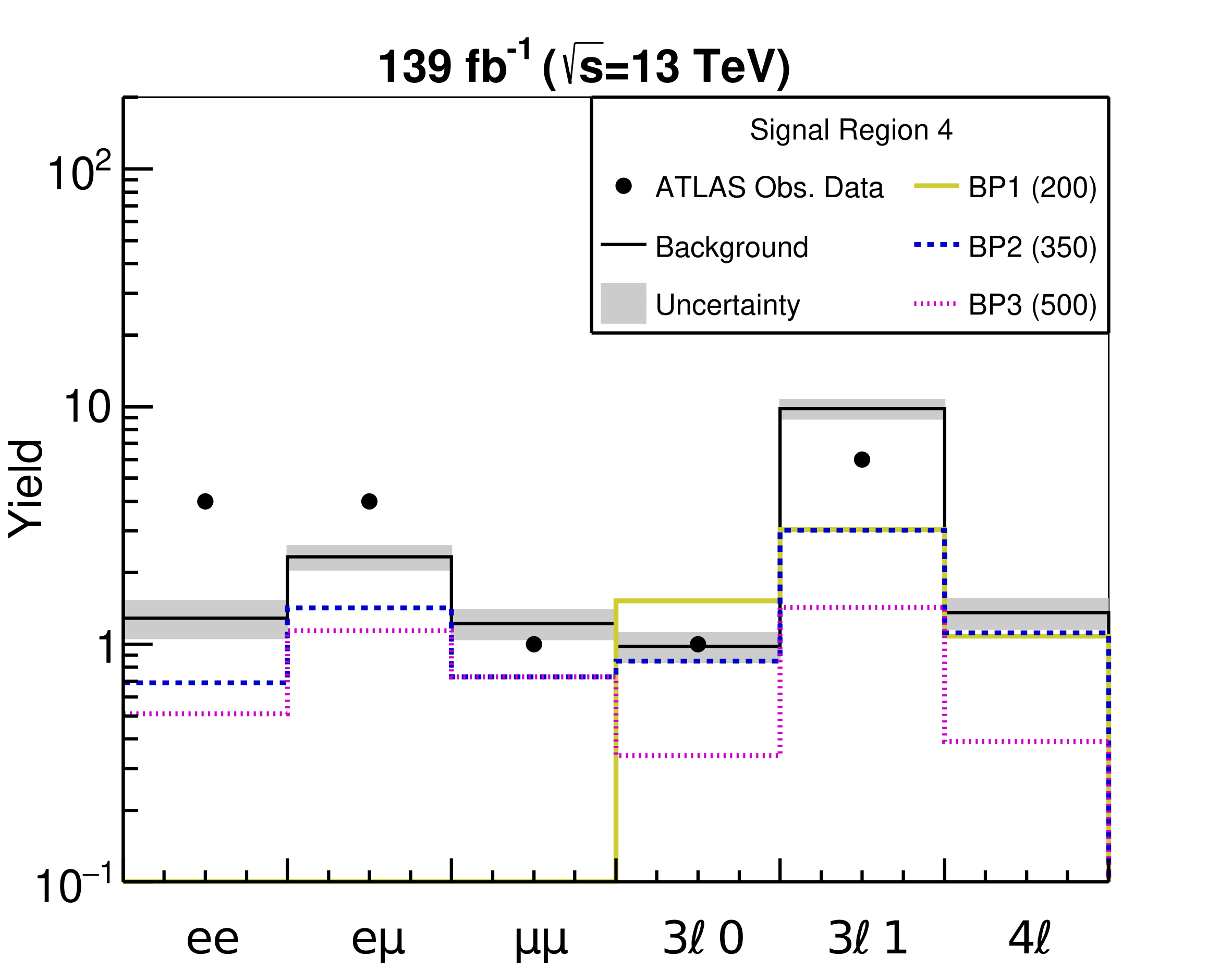}
\caption{The expected SM background events, the observed events and the expected signal events for four different SRs. For each SR, the yields are shown for all the channels. The signal predictions are shown for three benchmark points listed in Table~\ref{atlas_BP}.}
\label{atlas_dist}
\end{figure}
\end{widetext}

%=============================================================================
\subsection{\label{sec:95cl0}95\% CL lower limit on $m_{H^{\pm \pm}}$}
%=============================================================================
In view of the observations being consistent with the SM background expectations, it is tempting to derive limits on $m_{H^{\pm \pm}}$ using the LHC searches. In what follows, we present stringent limits with 95\% CL on $m_{H^{\pm \pm}}$ for a wide range of $v_t$ and $\Delta m$. In deriving the present limits, we use the above-described CMS multilepton and ATLAS multiboson leading to multilepton searches.

\begin{widetext}

\begin{figure}[htb!]
\centering
\includegraphics[width=0.32\columnwidth]{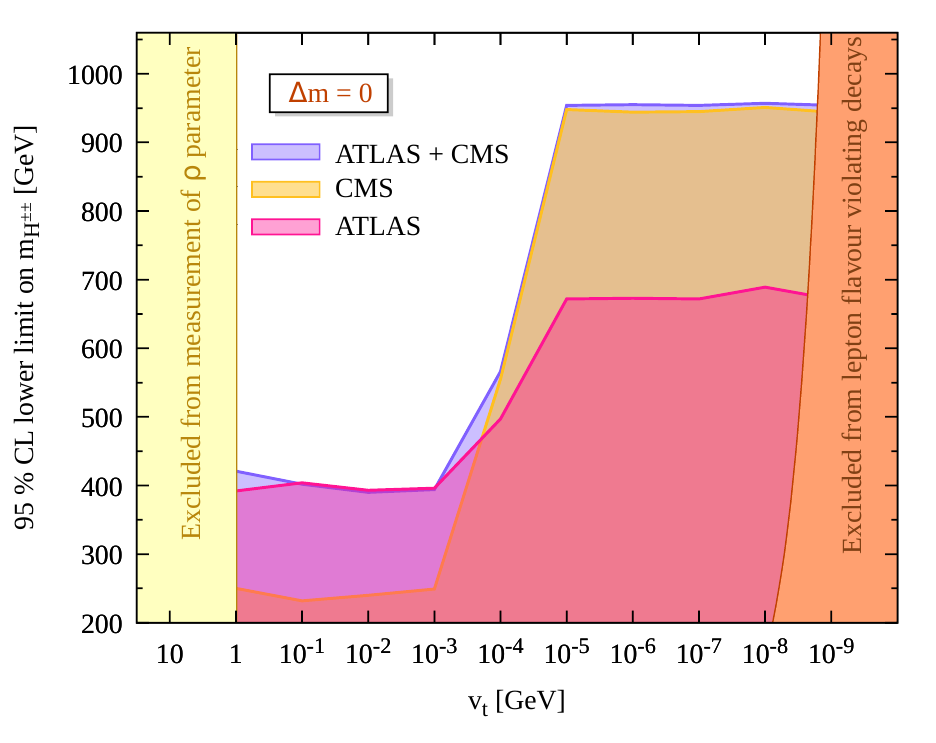}
\includegraphics[width=0.32\columnwidth]{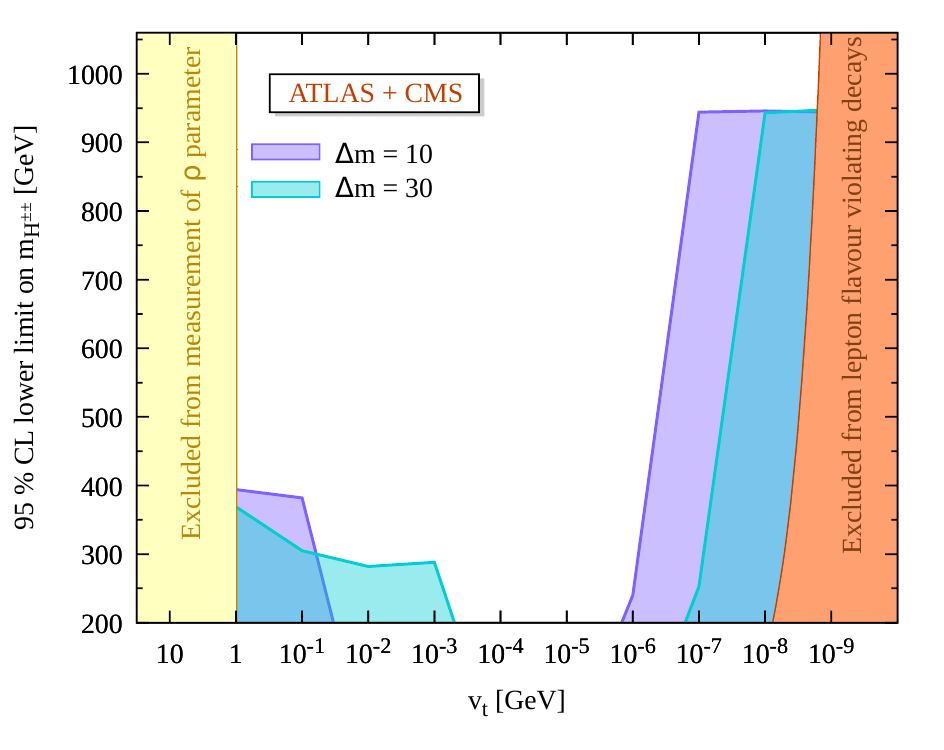}
\includegraphics[width=0.32\columnwidth]{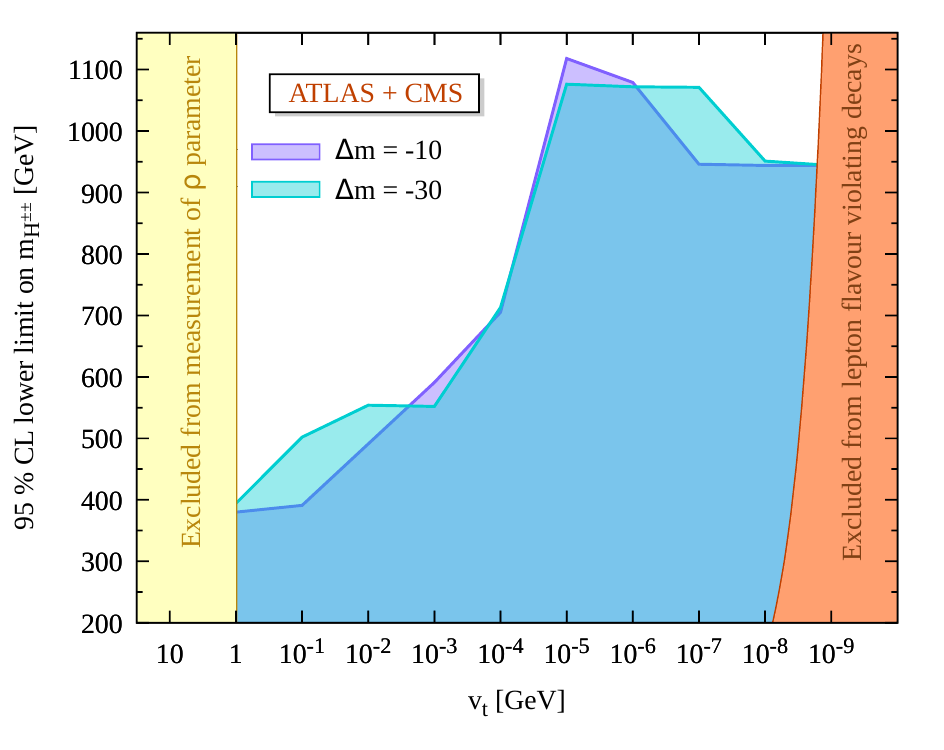}
\caption{Left: 95\% CL lower limits on $m_{H^{\pm \pm}}$ as a function of $v_t$ for $\Delta m=0$ assuming NH with $m_1 = 0.03$ eV. The shaded regions are excluded from different searches at the 13 TeV LHC. Middle(Right): Excluded regions from the CMS and ATLAS combined search for $\Delta m=10$ and $30$ ($-10$ and $-30$) GeV.}
\label{dm0}
\end{figure}

\end{widetext}

The left plot in Figure~\ref{dm0} shows 95\% CL lower limits on $m_{H^{\pm \pm}}$ as a function of $v_t$ for $\Delta m=0$ assuming NH neutrino mass spectrum with $m_1 = 0.03$ eV. The khaki shaded region (on the left) is excluded from the $\rho$ parameter measurement from the electroweak precision data, whereas the coral shaded region (on the right) is excluded from the lepton flavour violating decay constraints. The dark goldenrod and pink shaded regions are excluded, respectively, from the CMS multilepton and the ATLAS multiboson leading to multilepton searches. For small $v_t$, the triplet-like scalars with masses below 950 GeV are excluded from the CMS search. This exclusion limit is beyond those from the previous LHC searches \cite{CMS:2017pet,Aaboud:2017qph} by approximately 200--230 GeV. For large $v_t$, the above-described ATLAS search excludes the triplet-like scalars up to 400 GeV masses which is beyond the ATLAS limit in Ref.~\cite{Aad:2021lzu} by approximately 50 GeV.

Given that the CMS and ATLAS searches are mutually exclusive, it is reasonable to combine them. That said, as these two searches are primarily designed to target different regions in the parameter space, {\it viz.} small $v_t$ and large $v_t$, we expect only marginal improvement on the limits while combining them. The purple shaded region shows excluded parameter space when these two searches are combined. We find that, as expected, improvement on the limit is marginal. The middle (right) plot shows excluded regions from the CMS and ATLAS combined search for $\Delta m=10$ and $30$ ($-10$ and $-30$) GeV.

For very large/small $v_t$, the limits for $\Delta m=\pm 10,\pm 30$ GeV are similar to those for $\Delta m=0$ case. This is because the cascade decays are yet to kick off for very large/small $v_t$. This makes the non-degenerate scenario identical to the degenerate one. For moderate $v_t$ and large enough $\Delta m$, the cascade decays kick-off and swiftly dominates over the other decay modes (see Figure ~\ref{Br500}). In positive scenario, $H^{\pm \pm}$ and $H^\pm$ decay into off-shell $W^\pm$s (which gives rise to soft jets/leptons) and $H^0/A^0$ (which further decays invisibly into neutrinos or into $h^0h^0,ZZ/h^0Z$ depending on the value of $v_t$). For $H^0/A^0$ decaying into neutrinos, there are hardly visible objects in the final state, so much as the monojet search by ATLAS \cite{Aad:2021egl} and the soft leptons search by CMS \cite{CMS:2021xji,Sirunyan:2018iwl} fall short in constraining this part of the parameter space, see the middle plot in Figure~\ref{dm0}. On the contrary, for $H^0/A^0$ decaying into $h^0h^0,ZZ/h^0Z$, the signal cross-section is small compared to the overwhelming background from either QCD jets or Drell-Yan processes. This makes such a scenario challenging to probe. Note that for $v_t \sim \mathcal{O}(10^{-2})$--$\mathcal{O}(10^{-3})$ GeV, the ATLAS search manages to put some bounds in the $\Delta m=30$ case, but it fails in the $\Delta m=10$ case. This is because for larger $\Delta m$, some of the leptons from the off-shell $W^\pm$'s pass the object reconstruction and selection criteria to contribute to the signal yields, whereas the leptons are too soft to do so for smaller $\Delta m$. As one approaches towards small $v_t$, the leptonic decays retrieve their dominance over the cascade one, and give rise to multilepton final states; this occurs at $v_t \sim \mathcal{O}(10^{-6})$ and $\mathcal{O}(10^{-7})$ GeV, respectively, for $\Delta m=10$ and 30 GeV. This has been reflected in the middle plot. On the contrary, in negative scenario, $H^\pm$ and $H^0/A^0$ decay into off-shell $W^\pm$'s and $H^{\pm \pm}$, thereby enhancing the effective production cross-section for $H^{\pm \pm}$. Therefore, in such a scenario, the limit gets enhanced compared to the degenerate case, see the right plot in Figure~\ref{dm0}. For $\Delta m=-10$($-30$) GeV, the exclusion limit extends up to 1115(1076) GeV compared to 955 GeV for $\Delta m=0$. Note that for a given $m_{H^{\pm \pm}}$, $H^\pm$ and $H^0/A^0$ are lighter in the $\Delta m=-10$ GeV case compared to those in the $\Delta m=-30$ GeV case. Thus, the signal cross-section is larger for $\Delta m=-10$ GeV than for $\Delta m=-30$ GeV. This explains the stronger limits for $\Delta m=-10$ GeV than $\Delta m=-30$ GeV.

%========================================================================================
\subsection{\label{sec:proposed}Proposed multilepton final states search for small $v_t$}
%========================================================================================
For small $v_t$ and $\Delta m=0$, the triplet-like scalars up to 950 GeV masses are excluded from the CMS multilepton search with 139 fb$^{-1}$ of data, see Figure~\ref{dm0}. Given the inappreciable signal cross-section for $m_{H^{\pm \pm}} > 1$ TeV and comparatively sizeable backgrounds in the afore-discussed CMS multilepton search, a similar search at high-luminosity is prophesied not to be sensitive enough in probing the triplet-like scalars much heavier than 1 TeV. Also, note that the said CMS search, which results in the most stringent limits in the small $v_t$ region, is not designed to probe the triplet-like scalars. That said, to probe them, one may envisage a multilepton search such that the background contributions cease to survive. In this section, we delineate a multilepton search that is optimised to probe the triplet-like scalars much heavier than 1 TeV in the small $v_t$ region. In what follows, we give a brief description about reconstruction and selection of various objects (jets, leptons, {\it etc.}), event selection and classification of selected events into mutually exclusive signal regions (SRs) for our proposed multilepton final states search.\\

\noindent {\bf Object reconstruction and selection:} Different physics objects, {\it viz.} jets, electrons, muons and missing transverse energy, are reconstructed in Delphes \cite{deFavereau:2013fsa}. Jets are reconstructed using the anti-kT algorithm \cite{Cacciari:2008gp} with a distance parameter $\Delta R=0.4$ as implemented in the FastJet package\cite{Cacciari:2011ma}. Reconstructed jets are required to have transverse momentum $p_T>30$ GeV within the central pseudorapidity range $|\eta|<2.5$. Electron (muon) candidates with $p_T > 10$ GeV and $|\eta|<2.5(2.4)$ are considered for further analysis. For the electron candidates within barrel (endcap), we demand a maximum 5-10\% (5-15\%) $p_T$-dependent relative isolation with $\Delta R=0.4$, \footnote{The relative isolation is defined as the scalar $p_T$ sum, normalized to the lepton $p_T$, of photons and hadrons within a cone of $\Delta R$ around the lepton. For electrons, this is required to be smaller than $0.0478+0.506/p_T$ ($0.0658+0.963/p_T$) within barrel (endcap) {\it i.e.,} $|\eta|<1.479$ ($|\eta|>1.479$) with $\Delta R=0.3$.} whereas we demand a maximum 15\% relative isolation with $\Delta R=0.4$ for muons. In addition, the following set of lepton displacement requirements on the transverse and longitudinal impact parameters, $d_z$ and $d_{xy}$, with respect to the primary vertex are enforced. For electron candidate within barrel (endcap), we demand $d_z<1$ mm and $d_{xy}<0.5$ mm ($d_z<2$ mm and $d_{xy}<1$ mm), whereas muon candidates require $d_z<1$ mm and $d_{xy}<0.5$ mm. Lepton isolation which trims hadronic activity inside the isolation cone along with impact parameter requirements come in handy in vanquishing the reducible backgrounds such as $Z$+jets and $t\overline{t}$+jets, where a jet is misidentified as lepton or additional leptons originate from heavy quark decays. Finally, the missing transverse momentum $p_T^{\rm miss}$ is computed as the magnitude of the negative vector sum of the transverse momenta of all reconstructed particle-flow objects in an event.

Overlaps between reconstructed objects resulting in ambiguities among them lead to object double counting. To sidestep that, all selected jets within a cone of $\Delta R<0.4$ of a selected lepton are thrown away. In addition, all selected electrons within a cone of $\Delta R<0.05$ of a selected muon are discarded as these are likely due to bremsstrahlung interactions of the muon with the inner detector material. Some of the jets, especially those on the tail of the detector response, and single pions could mimic lepton signatures and could be misidentified as leptons. Though the composition of the fake-lepton background differs substantially among the analysis channels, without going into the intricacy of modelling the fake-lepton contributions, we straightforwardly take the probability of 0.1--0.3\% \cite{ATLAS:2016iqc} for a jet to be misidentified as a lepton. Furthermore, bremsstrahlung interactions of the electrons with the inner detector material could lead to charge misidentification. The radiated photon converts to $e^-e^+$ pair near the primary electron trajectory leading to charge misidentification ambiguity. Also, the photon could traverse the inner detector without creating any track. In such a case, the electron usually has a short lever arm on its curvature. This could lead to incorrect determination of the electron charge. We adopt the charge misidentification probability from Ref.~\cite{ATLAS:2017xqs}: $P(p_T,\eta)=\sigma(p_T) \times f(\eta)$, where $\sigma(p_T)$ is found to be 0.02--0.1 and $f(\eta)$ is found to be 0.03--1 such that $P(p_T,\eta)$ ranges from 0.02\% to 10\%. Note that the high-$p_T$ electrons are more likely to be affected by charge misidentification as they have almost straight tracks, thereby making the curvature measurement very challenging. Also, the electrons with larger $\eta$ have a larger misidentification probability as they traverse through a higher amount of inner detector material.\\

\noindent {\bf Event selection and signal region definition:} Events with three or more light leptons are considered for this search. Events containing a lepton pair with $\Delta R<0.4$ or a same-flavour lepton pair with invariant mass below 12 GeV are vetoed. This subdues background contributions from final-state radiations as well as low-mass resonances --- Drell-Yan processes and neutral mesons. Furthermore, events containing a same-flavour lepton pair with an invariant mass within the nominal Z-boson mass window, {\it i.e.} $M_Z \pm 15$ GeV are discarded. \footnote{Note that we have relaxed the opposite charge condition for the same-flavour lepton pair to suppress the background contributions due to charge misidentification.} This suppresses background contributions from the $Z\to \ell \ell^* \to \ell \ell \gamma (\to \ell \ell)$ process as well as the $WZ$ production. Events with exactly three light leptons (3L) in one category and four or more light leptons (4L) in another category are considered for further analysis.

Noting that the triplet-like scalars, which are to be probed, are heavier than 1 TeV, we persuade to exploit the relatively high momenta of their decay products. Before continuing, let us briefly reckon the processes contributing to the 3L and 4L signal events. For 3L events, the dominant contribution arises either from the $H^{\pm \pm}H^\mp \to \ell^\pm \ell^\pm \ell^\mp \nu$ process or from the the $H^{++}H^{--} \to \ell^\pm \ell^\pm \ell^\mp \tau^\mp$ process with $\tau^\pm$ decaying hadronically. Therefore, the invariant mass distribution of the same-sign lepton pair is expected to peak at $m_{H^{\pm \pm}}$. One would expect high-$p_T$ leptons, handsome $p_T^{\rm miss}$ and no high-$p_T$ jet (except for those coming from ISR and FSR) in the final states for the former. For the latter, one would expect final states with high-$p_T$ leptons, small $p_T^{\rm miss}$ and at least one handsome-$p_T$ jet. Then, the dominant contribution to the 4L signal events comes from the $H^{++}H^{--} \to \ell^+ \ell^+ \ell^- \ell^-$ process. The invariant mass distributions of both the same-sign lepton pairs are expected to peak at $m_{H^{\pm \pm}}$. Once again, one would expect high-$p_T$ leptons and small $p_T^{\rm miss}$ in the final states.\\

\noindent {\bf SM Backgrounds:} A number of SM processes which could mimic the multilepton final states are considered as relevant backgrounds in this analysis. The relevant backgrounds includes $ZZ$, $WZ$, $WW$, $Zh$, $Wh$, $t\bar{t}$, $t\bar{t}W$, $t\bar{t}Z$, $t\bar{t}h$, $WWW$, $WWZ$, $WZZ$, $ZZZ$, $ZZh$, $WWh$, $t\bar{t}t(\bar{t})$, $t\bar{t}t\bar{t}$ and Drell-Yan processes. These backgrounds can be classified into two classes --- reducible and irreducible backgrounds. The reducible backgrounds are from the SM processes like $Z/\gamma^*+$jets, $t\bar{t}+$jets, {\it etc.}, where a jet is misidentified as lepton or additional leptons originate from heavy quark decays. The irreducible ones are from diboson and triboson production and processes like $t\bar{t}W$, $t\bar{t}Z$ and Higgs boson production, {\it etc}. Note that final state events with $n$ leptons also contribute to those with $n-1$ leptons when one of the leptons falls outside the detector coverage (in the high rapidity region) or is too soft to pass the object reconstruction and selection criteria or gets misidentified by the detector. All the background events are generated in association of up to two jets using MadGraph \cite{Alwall:2011uj,Alwall:2014hca} at the leading order using the 5 flavour scheme followed by MLM matching in PYTHIA \cite{Sjostrand:2014zea}, and the corresponding cross-sections are taken at least upto NLO \cite{Campbell:1999ah,Catani:2007vq,Campanario:2008yg,Balossini:2009sa,Bredenstein:2009aj,Catani:2009sm,Campbell:2011bn,Bevilacqua:2012em,Garzelli:2012bn,Nhung:2013jta,Kidonakis:2015nna,Muselli:2015kba,Shen:2015cwj}.

We plot different kinematic distributions for 3L events in Figure~\ref{3L} for a benchmark point $\bf BP1$, defined as $m_{H^{\pm \pm}}=1$ TeV, $v_t \sim 10^{-8}$ GeV and $\Delta m=0$. The first two plots of the top panel show the transverse momentum distributions of the leading and subleading lepton in the same-charge lepton pair. The effective mass, defined as $m_{\rm eff}=H_T+L_T+p_T^{\rm miss}$ with $H_T(L_T)$ being the scalar sum of transverse momenta of all the jets (leptons), distribution is shown in the rightmost plot of the same panel. The bottom panel shows distributions of the missing transverse momentum and the invariant mass of the same-sign lepton pair ($m_{\ell \ell}^{\rm sc}$).\footnote{In an ideal scenario, the invariant mass distribution in the bottom left plot would have a sharp peak at 1 TeV. However, momentum smearing of the reconstructed objects due to finite resolution of the detectors results in much broader peaks around 1 TeV.} These kinematic distributions demonstrate that relatively stronger cuts on the same-sign leptons' $p_T$ and the $m_{\rm eff}$ appreciably reduce the relevant backgrounds. 

For 3L events, we require one same-charge lepton pair. The leading (subleading) lepton in the pair is required to have $p_T>300(100)$ GeV. We discard events with $m_{\rm eff}<1500$ GeV. To enhance the sensitivity of this search, the selected events are categorised into two mutually exclusive SRs, namely $3L0J$\footnote{Three leptons events with no reconstructed jet with $p_T >30$ GeV are considered in the $3L0J$ SR.} and $3L1J$, based on whether or not at least one selected jet exists in the event. $3L1J$ events are further classified as $3L1J$-1 and $3L1J$-2 based on whether $p_T^{\rm miss}$ is larger or smaller than 150 GeV. The $3L0J$ events with $p_T^{\rm miss}<150$ GeV or $m_{\ell \ell}^{\rm sc}<800$ GeV are thrown away to get rid of the sizeable SM backgrounds. Furthermore, we reject $3L1J$-1 events with $p_T^{\rm miss}/{H_T}<1.0$. The cut on $p_T^{\rm miss}/{H_T}$ turns out to be remarkably effectual in diminishing the leftover backgrounds. Finally, to supplement the sensitivity of this search, the selected events in $3L1J$-1 and $3L1J$-2 SRs are divided into six bins each in the [600:1800] GeV range using $m_{\ell \ell}^{\rm sc}$ as the primary kinematic discriminant.\footnote{The overflow (underflow) events are contained in the last (first) bin in each signal region.}

\begin{comment}
\begin{table}[htb!]
\centering
\scalebox{0.85}{
\begin{tabular}{|c|c|c|c|c|c|c|c|c|}
\hline 
SR & $p_T(\ell^{\rm sc}_0)$ & $p_T(\ell^{\rm sc}_1)$ & $m_{\rm eff}$ & $n_{\rm jet}$ & $p_T^{\rm miss}$ & $m_{\ell \ell}^{\rm sc}$ & $\frac{p_T^{\rm miss}}{m_{\rm eff}}$ & $\frac{p_T^{\rm miss}}{H_T}$ \\
\hline 
$3L0J$ & $>300$ & $>100$ & $>1500$ & 0 & $>150$ & $>800$ & $<0.5$ & -- \\ 
\hline 
$3L1J$-1 & $>300$ & $>100$ & $>1500$ & $\geq 1$ & $>150$ & -- & $<0.5$ & $>1.0$ \\ 
\hline 
$3L1J$-2 & $>300$ & $>100$ & $>1500$ & $\geq 1$ & $<150$ & -- & $<0.1$ & -- \\ 
\hline 
\end{tabular} 
}
\caption{\label{3LSR}3L signal regions. $p_T(\ell^{\rm sc}_0)$ and $p_T(\ell^{\rm sc}_1)$, respectively, denote transverse momentum of the leading and subleading same-charge lepton. $n_{\rm jet}$ denotes jet multiplicity. $p_T(\ell^{\rm sc}_{0,1})$, $m_{\rm eff}$, $p_T^{\rm miss}$ and $m_{\ell \ell}^{\rm sc}$ are in GeV.}
\end{table}
\end{comment}

\begin{widetext}

\begin{figure}[htb!]
\centering
\includegraphics[scale=0.29]{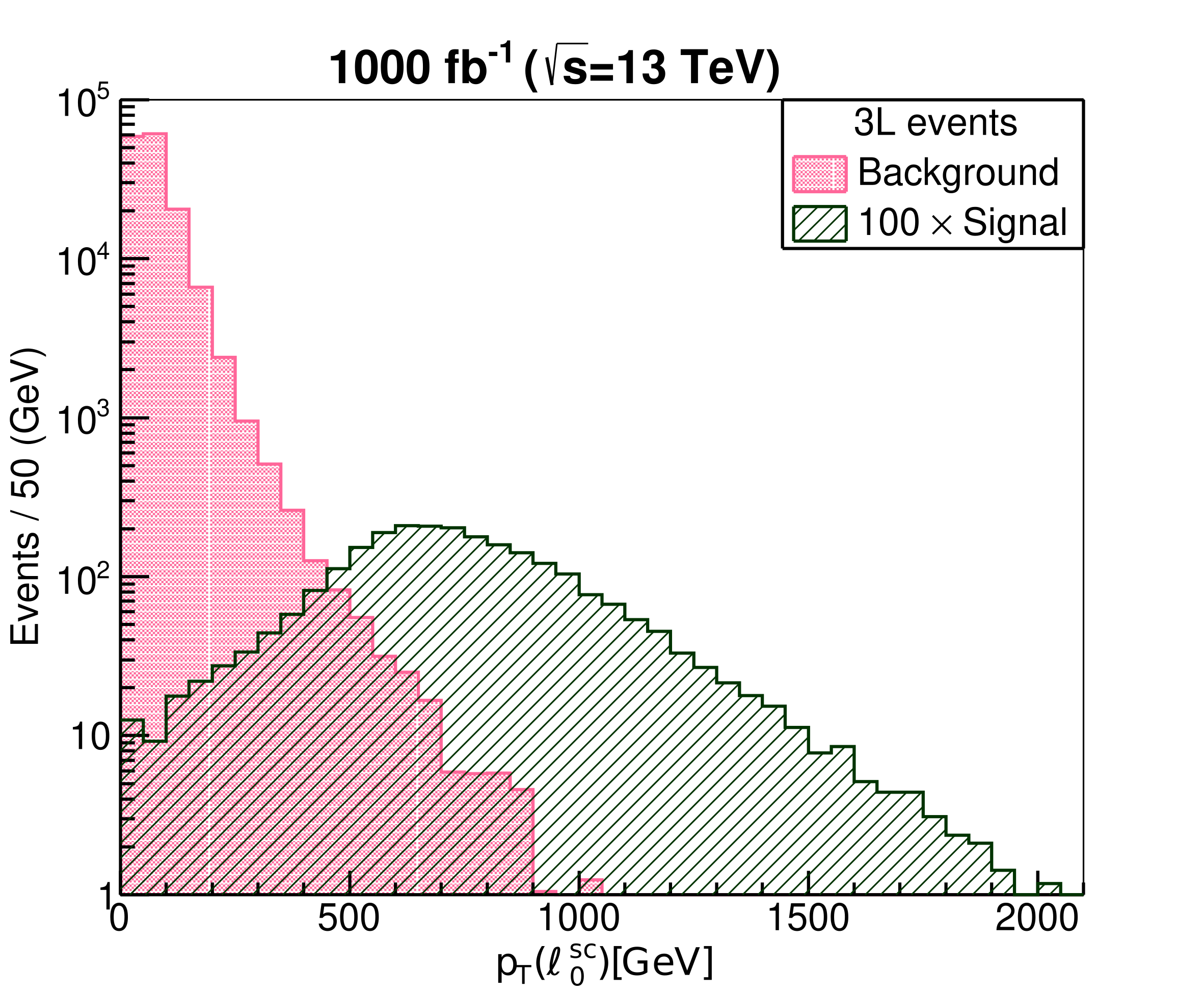}
\includegraphics[scale=0.29]{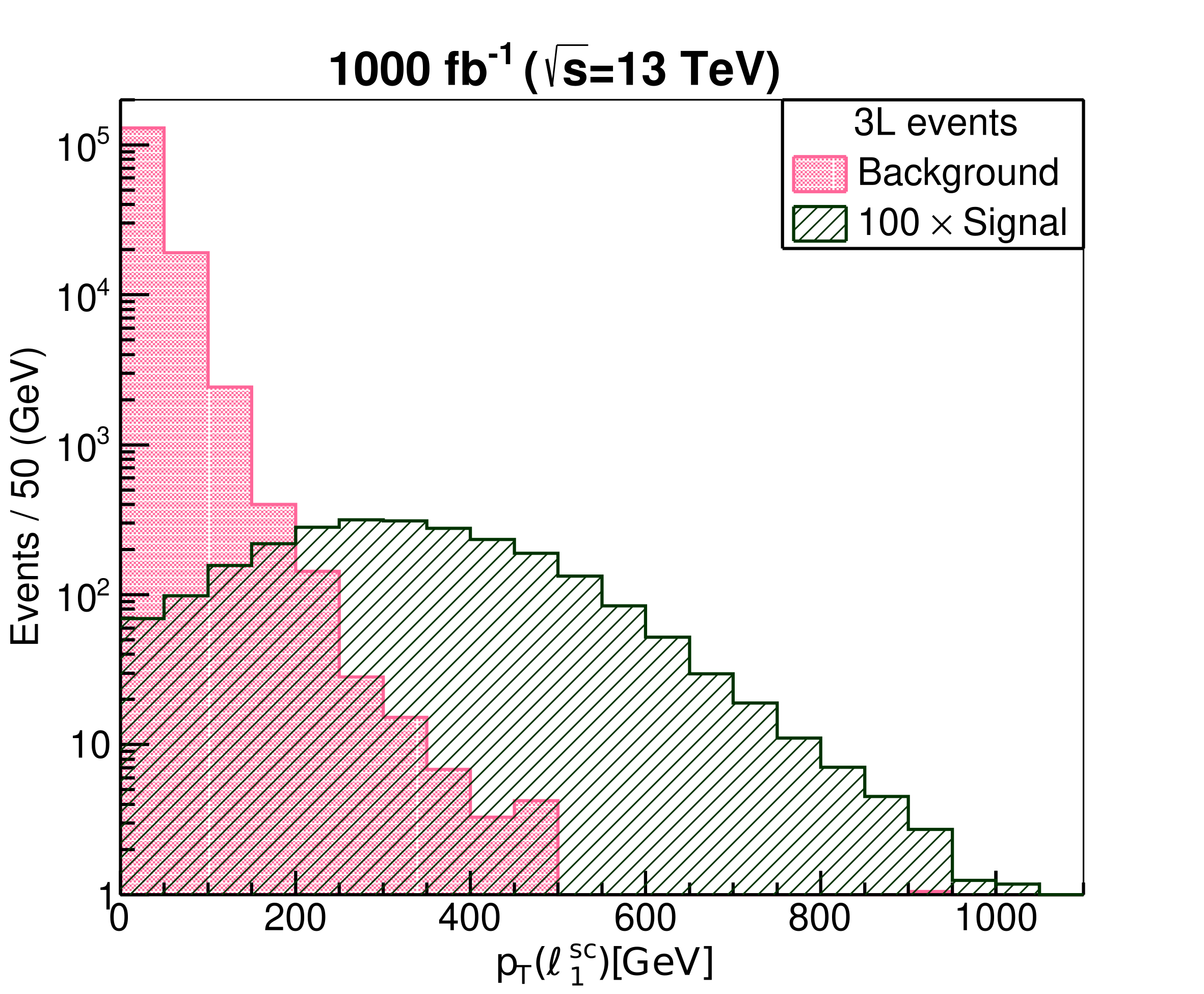}
\includegraphics[scale=0.29]{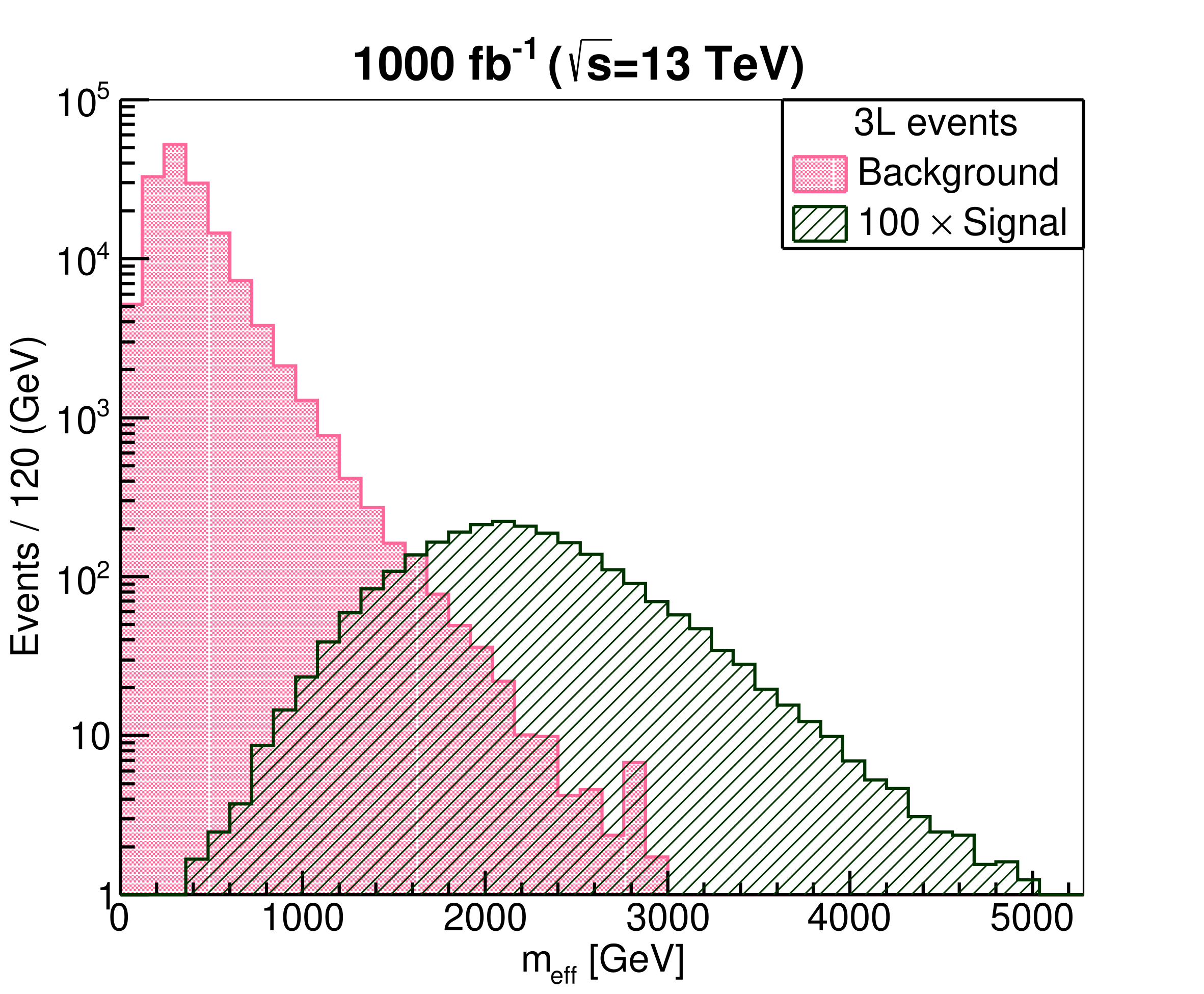}
\includegraphics[scale=0.29]{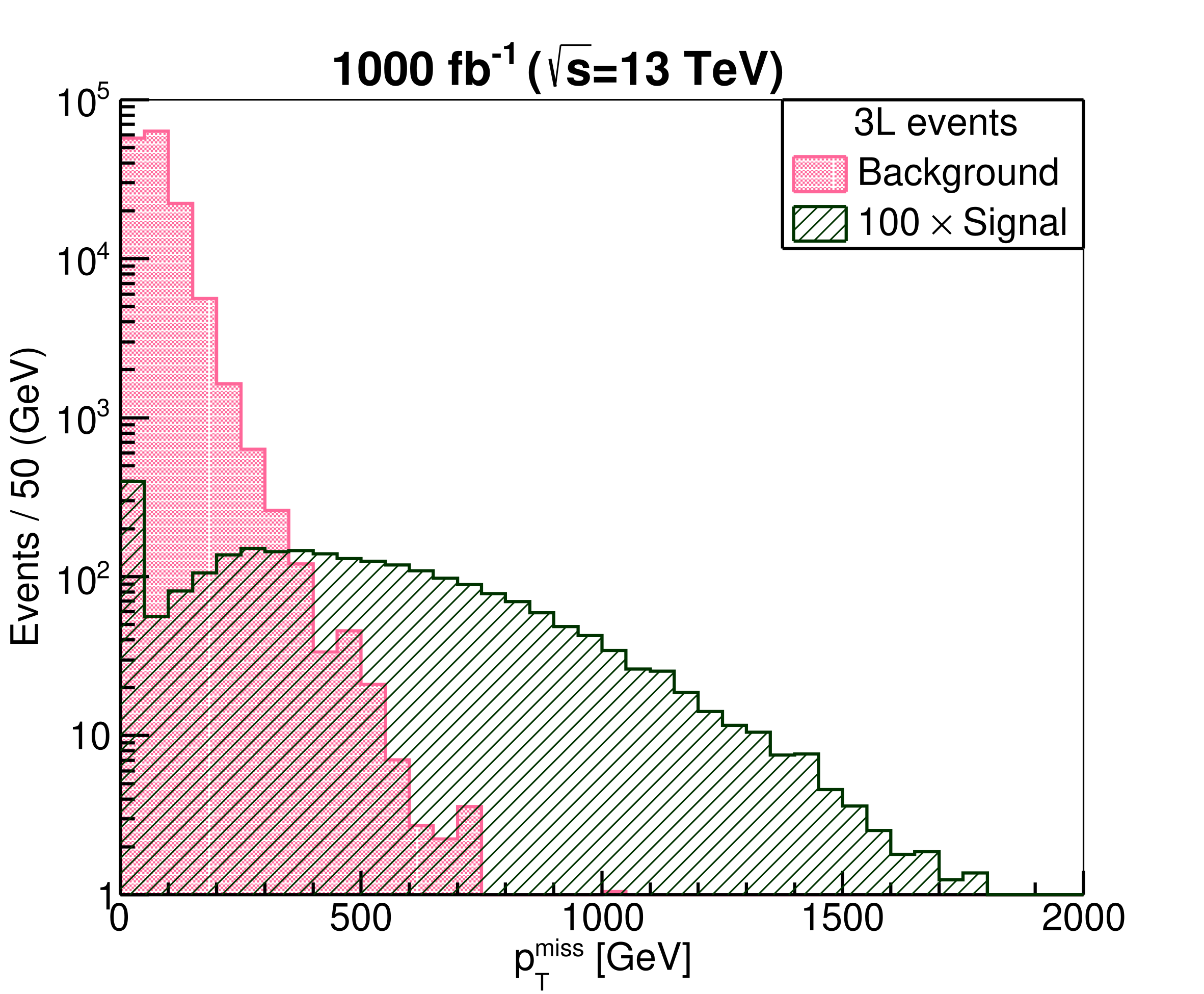}
\includegraphics[scale=0.29]{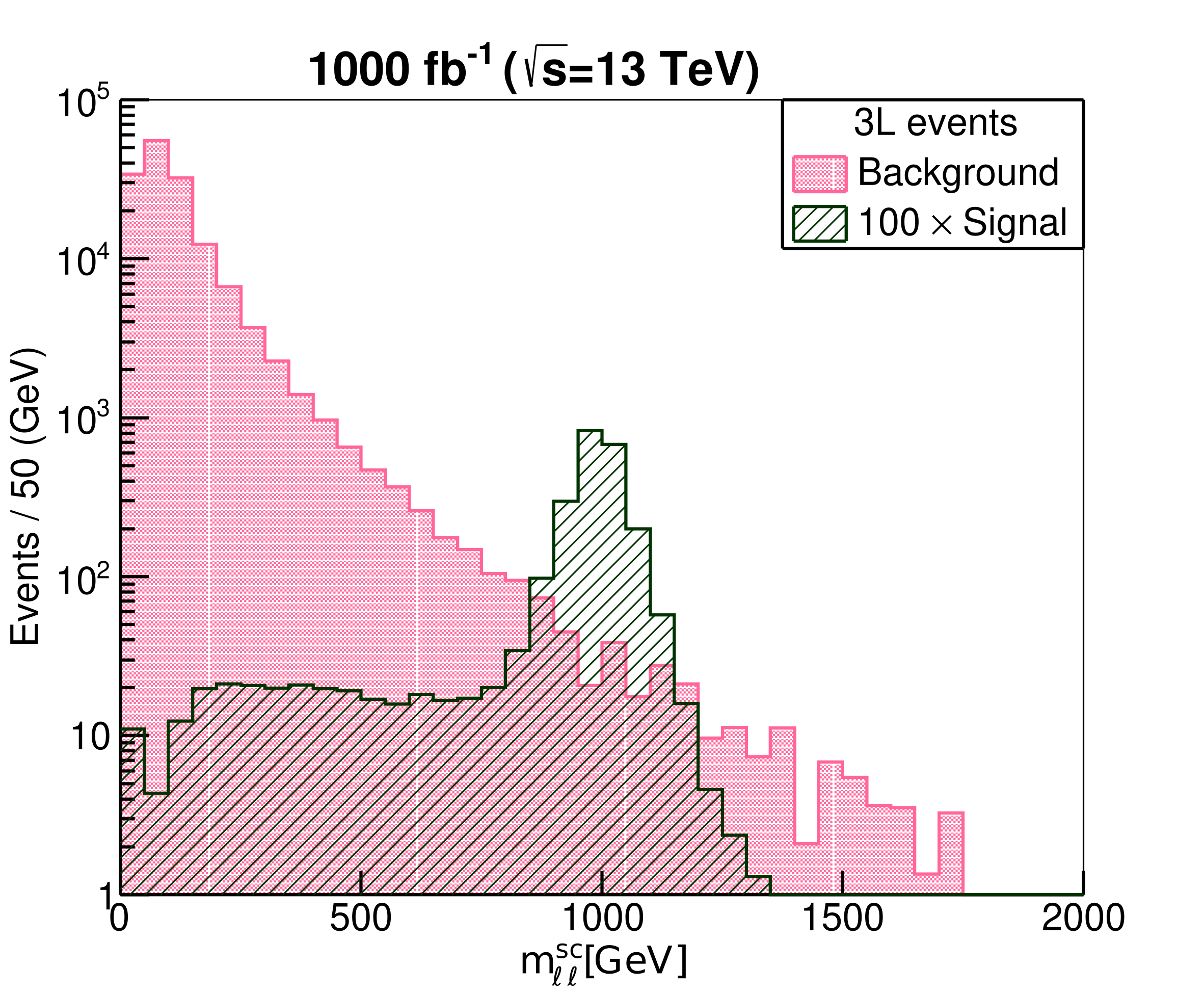}
\caption{Kinematic distributions for 3L events for $\bf BP1$. Top panel: the $p_T$-distributions of the leading (leftmost) and subleading (middle) same-charge lepton, and the $m_{\rm eff}$ distribution (rightmost). Bottom panel: the $p_T^{\rm miss}$ (left) and the $m_{\ell \ell}^{\rm sc}$ (right) distributions. The events are weighted at 1000 fb$^{-1}$ luminosity at the 13 TeV LHC.}
\label{3L}
\end{figure}

\begin{figure}[htb!]
\centering
\includegraphics[scale=0.29]{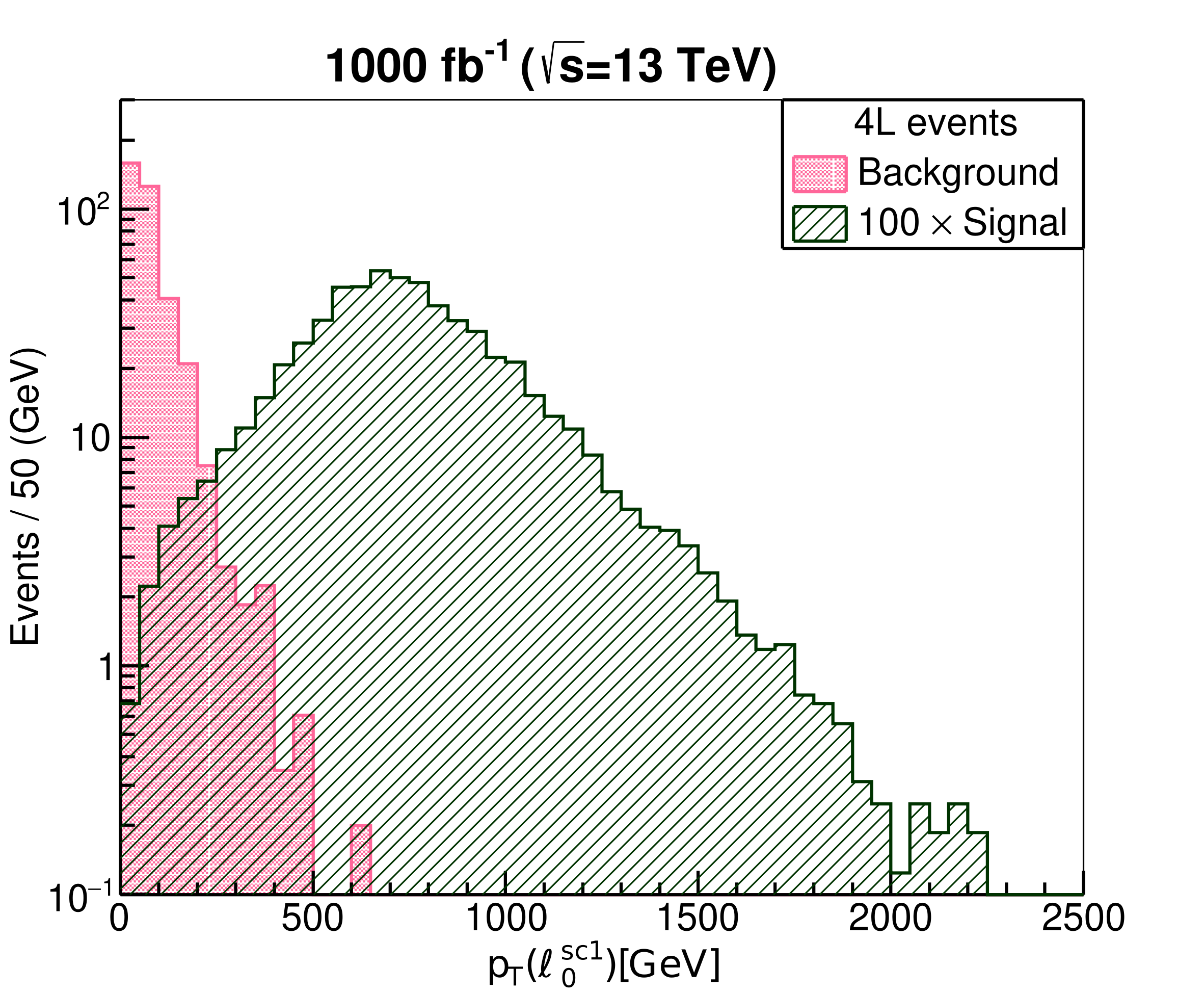}
\includegraphics[scale=0.29]{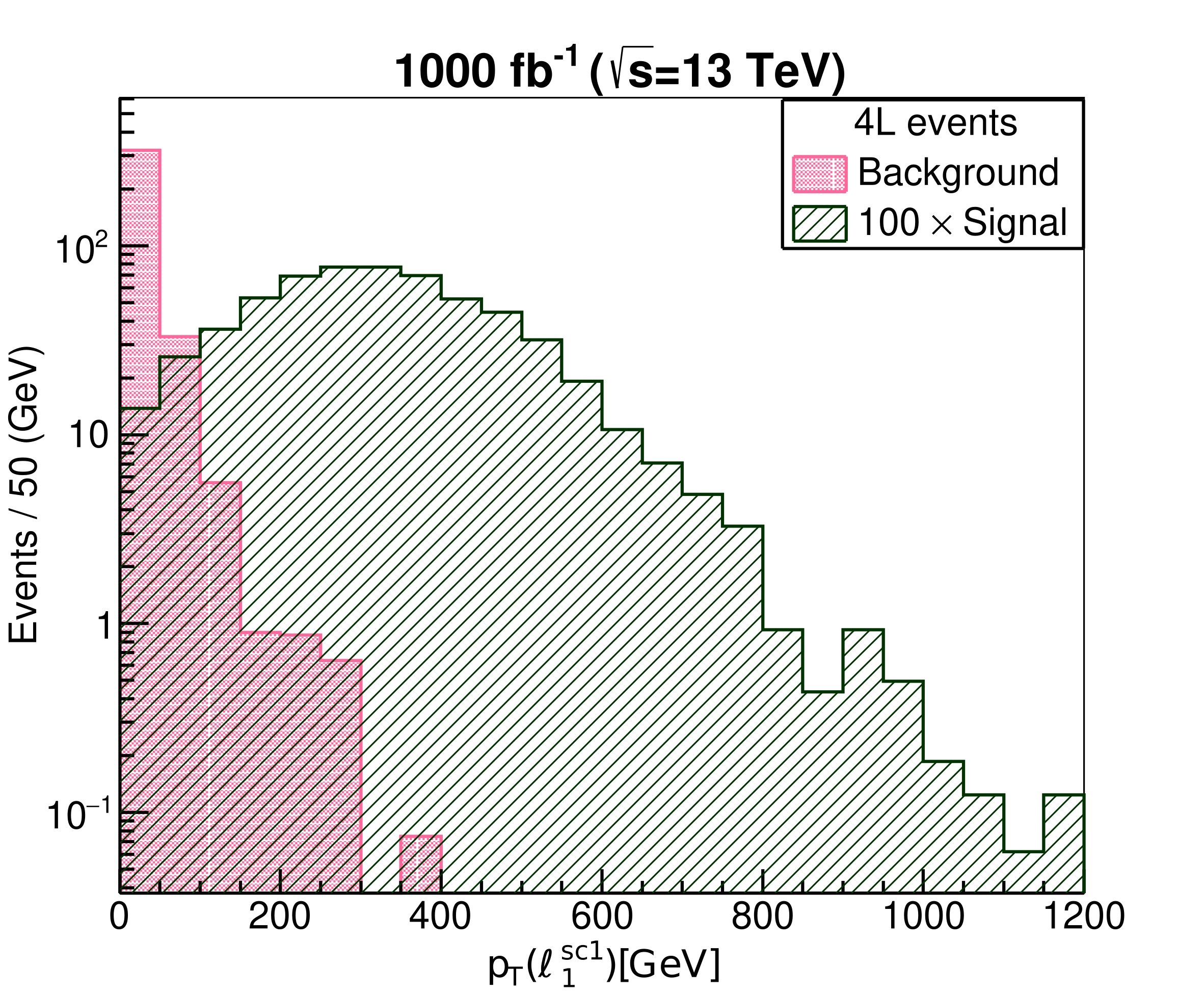}
\includegraphics[scale=0.29]{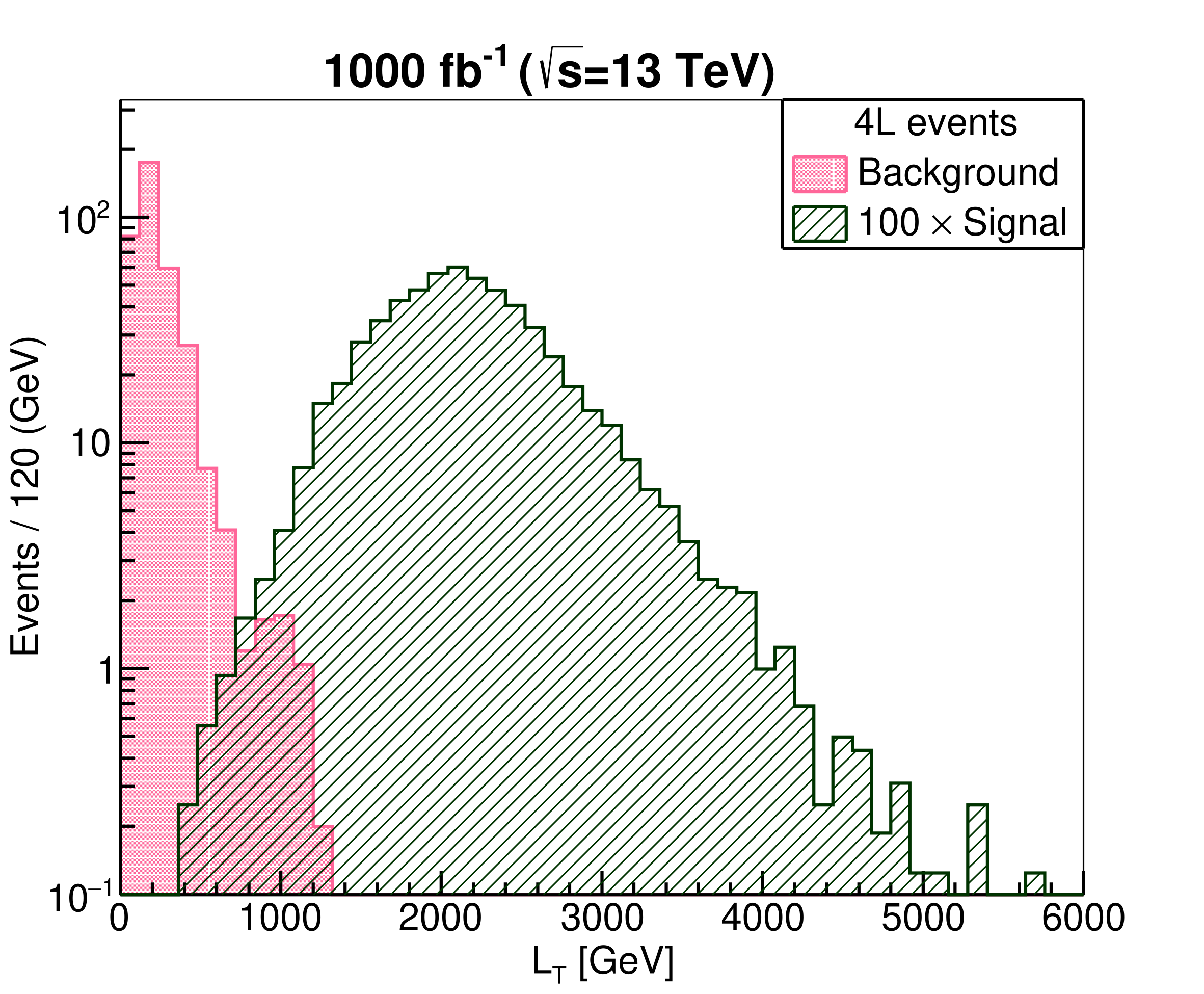}
\caption{Kinematic distributions for 4L events for $\bf BP1$. Left (middle): the $p_T$-distributions of the leading (subleading) positive-charge lepton. Right: the $L_T$ distribution. The events are weighted at 1000 fb$^{-1}$ luminosity at the 13 TeV LHC.}
\label{4L} 
\end{figure}
\end{widetext}

Different kinematic distributions for 4L events are plotted in Figure~\ref{4L} for $\bf BP1$. The leftmost and middle plot in the top panel shows the transverse momentum distributions of the leading and subleading lepton in the positive-charge lepton pair. The leptons in the negative-charge lepton pair have similar $p_T$-distributions, we avert to show them for brevity. The distributions of the scalar $p_T$ sum of the leptons is shown in the rightmost plot in the same panel. It is evident from these kinematic distributions that relatively stronger cuts on the same-sign leptons' $p_T$ comes in handy in subjugating the relevant backgrounds. Further, a cut on $L_T$ turns out to be efficacious in vanquishing the remaining background. 

For 4L events, we require two same-charge lepton pairs. The leading (subleading) lepton in both the pairs are required to have $p_T>300(100)$ GeV. The events with $L_T<1500$ GeV are vetoed. Further, we require $r=|m_{\ell \ell}^{\rm sc1} - m_{\ell \ell}^{\rm sc2}|/(m_{\ell \ell}^{\rm sc1} + m_{\ell \ell}^{\rm sc2}) < 0.1$, where $m_{\ell \ell}^{\rm sc1}$ and $m_{\ell \ell}^{\rm sc2}$ are the invariant masses of the same-charge lepton pairs. The last cut ensures correct pairing of the leptons. %4L signal region is summarised in Table~\ref{4LSR}.

\begin{comment}
\begin{table}[htb!]
\centering
\scalebox{1.0}{
\begin{tabular}{|c|c|c|c|c|c|c|}
\hline 
SR & $p_T(\ell^{\rm sc1}_0)$ & $p_T(\ell^{\rm sc1}_1)$ & $p_T(\ell^{\rm sc2}_0)$ & $p_T(\ell^{\rm sc2}_1)$ & $L_T$ & $\frac{|m_{\ell \ell}^{\rm sc1} - m_{\ell \ell}^{\rm sc2}|}{m_{\ell \ell}^{\rm sc1} + m_{\ell \ell}^{\rm sc2}}$ \\
\hline 
$4L$ & $>300$ & $>100$ & $>300$ & $>100$ & $>1500$ & $<0.1$ \\ 
\hline 
\end{tabular} 
}
\caption{\label{4LSR}4L signal region. $p_T(\ell^{\rm sc1,sc2}_0)$ and $p_T(\ell^{\rm sc1,sc2}_1)$, respectively, denote transverse momenta of the leading and subleading leptons in the same-charge lepton pairs. $p_T(\ell^{\rm sc1,sc2}_{0,1})$ and $L_T$ are in GeV.}
\end{table}
\end{comment}

\begin{figure}[htb!]
\centering
\includegraphics[scale=0.42]{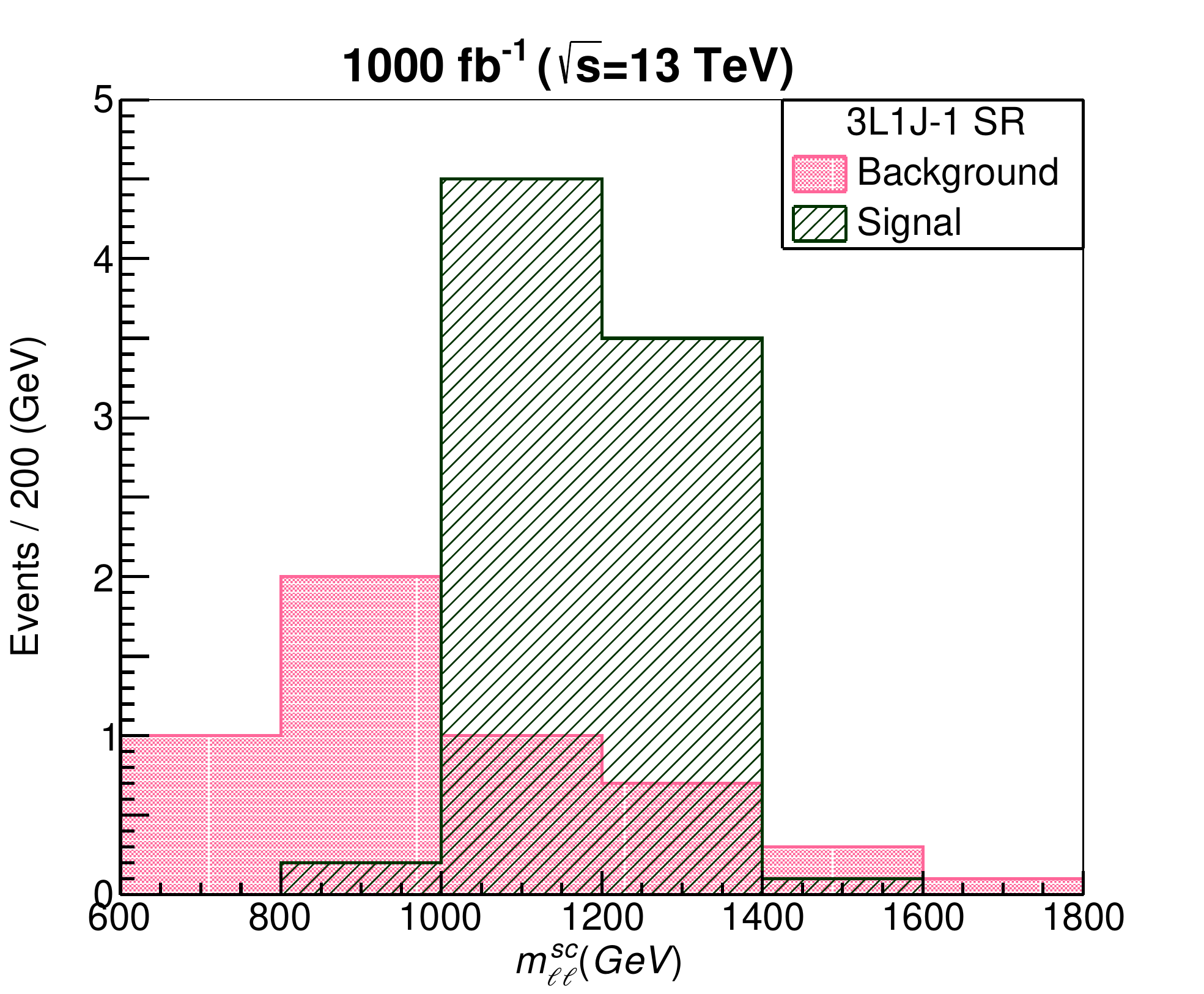}
\caption{$m_{\ell \ell}^{\rm sc}$ distributions of signal and background events in $3L1J$-1 signal regions after passing various selection cuts for $\bf BP1$ for 1000 fb$^{-1}$ of luminosity data at the 13 TeV LHC.}
\label{3L_after_cut} 
\end{figure}

Number of expected signal and background events in different signal regions after passing various selection cuts for $\bf BP1$ for 1000 fb$^{-1}$ of luminosity data at the 13 TeV LHC is are given in Table~\ref{cut_flow} and Table~\ref{cut_flow2}. The $3L0J$ and $4L$ signal regions are free from any background, whereas some backgrounds remain after all the selection cuts in the other two signal regions. However, these remaining backgrounds are mostly distributed in the lower $m_{\ell \ell}^{\rm sc}$ region unlike the signal events which are distributed in a narrow central $m_{\ell \ell}^{\rm sc}$ region, see Figure~\ref{3L_after_cut}. Figure~\ref{3L_after_cut} shows $m_{\ell \ell}^{\rm sc}$ distributions of signal and background events in $3L1J$-1 signal region after passing various selection cuts for $\bf BP1$ for 1000 fb$^{-1}$ of luminosity data at the 13 TeV LHC. For brevity, we avert to show similar distribution for the $3L1J$-2 signal region. This simple binning of the selected events enhances sensitivity of these two signal regions.

\begin{table}[ht]
\centering
\scalebox{1.0}{
\begin{tabular}{|c|c|c|c|}
\hline
SR & Selection cuts & Background & Signal\\
\hline
\multirow{4}{*}{(3L events)} & Basic & 151444 & 24.9 \\
& $p_T(\ell_0^{\rm sc}) > 300$ & 1277 & 23.7 \\
& $p_T(\ell_1^{\rm sc}) > 100$ & 501 & 22.7 \\
& $m_{\rm eff} > 1500$ & 46.0 & 20.8 \\
\hline
\multirow{3}{*}{$3L0J$} & $n_J=0$ \& $p_T^{\rm miss}>150$ & 1.0 & 6.5 \\
& $m_{\ell \ell}^{\rm sc}>800$ & 0.0 & 6.2\\
\hline
\multirow{3}{*}{$3L1J$-1} & $n_J\geq 1$ \& $p_T^{\rm miss} > 150$ & 22.0 & 10.3 \\
& $p_T^{\rm miss}/H_T > 1.0$ & 5.1 & 8.3 \\
\hline
\multirow{1}{*}{$3L1J$-2} & $n_J \geq 1$ \& $p_T^{\rm miss} < 150$ & 23.0 & 3.0 \\
\hline
\end{tabular} 
}
\caption{\label{cut_flow} Number of expected background and signal events in different 3L signal regions after passing various selection cuts for $\bf BP1$ for 1000 fb$^{-1}$ of luminosity data at the 13 TeV LHC. All the dimensionfull cuts are in GeV.}
\end{table}

\begin{table}[ht]
\centering
\scalebox{1.0}{
\begin{tabular}{|c|c|c|}
\hline
Selection cuts & Background & Signal \\
\hline
Basic & 350 & 6.0 \\
$p_T(\ell_0^{\rm sc1}) > 300$ & 5.3 & 5.8 \\
$p_T(\ell_1^{\rm sc1}) > 100$ & 2.0 & 5.5 \\
$p_T(\ell_0^{\rm sc2}) > 300$ & 0.7 & 5.3 \\
$p_T(\ell_1^{\rm sc2}) > 100$ & 0.1 & 5.1 \\
$L_T > 1500$ \& $r<0.1$ & 0.0 & 4.9 \\
\hline
\end{tabular} 
}
\caption{\label{cut_flow2} Number of expected background and signal events in 4L signal region after passing various selection cuts for $\bf BP1$ for 1000 fb$^{-1}$ of luminosity data at the 13 TeV LHC. All the dimensionfull cuts are in GeV.}
\end{table}

%=============================================================================
\subsection{\label{sec:95cl} Future 95\% CL lower limit on $m_{H^{\pm \pm}}$}
%=============================================================================
In this section, we present our forecasted 95\% CL lower limits on $m_{H^{\pm \pm}}$ by using the ATLAS search \cite{Aad:2021lzu} scaled at high-luminosity \footnote{Note that the ATLAS search \cite{Aad:2021lzu} is optimised for probing the large $v_t$ region where the type-II see-saw anchors decay into bosons. Therefore, we use the same search strategy to forecast the future reach of the LHC in probing this part of the parameter space.} as well as our proposed search described in Section~\ref{sec:proposed}. We simply presume that not only the detector efficiencies and acceptances but also the background uncertainties remain the same while scaling the ATLAS search at high luminosity. Given that both statistical and systematic contributions to the background uncertainties are expected to be reduced with increasing volume of LHC data, our forecasted future limits are conservative. Also, while estimating significance for a given signal and background distributions, less than one background event at 3000 fb$^{-1}$ \footnote{All the relevant background events are generated in association of up to two jets using MadGraph \cite{Alwall:2011uj,Alwall:2014hca} at the leading order using the 5 flavour scheme followed by MLM matching in PYTHIA \cite{Sjostrand:2014zea} for an integerated luminosity of 3000 fb$^{-1}$ or more, and the corresponding cross-sections are taken at least upto NLO \cite{Campbell:1999ah,Catani:2007vq,Campanario:2008yg,Balossini:2009sa,Bredenstein:2009aj,Catani:2009sm,Campbell:2011bn,Bevilacqua:2012em,Garzelli:2012bn,Nhung:2013jta,Kidonakis:2015nna,Muselli:2015kba,Shen:2015cwj}.} is replaced by one background event. This renders our estimated limits to be conservative further. For the proposed search, we assume an overall 20\% total uncertainty on the estimated background.

\begin{figure}[htb!]
\centering
\includegraphics[scale=0.9]{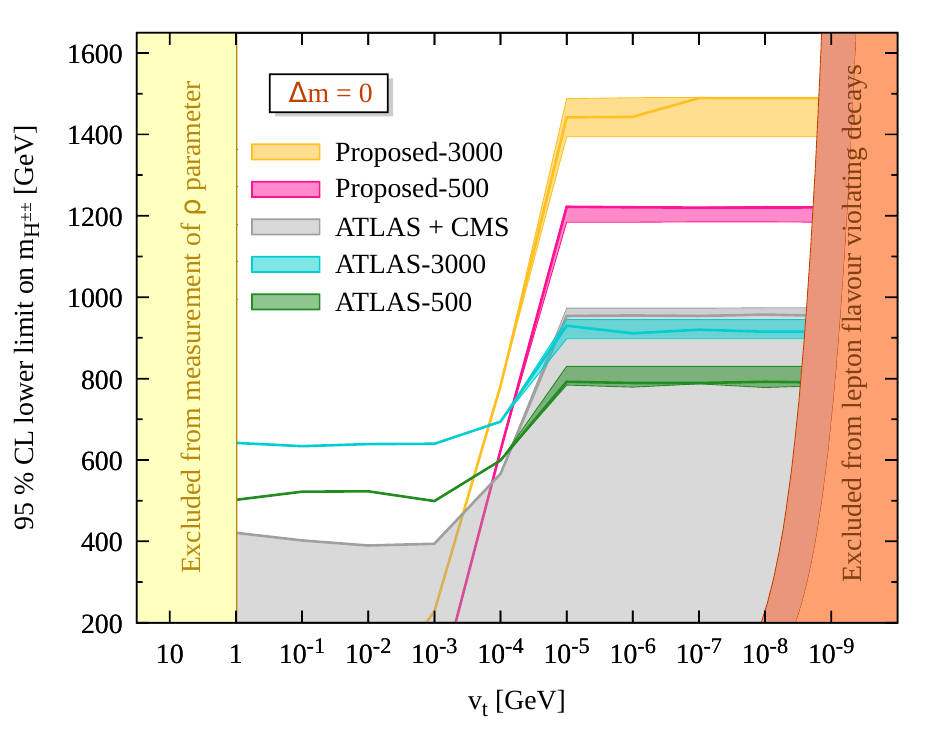}
\caption{Left: 95\% CL lower limits on $m_{H^{\pm \pm}}$ as a function of $v_t$ for $\Delta m=0$. The gray shaded region is excluded from the existing CMS and ATLAS combined search at the 13 TeV LHC. The bands resulted from different possible neutrino mass hypotheses. See text for details.}
\label{dm-0}
\end{figure}

The grey shaded region in Figure~\ref{dm-0} is excluded from the existing ATLAS and CMS combined search, see Section~\ref{sec:95cl0}. The regions below the green and cyan curves are expected to be excluded from the ATLAS search scaled at 500 and 3000 fb$^{-1}$ of luminosity, respectively. Our proposed search is expected to probe the regions below the goldenrod and pink curves, respectively, at 500 and 3000 fb$^{-1}$ of luminosity. For small (large) $v_t$, the future reach extends up to 1220 and 1490 (520 and 640) GeV, respectively, for 500 and 3000 fb$^{-1}$ of luminosity.

We consider both the NH and IH neutrino mass spectrum while varying the lightest neutrino mass in accordance with the bound from cosmology, $\sum_i m_i <0.12$ eV. The effect of different possible neutrino mass hypotheses on the limits are reflected as bands for small $v_t$ regions. This is because, for small $v_t$, the triplet-like scalars decay leptonically, and these decays are driven by the Yukawa couplings, which, in turn, are determined by the neutrino oscillation parameters up to $v_t$. For large $v_t$, the triplet-like scalars decay into diboson and hadrons, and these decays are independent of the Yukawa couplings and the neutrino oscillation parameters. The solid curves within the bands correspond to NH with $m_1=0.03$ eV.

\begin{widetext}

\begin{figure}[htb!]
\centering
\includegraphics[width=0.45\columnwidth]{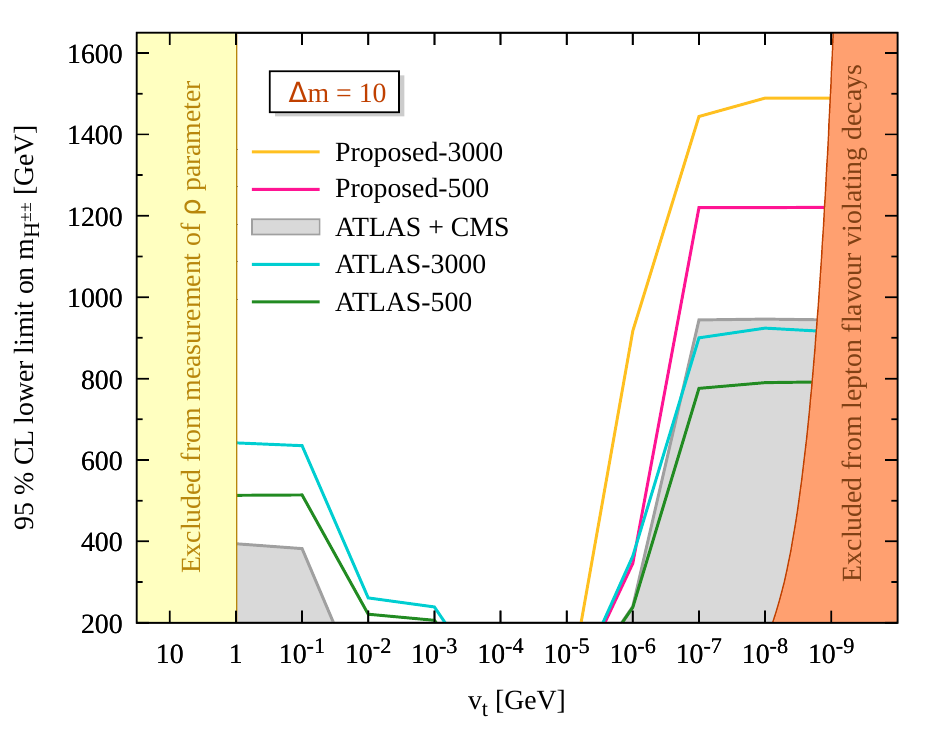}
\includegraphics[width=0.45\columnwidth]{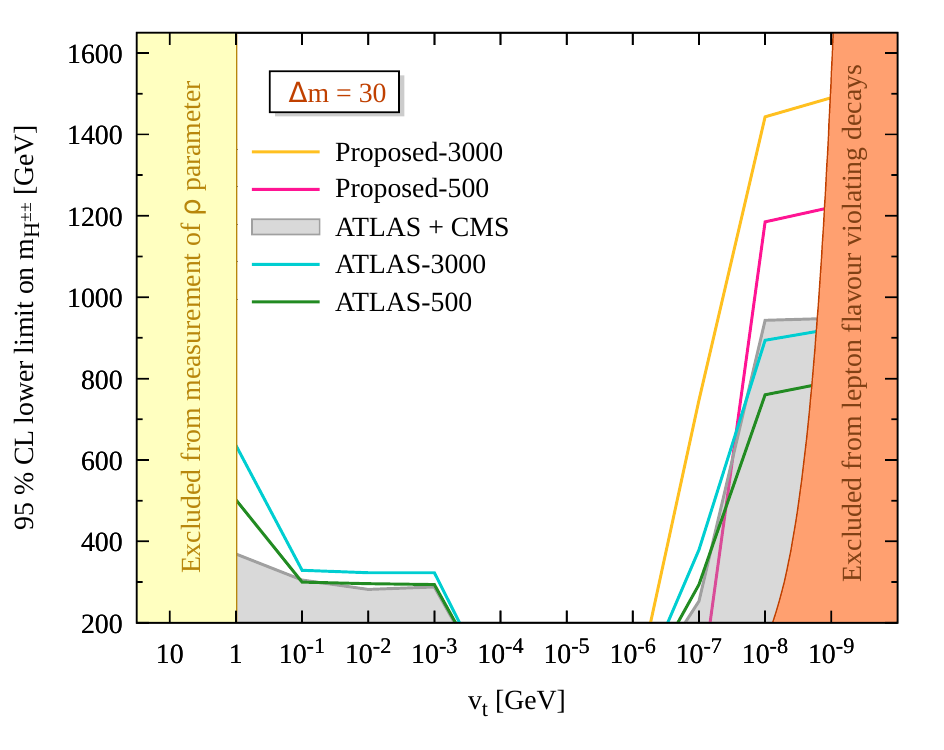}

\includegraphics[width=0.45\columnwidth]{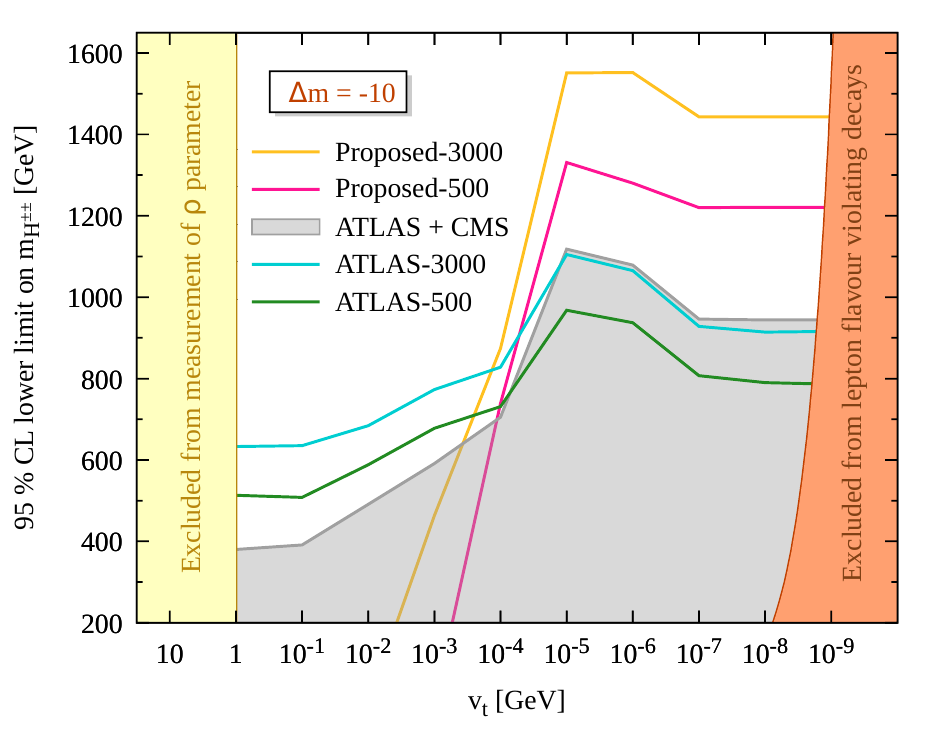}
\includegraphics[width=0.45\columnwidth]{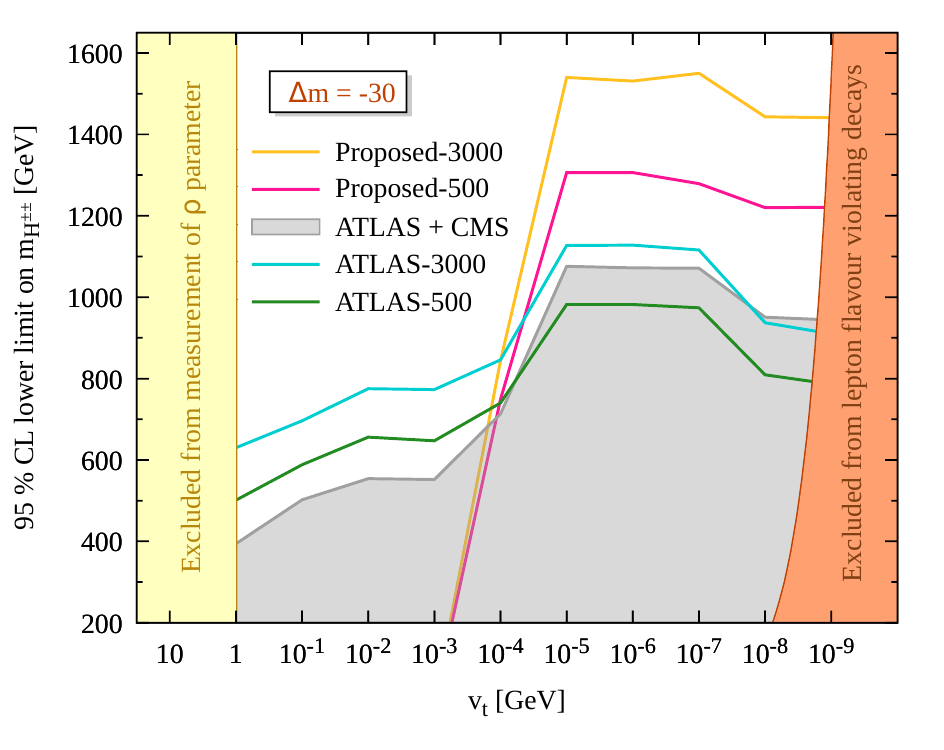}
\caption{95\% CL lower limits on $m_{H^{\pm \pm}}$ as a function of $v_t$ assuming NH with $m_1 = 0.03$ eV for $\Delta m=10$ GeV (top left), $\Delta m=30$ GeV (top right), $\Delta m=-10$ GeV (bottom left) and $\Delta m=-30$ GeV (bottom right). See text for details.}
\label{dm-all}
\end{figure}
\end{widetext}

The plots in Figure~\ref{dm-all} show 95\% CL future sensitivity of the LHC to probe as a function of $v_t$ assuming NH with $m_1 = 0.03$ eV for four different values of $\Delta m$ --- $\Delta m=10$ GeV (top left), $\Delta m=30$ GeV (top right), $\Delta m=-10$ GeV (bottom left) and $\Delta m=-30$ GeV (bottom right). The color codings are same as those in Figure~\ref{dm-0}. For $\Delta m=-10(-30)$ GeV and moderate $v_t$, the expected reach from our proposed search extends up to 1330(1310) and 1555(1550) GeV, respectively, at 500 and 3000 fb$^{-1}$ of luminiosity.

%For very large/small $v_t$, the limits for $\Delta m=\pm 10,\pm 30$ GeV are similar to those for $\Delta m=0$ case. This is because the cascade decays are yet to kick off for very large/small $v_t$. This makes the non-degenerate scenario identical to the degenerate one. For moderate $v_t$ and large enough $\Delta m$, the cascade decays kick off and swiftly dominates over the other decay modes. In positive scenario, $H^{\pm \pm}$ and $H^\pm$ decay into off-shell $W^\pm$'s and $H^0/A^0$, which further decay invisibly into neutrinos so that there are hardly visible objects in the final state. This makes such a scenario impossible to probe, even the monojet search by ATLAS \cite{Aad:2021egl} and the soft leptons search by CMS \cite{CMS:2021xji,Sirunyan:2018iwl} fails miserably to constrain this part of the parameter space. On the other hand, in negative scenarios, $H^\pm$ and $H^0/A^0$ decay into off-shell $W^\pm$'s and $H^{\pm \pm}$, thereby enhancing the effective production cross-section for $H^{\pm \pm}$. Therefore, in such a scenario, the limit is expected to be enhanced compared to the degenerate case. 

%=============================================================================
\subsection{\label{sec:95cl}Summary and outlook}
%=============================================================================
The type-II see-saw mechanism based on the annexation of the Standard Model by weak gauge triplet scalar field proffers a natural explanation for the very minuteness of neutrino masses. Because of the presence of the doubly charged scalar bosons and their illustrious signatures, a number of collider searches have been carried out at the LHC by CMS and ATLAS to look for the same. In view of the observations being consistent with the SM background expectations, these searches derived stringent limits with 95\% CL on $m_{H^{\pm \pm}}$. Most of these limits are derived in the context of simplified scenarios without reckoning the footprints of the low-energy neutrino parameters. Furthermore, these limits are often conservative as these searches do not incorporate all the production channels for the triplet-like scalars. Above all, in the non-degenerate scenario, the cascade decays are entitled to play a notable role in the phenomenology, thereby making the phenomenology for the non-degenerate scenario substantially contrasting than that for the degenerate one. Patently, the aforesaid limits are not befitting to the entire parameter space, rather valid only for a constrained parameter space of the model. Bearing this discussion in mind, we perform a comprehensive study for a wide range of the model parameter space parametrised by $v_t$, $\Delta m$ and $m_{H^{\pm \pm}}$. Considering all the Drell-Yan production mechanisms for the triplet-like scalars and taking into account the all-encompassing complexity of their decays, we derive the most stringent $95\%$ CL lower limit on $m_{H^{\pm \pm}}$ for a vast range of $v_t$-$\Delta m$ parameter space by implementing already existing direct collider searches by CMS and ATLAS. Further, we forecast future limits by extending the same ATLAS search at high-luminosity, and we propose a search strategy that yields improved limits for a part of the parameter space. To the extent of our apprehension, such a study of up-to-the-minute collider limits for a vast range of parameter space is still lacking. This work is intended to fill this gap.

For large (small) $v_t$ and $\Delta m=0$, doubly charged scalar up to masses 420(955) GeV are already excluded from the existing ATLAS and CMS combined search; this exclusion limit is beyond those from the previous LHC searches \cite{CMS:2017pet,Aaboud:2017qph,Aad:2021lzu} by approximately 50(200--230) GeV. For moderate $v_t$ and $\Delta m=-10$($-30$) GeV, the present exclusion limit extends up to 1115(1076) GeV. For large (small) $v_t$, the ATLAS (our proposed) search scaled at 3000 fb$^{-1}$ of luminosity is expected to probe them with masses below 640(1490) GeV for $\Delta m=0$. For $\Delta m=-10(-30)$ GeV and moderate $v_t$, the expected reach from our proposed search extends up to 1555(1550) GeV at 3000 fb$^{-1}$ of luminosity. Furthermore, we find that for moderate $v_t$ and large enough positive $\Delta m$, the LHC searches fail to constrain the triplet-like scalars insomuch as $H^0/A^0$ decays invisibly into neutrinos or into $h^0h^0,ZZ/h^0Z$ depending on the value of $v_t$. For $H^0/A^0$ decaying into neutrinos, there are hardly visible objects in the final state, so much as the monojet search by ATLAS \cite{Aad:2021egl} and the soft leptons search by CMS \cite{CMS:2021xji,Sirunyan:2018iwl} fall short in constraining this part of the parameter space. On the contrary, for $H^0/A^0$ decaying into $h^0h^0,ZZ/h^0Z$, the signal cross-section is small compared to the overwhelming background either from QCD jets or Drell-Yan processes. This makes such a scenario challenging to probe. In closing this section, we mention that $ee$ colliders could have better prospects for probing such a nightmare scenario which we left for future work.

\begin{widetext}

%=============================================================================
\appendix
\section{\label{sec:app} Partial decay widths of the triplet-like scalars}
%============================================================================= 

\noindent The decay rates for different possible decay modes of the triplet-like physical scalar bosons, $H^{\pm \pm}$, $H^\pm$, $H^0$ and $A^0$, are given in the following \cite{Rizzo:1980gz,Keung:1984hn,Cahn:1990xc,Djouadi:1997rp,Perez:2008ha,Aoki:2011pz}.

\noindent {\bf Decay rates of $H^{\pm\pm}$:}
\noindent The following decay rates for $H^{\pm \pm}$ can be procured: 
\begin{align}
&\Gamma(H^{\pm\pm} \to \ell_i^\pm \ell_j^\pm)
=\frac{\kappa |Y^\nu_{ij}|^2m_{H^{\pm \pm}}}{8\pi}\left(1-\frac{m_i^2}{m_{H^{\pm \pm}}^2}-\frac{m_j^2}{m_{H^{\pm \pm}}^2}\right)\left[\lambda\left(\frac{m_i^2}{m_{H^{\pm \pm}}^2},\frac{m_j^2}{m_{H^{\pm \pm}}^2}\right)\right]^{1/2},
\\
&\Gamma(H^{\pm\pm} \to W^\pm W^\pm)=\frac{g^4v_t^2m_{H^{\pm \pm}}^3}{64\pi m_W^4}\left(1-3\frac{m_W^2}{m_{H^{\pm \pm}}^2}+12\frac{m_W^4}{m_{H^{\pm \pm}}^4}\right)\beta\left(\frac{m_W^2}{m_{H^{\pm \pm}}^2}\right),
\\
&\Gamma(H^{\pm\pm} \to W^\pm W^{\pm *}) =\frac{3g^6v_t^2m_{H^{\pm \pm}}}{512\pi^3m_W^2}F\left(\frac{m_W^2}{m_{H^{\pm \pm}}^2}\right),
\\
&\Gamma(H^{\pm\pm} \to H^\pm W^\pm )=\frac{g^2m_{H^{\pm \pm}}^3\cos^2\beta_\pm}{16\pi m_W^2}\left[\lambda\left(\frac{m_W^2}{m_{H^{\pm \pm}}^2},\frac{m_{H^\pm}^2}{m_{H^{\pm \pm}}^2}\right)\right]^{3/2},
\\
&\Gamma(H^{\pm\pm} \to H^\pm W^{\pm *})=
\frac{9g^4m_{H^{\pm \pm}}\cos^2\beta_\pm}{128\pi^3} G\left(\frac{m_{H^\pm}^2}{m_{H^{\pm \pm}}^2},\frac{m_W^2}{m_{H^{\pm \pm}}^2}\right),
\end{align}
where $m_i$ with $i=e,\mu,\tau$ denotes the charged lepton mass and $\kappa=2(1)$ for $i\neq j(i=j)$. \\

\noindent {\bf Decay rates of $H^\pm$:}
\noindent The decay rates for $H^{\pm}$ can be obtained as
\begin{align}
&\Gamma(H^\pm \to q\bar{q}')=\frac{3m_{H^\pm}^3\sin^2\beta_\pm}{8\pi v_d^2} 
\left[\left(\frac{m_q^2}{m_{H^\pm}^2}+\frac{m_{q'}^2}{m_{H^\pm}^2}\right)\left(1-\frac{m_q^2}{m_{H^\pm}^2}-\frac{m_{q'}^2}{m_{H^\pm}^2}\right)-4\frac{m_q^2}{m_{H^\pm}^2}\frac{m_{q'}^2}{m_{H^\pm}^2}\right] \left[\lambda\left(\frac{m_q^2}{m_{H^\pm}^2},\frac{m_{q'}^2}{m_{H^\pm}^2}\right)\right]^{1/2},
\\
&\Gamma(H^\pm \to \ell_i^\pm\nu_j)=\frac{m_{H^\pm}}{8\pi v_d^2} \left( \delta_{ij}m_i^2\sin^2\beta_\pm +|Y^\nu_{ij}|^2v_d^2\cos^2\beta_\pm \right) \left(1-\frac{m_i^2}{m_{H^\pm}^2}\right)^2,
\\
&\Gamma(H^\pm \to W^\pm Z) =\frac{g^4v_t^2\cos^2\beta_\pm}{32\pi \cos^2\theta_w m_{H^\pm}}\left[\lambda\left(\frac{m_W^2}{m_{H^\pm}^2},\frac{m_Z^2}{m_{H^\pm}^2}\right)\right]^{1/2}\left[2+\frac{m_{H^\pm}^4}{4m_W^2m_Z^2}\left(1-\frac{m_W^2}{m_{H^\pm}^2}-\frac{m_Z^2}{m_{H^\pm}^2}\right)^2\right],
\\
&\Gamma(H^\pm \to W^\pm Z^*)=\frac{3g^6v_t^2\cos^2\beta_\pm}{1024\pi^3 \cos^2\theta_w m_{H^\pm}} H\left(\frac{m_W^2}{m_{H^\pm}^2},\frac{m_Z^2}{m_{H^\pm}^2}\right)\left(7-\frac{40}{3}\sin^2\theta_W+\frac{160}{9}\sin^4\theta_W\right),
\\
&\Gamma(H^\pm \to W^{\pm *} Z)=\frac{9g^6v_t^2\cos^2\beta_\pm}{512\pi^3 \cos^2\theta_w m_{H^\pm}} H\left(\frac{m_Z^2}{m_{H^\pm}^2},\frac{m_W^2}{m_{H^\pm}^2}\right),
\\
&\Gamma(H^\pm \to \hat{\varphi} W^\pm ) =\frac{g^2m_{H^\pm}^3}{64\pi m_W^2}\xi_{H^\pm W^\mp \hat{\varphi}}^2\left[\lambda\left(\frac{m_W^2}{m_{H^\pm}^2},\frac{m_{\hat{\varphi}}^2}{m_{H^\pm}^2}\right)\right]^{3/2},
\\
&\Gamma(H^\pm \to \hat{\varphi} W^{\pm *}) =\frac{9g^4m_{H^\pm}}{512\pi^3}\xi_{H^\pm W^\mp \hat{\varphi}}^2G\left(\frac{m_{\hat{\varphi}}^2}{m_{H^\pm}^2},\frac{m_W^2}{m_{H^\pm}^2}\right),
\\
&\Gamma(H^\pm \to H^{\pm \pm} W^\mp) =\frac{g^2m_{H^\pm}^3\cos^2\beta_\pm}{16\pi m_W^2} \left[\lambda\left(\frac{m_W^2}{m_{H^\pm}^2},\frac{m_{H^{\pm \pm}}^2}{m_{H^\pm}^2}\right)\right]^{3/2}
\\
&\Gamma(H^\pm \to H^{\pm \pm} W^{\mp *})=
\frac{9g^4m_{H^\pm}\cos^2\beta_\pm}{128\pi^3} G\left(\frac{m_{H^{\pm \pm}}^2}{m_{H^\pm}^2},\frac{m_W^2}{m_{H^\pm}^2}\right),
\end{align}

\noindent where $\xi_{H^\pm W^\mp \hat{\varphi}}=\cos\alpha\sin\beta_\pm-\sqrt{2}\sin\alpha\cos\beta_\pm,~ \sin\alpha\sin\beta_\pm+\sqrt{2}\cos\alpha\cos\beta_\pm,~ \sin\beta_0\sin\beta_\pm+\sqrt{2}\cos\beta_0\cos\beta_\pm $, respectively, for $\hat{\varphi}= h^0,H^0$ and $A^0$; $\theta_W$ is the weak mixing angle.\\

\noindent {\bf Decay rates of $H^0$:}
The decay rates for $H^0$ are given by 
\begin{align}
&\Gamma(H^0\to f\bar{f})=\frac{N_c^fm_f^2m_{H^0}}{8\pi v_d^2}\sin^2\alpha\left[\beta\left(\frac{m_f^2}{m_{H^0}^2}\right)\right]^3,
\\
&\Gamma(H^0 \to \nu \nu)=\frac{\kappa m_{H^0} \cos^2\alpha}{8\pi}\sum_{i,j=1}^3|Y^\nu_{ij}|^2,
\\
&\Gamma(H^0\to W^+W^-)=\frac{g^4 m_{H^0}^3}{256\pi m_W^4}\left(v_d\sin\alpha -2v_t \cos\alpha\right)^2\left(1-4\frac{m_W^2}{m_{H^0}^2}+12\frac{m_W^4}{m_{H^0}^4}\right)\beta\left(\frac{m_W^2}{m_{H^0}^2}\right),
\\
&\Gamma(H^0\to ZZ)=\frac{g^4 m_{H^0}^3}{512\pi m_W^4}\left(v_d\sin\alpha -4v_t \cos\alpha\right)^2\left(1-4\frac{m_Z^2}{m_{H^0}^2}+12\frac{m_Z^4}{m_{H^0}^4}\right)\beta\left(\frac{m_Z^2}{m_{H^0}^2}\right),
\\
&\Gamma(H^0\to WW^*)=\frac{3 g^6 m_{H^0}}{2048\pi^3m_W^2}(v_d\sin\alpha-2v_t \cos\alpha)^2F\left(\frac{m_W^2}{m_{H^0}^2}\right),
\\
&\Gamma(H^0\to ZZ^*)=\frac{g^6 m_{H^0}}{8192\pi^3\cos^4\theta_w m_W^2}(v_d\sin\alpha-4v_\Delta \cos\alpha)^2\left(7-\frac{40}{3}\sin^2\theta_W+\frac{160}{9}\sin^4\theta_W\right)F\left(\frac{m_Z^2}{m_{H^0}^2}\right),
\\
&\Gamma(H^0\to h^0h^0)=\frac{\lambda_{H^0h^0h^0}^2 v_d^2}{8\pi m_{H^0}}\beta\left(\frac{m_{h^0}^2}{m_{H^0}^2}\right),
\\
&\Gamma(H^0 \to H^\pm W^\mp ) =\frac{g^2m_{H^0}^3}{64\pi m_W^2}\xi_{H^\pm W^\mp H^0}^2\left[\lambda\left(\frac{m_W^2}{m_{H^0}^2},\frac{m_{H^\pm}^2}{m_{H^0}^2}\right)\right]^{3/2},
\\
&\Gamma(H^0 \to H^\pm W^{\mp *}) =\frac{9g^4m_{H^0}}{512\pi^3}\xi_{H^\pm W^\mp H^0}^2G\left(\frac{m_{H^\pm}^2}{m_{H^0}^2},\frac{m_W^2}{m_{H^0}^2}\right),
\end{align}
\noindent where and $N_c^f$ is the color factor with $N_c^q=3$ and $N_c^\ell=1$, and $\lambda_{H^0h^0h^0}=\frac{1}{4v_d^3}\Big[2v_t\left\{-2m_\Delta^2+v_d^2(\lambda_1+\lambda_4)\right\} \cos^3\alpha+v_d^3\left\{-3\lambda+4(\lambda_1+\lambda_4)\right\} \cos^2\alpha\sin\alpha +4v_t\big\{2m_\Delta^2+v_d^2(3\lambda_2+3\lambda_3-\lambda_1-\lambda_4)\big\}\cos\alpha\sin^2\alpha-2v_d^3(\lambda_1+\lambda_4)\sin^3\alpha\Big]$.

\noindent The functions $\lambda(x,y)$, $\beta(x)$, $F(x)$, $G(x,y)$ and $H(x,y)$ has the following form:
\begin{align}
&\lambda(x,y)=(1-x-y)^2-4xy~,
\\
&\beta(x)=\sqrt{\lambda(x,x)}=\sqrt{1-4x}~,
\\
&F(x)=\frac{3(1-8x+20x^2)}{\sqrt{4x-1}}\cos^{-1}\left(\frac{3x-1}{2x^{3/2}}\right) - \frac{(1-x)(2-13x+47x^2)}{2x}-\frac{3}{2}(1-6x+4x^2)\log x~, 
\\
&G(x,y)=\frac{1}{12y}\Bigg[2\left(-1+x\right)^3-9\left(-1+x^2\right)y+6\left(-1+x\right)y^2 -6\left(1+x-y\right)y\sqrt{-\lambda(x,y)}\Bigg\{\tan^{-1}\left(\frac{1-x+y}{\sqrt{-\lambda(x,y)}}\right) \notag
\\
&+\tan^{-1}\left(\frac{1-x-y}{\sqrt{-\lambda(x,y)}}\right)\Bigg\}-3\left(1+\left(x-y\right)^2-2y\right)y\log x\Bigg]~,
\\
&H(x,y)=\frac{\tan^{-1}\left(\frac{1-x+y}{\sqrt{-\lambda(x,y)}}\right) +\tan^{-1}\left(\frac{1-x-y}{\sqrt{-\lambda(x,y)}}\right)}{4x \sqrt{-\lambda(x,y)}} \Big\{-3x^3+(9y+7)x^2-5(1-y)^2x+(1-y)^3\Big\} \notag
\\
&+\frac{1}{24xy}\Big\{(-1+x)(2+2x^2+6y^2-4x-9y+39xy)-3y(1-3x^2+y^2-4x-2y+6xy)\log x\Big\}~.
\end{align}

\noindent {\bf Decay rates of $A^0$:}
The decay rates for $A^0$ are given by 
\begin{align}
&\Gamma(A^0\to f\bar{f})=\frac{N_c^fm_f^2m_{A^0}}{8\pi v_d^2}\sin^2\beta_0~ \beta\left(\frac{m_f^2}{m_{A^0}^2}\right),
\\
&\Gamma(A^0 \to \nu \nu)=\frac{m_{A^0} \cos^2\beta_0}{8\pi}\sum_{i,j=1}^3|Y^\nu_{ij}|^2,
\\
&\Gamma(A^0 \to h^0Z)=\frac{g^2m_{A^0}^3}{64\pi m_W^2} (\cos\alpha \sin\beta_0 -2\sin\alpha \cos\beta_0)^2 \left[\lambda\left(\frac{m_{h^0}^2}{m_{A^0}^2},\frac{m_Z^2}{m_{A^0}^2}\right)\right]^{3/2}
\\
&\Gamma(A^0 \to h^0Z^*)=\frac{3g^4m_{A^0}}{1024\pi^3 \cos^4\theta_w} (\cos\alpha \sin\beta_0 -2\sin\alpha \cos\beta_0)^2 \left(7-\frac{40}{3}\sin^2\theta_W+\frac{160}{9}\sin^4\theta_W\right) G\left(\frac{m_{h^0}^2}{m_{A^0}^2},\frac{m_Z^2}{m_{A^0}^2}\right)
\\
&\Gamma(A^0 \to H^\pm W^\mp ) =\frac{g^2m_{A^0}^3}{64\pi m_W^2}\xi_{H^\pm W^\mp A^0}^2\left[\lambda\left(\frac{m_W^2}{m_{A^0}^2},\frac{m_{H^\pm}^2}{m_{A^0}^2}\right)\right]^{3/2},
\\
&\Gamma(A^0 \to H^\pm W^{\mp *}) =\frac{9g^4m_{A^0}}{512\pi^3}\xi_{H^\pm W^\mp A^0}^2G\left(\frac{m_{H^\pm}^2}{m_{A^0}^2},\frac{m_W^2}{m_{A^0}^2}\right).
\end{align}

\end{widetext}

\end{document}